\documentclass[iop,apj,revtex4]{emulateapj}
\pdfoutput=1
\usepackage[]{natbib}
\bibpunct{(}{)}{;}{a}{}{,}
\usepackage{epstopdf}
\usepackage{url}
\usepackage[colorlinks, citecolor=blue]{hyperref}
\usepackage{amsmath,gensymb,latexsym,amssymb}

\shorttitle{Total Gas Masses in $z\sim2$--$3$ Galaxies}
\shortauthors{Sharon et al.}
\submitted{Accepted for publication in ApJ 6/1/2016}
\begin{document}
\title{A Total Molecular Gas Mass Census in $z\sim2$--$3$ Star-forming Galaxies:\\ Low-$J$ CO Excitation Probes of Galaxies' Evolutionary States}


\author{Chelsea E. Sharon\altaffilmark{1}, Dominik A. Riechers\altaffilmark{1}, Jacqueline Hodge\altaffilmark{2}, Chris L. Carilli\altaffilmark{3,}\altaffilmark{4}, Fabian Walter\altaffilmark{5}, \newline Axel Wei\ss\altaffilmark{6}, Kirsten K. Knudsen\altaffilmark{7}, \& Jeff Wagg\altaffilmark{8,}\altaffilmark{9}}

\altaffiltext{1}{Department of Astronomy, Cornell University, Ithaca, NY, 14853}
\altaffiltext{2}{Leiden Observatory, Univeriteit Leiden, NL-2300 RA Leiden, Netherlands}
\altaffiltext{3}{National Radio Astronomy Observatory, Socorro, NM 87801, USA}
\altaffiltext{4}{Cavendish Astrophysics Group, University of Cambridge, Cambridge, CB3 0HE, UK }
\altaffiltext{5}{Max-Planck-Institut f\"ur Astronomie, K\"onigstuhl 17, D-69117 Heidelberg, Germany}
\altaffiltext{6}{Max-Planck-Institut f\"ur Radioastronomie, Auf dem H\"ugel 69, D-53121 Bonn, Germany}
\altaffiltext{7}{Department of Earth and Space Sciences, Chalmers University of Technology, Onsala Space Observatory, SE-439 92 Onsala, Sweden}
\altaffiltext{8}{Square Kilometre Array Organization, Jodrell Bank Observatory, Lower Withington, Macclesfield, Cheshire SK11 9DL, UK}
\altaffiltext{9}{Cavendish Laboratory, 19 J. J. Thomson Avenue, Cambridge CB3 0HE, UK}

\begin{abstract}
We present \mbox{CO(1--0)} observations obtained at the Karl G. Jansky Very Large Array (VLA) for 14 $z\sim2$ galaxies with existing \mbox{CO(3--2)} measurements, including 11 galaxies which contain active galactic nuclei (AGN) and three submillimeter galaxies (SMGs). We combine this sample with an additional 15 $z\sim2$ galaxies from the literature that have both \mbox{CO(1--0)} and \mbox{CO(3--2)} measurements in order to evaluate differences in CO excitation between SMGs and AGN host galaxies, measure the effects of CO excitation on the derived molecular gas properties of these populations, and to look for correlations between the molecular gas excitation and other physical parameters. With our expanded sample of \mbox{CO(3--2)}/\mbox{CO(1--0)} line ratio measurements, we do not find a statistically significant difference in the mean line ratio between SMGs and AGN host galaxies as found in the literature, instead finding $r_{3,1}=1.03\pm0.50$ for AGN host galaxies and $r_{3,1}=0.78\pm0.27$ for SMGs (or $r_{3,1}=0.90\pm0.40$ for both populations combined). We also do not measure a statistically significant difference between the distributions of the line ratios for these populations at the $p=0.05$ level, although this result is less robust. We find no excitation dependence on the index or offset of the integrated Schmidt-Kennicutt relation for the two CO lines, and obtain indices consistent with $N=1$ for the various sub-populations. However, including low-$z$ ``normal" galaxies increases our best-fit Schmidt-Kennicutt index to $N\sim1.2$. While we do not reproduce correlations between the CO line width and luminosity, we do reproduce correlations between CO excitation and star formation efficiency. 
\end{abstract}

\keywords{galaxies:active --- galaxies: high-redshift --- galaxies: ISM --- galaxies: starburst --- ISM: molecules}

\section{Introduction}
\smallskip

Understanding the interaction between the growth of galaxies and their central supermassive black holes has been a key question in the field of galaxy evolution. The $M_\text{BH}$-$\sigma_{\rm bulge}$ relation at $z=0$ \citep[e.\/g.\/,][]{ferrarese2000, gebhardt2000}, the concurrent peaks of both active galactic nuclei (AGN) and star formation activity at $z\sim2$ \citep[e.\/g.\/,][]{madau1996,cowie2003,lafranca2005,hopkins2006,bongiorno2007,reddy2009}, and observations suggesting departure from the Magorrian relation \citep{magorrian1998} at $z\gtrsim2$ \citep[e.\/g.\/,][]{walter2004, alexander2008,coppin2008,riechers2009a,kimball2015} imply close coordination between galaxy evolution and the growth of central supermassive black holes (although see \citet{kormendy2013} for a review). While high-$z$ quasars are generally rare, a significant fraction have far-infrared (FIR) luminosities $>10^{13}$\,$L_{\odot}$ \citep[e.\/g.\/,][]{wang2008,leipski2013,leipski2014} which are mostly due to high star formation rates ($>$1000\,$M_{\odot}$\,yr$^{-1}$) fed by molecular gas reservoirs on the order of  $10^{10}$--10$^{11}$\,$M_\odot$ \citep[e.\/g.\/,][and references therein]{carilli2013b}. Among FIR-bright galaxies near the peak of cosmic growth, submillimeter galaxies (SMGs; \citealt{casey2014} and references therein) are substantially more common than FIR-luminous quasars, but have comparable $L_{\rm FIR}$, star formation rates (SFRs), molecular gas masses, and dynamical masses \citep[e.\/g.\/,][]{genzel2003, tecza2004, tacconi2006, tacconi2008, riechers2008a, riechers2008b, ivison2010a, hainline2011,hodge2012}. Both SMGs and optically-selected AGN (quasars) are suspected to trace massive structure formation at high redshift since both populations have similar clustering properties \citep[e.\/g.\/,][]{blain2004,hickox2012}. SMGs, however, typically have $\sim1.5$--$2$ orders of magnitude less massive black holes ($\sim10^7$ vs.\ 10$^9$\,$M_{\odot}$; e.\/g.\/, \citealt{alexander2005,alexander2008}). This difference suggests that both high-$z$ populations deviate from the $M_{\rm BH}$--$M_{\star}$ relation for nearby spheroidal galaxies \citep[e.\/g.\/,][]{tremaine2002,marconi2003}, but in different directions \citep[e.\/g.\/,][]{alexander2008,riechers2008b,coppin2008}.

Given the many similarities in physical properties between SMGs and AGN, attempts have been made to fit both populations into a unified picture \citep[e.\/g.\/,][]{granato2001, somerville2008, bonfield2011} analogous to the ``classical" merger-driven ultra/luminous infrared galaxy (U/LIRG)-quasar-transition hypothesis at low redshifts \citep[e.\/g.\/,][]{sanders1988}. However, evidence has been mixed regarding the frequency of mergers within the SMG population \citep[e.\/g.\/,][]{riechers2008a, riechers2008b, tacconi2008, dave2010, hayward2011, hayward2013, swinbank2011, hodge2012, hodge2013, hodge2015, aguirre2013, riechers2013b, riechers2014b, debreuck2014, sharon2015,narayanan2015}. Independent of merger state, theoretical studies have found it necessary to invoke AGN-powered feedback to end the starbursts of massive galaxies \citep[e.\/g.\/,][]{somerville2008}. However, both the relative importance of AGN feedback compared to stellar feedback \citep[e.\/g.\/,][]{bouche2010,dave2011a,dave2012,shetty2012,lilly2013,cicone2014} and exact feedback mechanism for AGN (outflows vs.~accretion suppression; e.\/g.\/, \citealt{croton2006,hopkins2006,gabor2011,cicone2014}) are still debated. In addition, recent work suggests that AGN may enhance star formation in galaxies' centers \citep[e.\/g.\/,][]{stacey2010,ishibashi2012,silk2013}. 

If AGN directly influence the star formation history of galaxies, their effects should be measured in the molecular interstellar medium (ISM) that fuels star formation. At low-$z$, molecular outflows have been observed in luminous AGN \citep[e.\/g.\/,][]{feruglio2010,cicone2014}, and very high excitation CO lines have been observed in some AGN \citep[e.\/g.\/,][]{weiss2007, hailey2012, spinoglio2012b}, but these components represent a small fraction in mass of the total molecular gas reservoirs in star-forming galaxies. 

Promising evidence for AGN affecting the bulk of galaxies' molecular ISM has been found in the initial \mbox{CO(1--0)} samples of $z\sim2$ SMGs and AGN host galaxies. The relatively recent availability of Ka-band receivers on the Karl G. Jansky Very Large Array (VLA) and the Robert C. Byrd Green Bank Telescope (GBT) has led to a small but growing number of CO detections in $z\sim2$--$3$ galaxies that are complete down to the lowest rotational line, \mbox{CO(1--0)}, which is crucial for tracing the coldest gas components (e.\/g.\/, \citealt{swinbank2010a, harris2010, harris2012, ivison2011, ivison2013, danielson2011, riechers2011c, riechers2011d, riechers2011e, ivison2012, thomson2012, fu2013, thomson2015, sharon2013, sharon2015}; for \mbox{CO(1--0)} detections at other redshift using other bands, see \citealt{carilli2002, greve2003, riechers2006, riechers2009a, riechers2011e, riechers2013b, hainline2006, dannerbauer2009, carilli2010, aravena2010a, aravena2014, aravena2016, emonts2011}). Observations revealed that SMGs appear to have a common \mbox{$L^\prime_{\rm CO(3-2)}$}/\mbox{$L^\prime_{\rm CO(1-0)}$} line luminosity ratio of $r_{3,1}\approx0.6$, indicative of a multi-phase molecular ISM including a previously unaccounted for cold gas reservoir (e.\/g.\/, \citealt{swinbank2010a, harris2010, ivison2011, danielson2011, thomson2012, bothwell2013}; cf.~\citealt{riechers2011d, sharon2013, sharon2015}). In contrast, AGN host galaxies at similar redshifts appear to have an entirely different line ratio, $r_{3,1}=1$, which could be indicative of only warm single-phase gas \citep{riechers2011f, thomson2012}. While sub-unity values of $r_{3,1}$ are also expected for subthermally excited gas or cold gas where there is a difference between the Planck and Rayleigh-Jeans temperatures, commonly observed higher-$J$ CO emission from systems with $r_{3,1}<1$ disfavor subthermal and low temperature single phase interpretations of low $r_{3,1}$ \citep[e.\/g.\/,][]{harris2010}. The apparent difference in \mbox{CO(3--2)}/\mbox{CO(1--0)} line ratios for SMGs and AGN host galaxies can be interpreted as supporting an evolutionary connection between the two populations such as high $r_{3,1}$ occurring in late stage mergers where the molecular gas has been funneled by gravitational torques to central high-density and/or AGN-dominated region, or high $r_{3,1}$ occurring once the bulk of the molecular gas reservoir has been ejected by AGN (and/or stellar) feedback. However, this single line ratio does not distinguish between ``direct" SMG-AGN evolutionary connections where high $r_{3,1}$ values are due to the influence of the central AGN, or ``indirect" models where galaxies' changing $r_{3,1}$ and AGN activity mutually track some other evolutionary process.

The molecular gas excitation conditions and their differences between galaxy populations has important consequences for characterizations of high-$z$ sources since the lines ideally used to trace the molecular gas mass, and thus star-forming potential, of galaxies are dependent on the gas physical conditions. The sub-unity $r_{3,1}\sim0.6$ found in (most) SMGs to-date means that molecular gas mass estimates based on the \mbox{CO(3--2)} line luminosity would be off by a factor of $\sim2$ without an appropriate excitation correction. If the average excitation differs between galaxy populations, incorrect assumptions about the excitation would bias comparisons between those populations, potentially affecting understanding of their evolutionary connections. Incorrect gas masses could also introduce biases in the observed Schmidt-Kennicutt relation \citep{schmidt1959, kennicutt1989}, an empirical correlation which probes the physical process responsible for star formation (e.\/g.\/, \citealt{bigiel2008} and references therein) and is frequently used as input for numerical simulations of galaxy formation \citep[e.\/g.\/,][]{springel2003, narayanan2008b, somerville2008, juneau2009}. Despite numerous studies of the Schmidt-Kennicutt relation, differences in methods (integrated vs.~spatially resolved studies; e.\/g.\/, \citealt{young1986, solomon1988, kennicutt1989, buat1989, kennicutt1998, gao2004, bouche2007, bigiel2008, krumholz2009, bigiel2010, daddi2010, genzel2010, wei2010, tacconi2013}), assumptions (particularly gas mass conversion factors; e.\/g.\/, \citealt{bigiel2008, daddi2010, genzel2010}), gas and star formation rate (SFR) tracers \citep[e.\/g.\/,][]{gao2004, narayanan2005, gracia-carpio2008, bussmann2008, iono2009, juneau2009, kennicutt2012}, and galaxy populations \citep[e.\/g.\/][]{gao2004, daddi2010, genzel2010, tacconi2013} leads to significant uncertainties in the relation's characteristics (such as the index of the power law) and interpretation. Different molecular emission lines are sensitive to different density regimes in the ISM \citep{krumholz2007, narayanan2008d, narayanan2011}, making the observed index of the Schmidt-Kennicutt relation dependent on the gas physical conditions. How the Schmidt-Kennicutt index varies between gas tracers with different critical densities therefore probes the underlying \emph{volumetric} star formation relation (the Schmidt law) set by the physics of star formation. The fidelity of different molecular gas tracers in capturing the Schmidt-Kennicutt relation is particularly important for comparisons between sources at different redshifts since atmospheric and instrumentation limitations have caused the molecular gas in most $z\sim2$--$3$ sources to be characterized with mid-$J$ CO lines ($J_{\rm upper}=3$, 4, 5) whereas the molecular gas in local galaxies is mostly characterized using \mbox{CO(1--0)} or \mbox{CO(2--1)}. While several studies have examined the change in the power law index of the Schmidt-Kennicutt relation with different CO lines with mixed results \citep[e.\/g.\/,][]{yao2003, bayet2009, greve2014, liu2015,kamenetzky2015}, there has been no systematic study on the effects of excitation on the Schmidt-Kenicutt relation at high redshift to date.

Here we present a systematic study of \mbox{CO(1--0)} emission for nearly all $z\sim2$--$3$ SMGs and AGN host galaxies with existing \mbox{CO(3--2)} detections at the time of the observations. We present new observations of the \mbox{CO(1--0)} line obtained at the Karl G. Jansky Very Large Array (VLA) for 14 objects, including 13 successful detections and one new upper limit. We describe the observations and discuss the sample in Sections~\ref{sec:obs} and \ref{sec:results}, respectively. In Section~\ref{sec:analysis}, we determine if the previously observed dichotomy in $r_{3,1}$ values between SMGs and AGN host galaxies hold up for the expanded sample, evaluate the effects of the excitation on the characterization of these galaxies' star formation properties (e.\/g.\/, the Schmidt-Kennicutt relation), and evaluate evidence for a SMG-quasar transition among $z\sim2$--$3$ galaxies. Our results are summarized in Section~\ref{sec:concl}.

We assume the flat WMAP9+BAO+$H_0$ mean $\Lambda$CDM cosmology throughout this paper, with $\Omega_\Lambda=0.712$ and $H_0=69.33\,{\rm km\,s^{-1}\,Mpc^{-1}}$ \citep{hinshaw2013}.

\section{Observations and Data Reduction}
\label{sec:obs}

The observed sample was selected from all known $z\sim2$--$3$ SMGs and AGN host galaxies with existing \mbox{CO(3--2)} measurements at the time of the observations, excluding those with existing \mbox{CO(1--0)} measurements, those being observed in \mbox{CO(1--0)} as part of other observing programs, and those with prohibitively long integration times (four sources). Eight galaxies from this observational program have already been published  \citep{riechers2011e,riechers2011f,riechers2011c}. Here we present an additional 14 objects: 11 AGN host galaxies (six lensed and five unlensed) and three SMGs (two lensed and one unlensed). The galaxies in our new observations have redshifts between $2.0590\leq z\leq3.2399$ and magnification factors as high as $\mu=173$. For our final analysis, we include 15 additional sources from the literature that have both \mbox{CO(1--0)} and \mbox{CO(3--2)} measurements: three lensed AGN host galaxies and 12 SMGs (eight lensed and four unlensed). These additional sources have a similar redshift range to that of our new sample, $2.2020\leq z\leq3.1999$, and have a maximum magnification of $\mu=45$. The (magnification-corrected) FIR luminosities of the complete sample (adopted from the literature and listed in Table~\ref{tab:measure}), including new and literature \mbox{CO(1--0)}-detected sources, are large, and in the regime of U/LIRGs and hyper-luminous infrared galaxies (HyLIRGs) with $11.2\leq\log(L_{\rm FIR}/L_\sun)\leq13.3$\footnote{We use FIR luminosities as reported in the literature, which are calculated using a variety of different methods in addition to using different wavelength regimes to define FIR. Since simple corrections between wavelength assumptions require either arbitrary choices of dust temperatures and modified black body indices or model SEDs (spectral energy distributions), and produce corrections that are within the scatter of the measurement techniques, we do not correct for these differences here.}.

The classification of these sources as SMGs or AGN host galaxies are entirely historical and uses their previous categorizations from the literature. We use the literature classifications in order to compare our results with previous work that studies the \mbox{CO(3--2)}/\mbox{CO(1--0)} line ratio differences between SMGs and AGN host galaxies using the same literature-based classifications. The SMGs and AGN have comparable FIR luminosities that are mostly ULIRG-like ($12.0\leq\log(L_{\rm FIR}/L_\sun)<13.0$). Both categories have two LIRGs ($11.0\leq\log(L_{\rm FIR}/L_\sun)<12.0$) each, and there are two SMG HyLIRGs ($\log(L_{\rm FIR}/L_\sun)\geq13.0$) and one AGN host galaxy HyLIRG. The AGN in this sample are either optically-selected quasars and radio-loud AGN, with the exception of F10214+4724.

Our observing program was carried out at the VLA over multiple observing periods from fall 2009 until fall 2015 (programs AR708, 11B-025, 11B-151, 12A-009, and 15B-329). Since this period includes the commissioning of the Ka-band receivers and the Wideband Interferometric Digital Architecture (WIDAR) correlator, both the correlator setups and number of available antennas varied between observations; the observations are summarized in Table~\ref{tab:obs}. Most observations were carried out in the D configuration (minimum and maximum baselines of $40.0\,{\rm m}$ and $1.03\,{\rm km}$, respectively), but some higher resolution C configuration observations were also taken (minimum and maximum baselines of $78.1\,{\rm m}$ and $3.39\,{\rm km}$, respectively); the number of antennas, array configuration, and resulting synthesized beam sizes are listed in Table~\ref{tab:obs} and accounts for antennas that still lacked Ka-band receivers at the time of the observations or antennas that were excluded from the analysis due to other technical problems.

For all observations except HS\,1611+4719 (the first object observed) and data obtained in fall 2015, we obtained the full polarization information with $2\,{\rm MHz}$ spectral resolution. For the fall 2015 data, we observed in dual polarization mode with $1\,{\rm MHz}$ channel widths\footnote{Although the second track from fall 2015 on B9138+666 has $2\,{\rm MHz}$/full polarization for the high frequency intermediate frequency (IF) channel pair and $1\,{\rm MHz}$/dual polarization for the lower frequency IF pair.}. The total contiguous bandwidth in the two intermediate frequency (IF) channel pairs was either $128\,{\rm MHz}$ or $1024\,{\rm MHz}$ each. For the wider bandwidth observations, there are eight sub-bands per IF pair, each with 64 channels ($128\,{\rm MHz}$ bandwidth) and no frequency overlap between sub-bands. The two IF pairs were spaced $1\,{\rm GHz}$ apart for all observations where the restrictions in Ka-band IF tuning allowed it. For the narrower bandwidth observations, the two IF pairs were tuned to overlap by $4\,{\rm MHz}$ (two channels) for all observations where the IF-pair tuning restrictions allowed it. For HS\,1611+4719, the channel size is $3.125\,{\rm MHz}$ and there are seven channels per IF pair (total bandwidth of $21.875\,{\rm MHz}$ per IF pair). The two IF pairs were tuned to provide zero frequency overlap and dual polarization. For the four cases where restrictions in the IF pair tuning did not allow a $1\,{\rm GHz}$ separation, the higher frequency IF pair was tuned to either the $146.969\,{\rm GHz}$ \mbox{CS(3--2)} line (for RX\,J0911+0551, J22174+0015, and B1359+154) or the $130.269\,{\rm GHz}$ \mbox{SiO(3--2)} line (for J04135+10277).

Observations alternated between the target source (integration times of $\sim3$--$9$ minutes) and a nearby quasar ($\sim1$ minute) that is used for phase and secondary flux calibration. One of 3C286, 3C138, 3C147, or 3C48 was observed for bandpass and primary flux calibrations. For three tracks the flux calibrator data was not recorded or the data are bad. In those cases we used the phase calibrator for flux calibration, manually setting the flux to the model values determined from whichever other track was observed nearest in time. While the observed quasars may be variable, for these particular sources we found that the flux densities at the chosen frequency varied between observing tracks by less than $10\%$, which is within the standard assumed value of the flux calibration uncertainty for the Ka-band. In order to check that our assumed fluxes are accurate, we verified that the noise in the images of each individual track and in the combined science image was appropriate using the radiometer equation. Pointing checks were carried out approximately once an hour on either the phase or flux calibrator. We set initial target integration times to achieve $5$--$7\sigma$ detections of the integrated lines as predicted by the \mbox{CO(3--2)} fluxes assuming a single thermalized gas phase. Although we adjusted integration times if we obtained adequate signal-to-noise with less time or if the sources were undetected, observations for some sources are incomplete and we have not hit the target S/N in all cases. We used three second correlator integration times per visibility data point for all observations. However, during the observations for nine of the 14 sources, the WIDAR correlator malfunctioned and recorded data from only the first second of each three second integration (for a factor of $\sqrt{3}$ reduction in S/N)\footnote{\href{http://www.vla.nrao.edu/astro/archive/issues/\#1009}{http://www.vla.nrao.edu/astro/archive/issues/\#1009}}. In order to salvage those data, we modified the time-stamps and exposure times in the measurement sets of the effected tracks using Common Astronomy Software Application (CASA)\footnote{\href{http://casa.nrao.edu}{http://www.casa.nrao.edu}} to accurately reflect how the data were obtained, although adjusting the timestamps does not appear to affect the resulting maps.

Calibrations were done using CASA version 4.2.2 and using the EVLA pipeline prototype version 1.3.1 (without Hanning smoothing). We slightly modified the pipeline so that only the first and last channel of spectral windows were flagged out (except for B1938+666). After running the pipeline, the calibrated data was visually inspected and in some cases additional antennas were flagged out (reflected in Table~\ref{tab:obs}). The pipeline script did not work for HS\,1611+4719, so that data was processed by hand in the same CASA version. All maps were made using natural weighting in CASA. We also performed self-calibration on the continuum emission for B1938+666 and MG 0414+0534, and baseline-based gain calibration for MG 0414+0534. For sources with signficant continuum emission (B1938+666, HE\,0230--2130, RX\,J1249--0559, HE\,1104+1805, J1543+5359, VCV\,J1409+5628, MG\,0414+0534, and B1359+154) we performed $uv$-plane continuum subtraction prior to the analysis of the integrated line maps. All other sources either had sufficiently weak or undetected continuum emission that did not require removal for analysis of the line maps.

\begin{turnpage} 
\begin{deluxetable}{lccccccccccc}
\tabletypesize{\tiny}
\tablewidth{0pt}
\tablecaption{Details of Observations \label{tab:obs}}
\tablehead{{Source}	& {Date} & {$t_{\rm int}$} & $N_{\rm ant}$ & {Config.}\tablenotemark{a} & {Beam}\tablenotemark{b} & {Bandwidth per} & {Band centers} & {Phase} & {Flux} & {$S_\nu$\tablenotemark{c}} & {$\nu$\tablenotemark{c}} \\
{} & {d/m/y} & {(hr)} & {} & {} & {} & IF pair (MHz) & (GHz) & {Calibrator} & {Calibrator} & (Jy) & (GHz) }
\startdata
B1938+666 & 9/10/2011 & 0.30\tablenotemark{d} & 26 & D & $2.^{\prime\prime}77\times2.^{\prime\prime}15$, $-54.03\degree$ & 1024 & 36.619, 37.619 & J2006+6424 & 3C286 & 0.5462 & 37.1710 \\
{} & 6/11/2015 & 0.90 & 24 & D & ($2.^{\prime\prime}81\times2.^{\prime\prime}17$, $-54.26\degree$) & {} & 36.7466, 37.7466 & {} & 3C48 & 0.5372 & 36.2987 \\
{} & 6/11/2015 & 0.90 & 24 & D & {} & {} & {} & {} & {} & 0.5285 & {} \\
\hline
HS 1002+4400 & 26/4/2010 & 1.74 & 19 & D & $2.^{\prime\prime}39\times1.^{\prime\prime}87$, $12.15\degree$ & 128 & 37.1043, 37.2283 & J0948+4039 & 3C286 & 1.0879 & 37.1043 \\
\hline
HE 0230-2130 & 8/10/2011 & 0.19\tablenotemark{d} & 21 & D & $3.^{\prime\prime}05\times1.^{\prime\prime}93$, $3.12\degree$ & 1024 & 35.469, 36.469 & J0204-1701 & 3C48 & 1.2141 & 35.0210 \\
{} & 29/10/2015 & 0.49 & 26 & D & ($3.^{\prime\prime}25\times2.^{\prime\prime}00$, $1.76\degree$) & {} & 35.469, 36.469 & {} & 3C147 & 1.6115 & 35.0218 \\
{} & 30/11/2015 & 0.49 & 21 & D & {} & {} & {} & {} & {} & 1.6192 & 35.0213 \\
{} & 9/12/2015 & 0.49 & 23 & D & {} & {} & {} & {} & {} & 1.5878 & 35.0212 \\
\hline
RX J1249-0559 & 3/1/2012 & 2.33 & 26 & D & $3.^{\prime\prime}27\times2.^{\prime\prime}24$, $-10.90\degree$ & 1024 & 34.565, 35.565 & J1246-0730 & 3C286 & 1.1484 & 34.1170 \\
\hline
HE 1104-1805 & 15/10/2011  & 0.44\tablenotemark{d} & 20 & D & $1.^{\prime\prime}69\times1.^{\prime\prime}12$, $-16.09\degree$ & 1024 & 33.634, 34.634 & J1048-1909 & 3C286 & 1.5576 & 33.1860 \\
{} & 26/4/2012 & 1.6 & 23 & C & ($1.^{\prime\prime}58\times1.^{\prime\prime}03$, $-8.38\degree$) &  &  & & & 1.7176 &  \\
{} & 3/12/2015 & 1.2 & 22 & D & {} & {} & 34.634, 35.634 & {} & {} & 1.2751 & 35.1863 \\
\hline
J1543+5359 & 6/4/2010 & 0.38 & 16 & D & $1.^{\prime\prime}08\times0.^{\prime\prime}97$, $26.72\degree$ & 128 & 34.1451, 34.2691 & J1549+5038 & 3C286 & 0.7466 & 34.1451 \\
{} & 31/12/2010 & 0.92 & 25 & C & ($1.^{\prime\prime}09\times0.^{\prime\prime}98$, $26.15\degree$) &  &  &  & & 0.5407 &  \\
\hline
HS 1611+4719 & 24/10/2009 & 3.07 & 14 & D & $2.^{\prime\prime}65\times2.^{\prime\prime}43$, $-27.61\degree$ & 21.875 & 33.9328, 33.9547 & J1620+4901 & 3C286 & 0.3180 & 33.9328 \\
{} & {} &{} & {} & {} & ($2.^{\prime\prime}51\times2.^{\prime\prime}44$, $-13.35\degree$) & {} & {} & {} & {} & {} & {} \\
\hline
J044307+0210 & 31/12/2010 & 1.25 & 24 & C & $1.^{\prime\prime}53\times1.^{\prime\prime}05$, $-12.35\degree$ & 128 & 32.7882, 32.9122 & J0442-0017 & 3C147 & 0.8889 & 32.7882 \\
{} & 19/10/2011 & 0.65\tablenotemark{d} & 25 & D & ($2.^{\prime\prime}16\times1.^{\prime\prime}60$, $-14.83\degree$) & 1024 & 32.914, 33.914 &  & 3C138 & 1.2130 & 33.4660 \\
{} & 27/10/2015 & 0.85 & 24 & D & {} & {} & {} & {} & 3C147 & 0.8306 & 32.4673 \\
\hline
VCV J1409+5628 & 7/7/2010 & 0.56 & 20 & D & $1.^{\prime\prime}74\times1.^{\prime\prime}42$, $57.33\degree$ & 128 & 32.1089, 32.2329 & J1419+5423 & 3C286 & 0.9828 & 32.1089 \\
{} & 16/10/2011 & 0.29\tablenotemark{d} & 25 & D & ($1.^{\prime\prime}62\times1.^{\prime\prime}33$, $60.43\degree$) & 1024 & 32.234, 33.234 &  &  & 0.3474\tablenotemark{e} & 32.7860 \\
{} & 5/12/2011 & 0.90 & 26 & D&  &  &  &  &  & 0.3368 & \\
{} & 29/1/2012 & 0.49 & 25 & C &  &  &  &  &  & 0.3580 &  \\
\hline
MG 0414+0534 & 24/10/2011 & 0.31\tablenotemark{d} & 26 & D & $3.^{\prime\prime}06\times2.^{\prime\prime}50$, $20.05\degree$ & 1024 & 31.741, 32.741 & J0433+0521 & 3C138 & 1.5478 & 32.2930 \\
{} & 28/10/2015 & 0.47 & 25 & D & ($3.^{\prime\prime}10\times2.^{\prime\prime}58$, $18.42\degree$) & {} & {} & {} & 3C147 & 2.9426 & 32.2942 \\
\hline
RX J0911+0551 & 8/2/2012 & 2.7 & 25 & C & $0.^{\prime\prime}88\times0.^{\prime\prime}70$, $38.34\degree$ & 1024 & 30.3025, 38.5248 & J0909+0121 & 3C286 & 1.6651 & 38.0768 \\
{} & 12/2/2012 & 2.7 & 26 & C &  &  &  &  &  & 1.4057 &  \\
\hline
{J04135+10277} & 24/8/2010 & 0.56 & 22 & D & $1.^{\prime\prime}12\times1.^{\prime\prime}01$, $11.92\degree$ & 128 & 29.9717, 32.1100 & J0409+1217 & 3C147 & 0.2258 & 29.9717 \\
{} & 13/11/2011 & 0.30\tablenotemark{d} & 26 & D & ($1.^{\prime\prime}01\times0.^{\prime\prime}90$, $11.30\degree$) & 1024 & 30.036, 33.807 &  & 3C138 & 0.3001 & 33.3590 \\
{} & 17/2/2012 & 1.32 & 26 & C & & & & & & 0.3021&  \\
\hline
J22174+0015 & 15/7/2010 & 3.92 & 21 & D & $3.^{\prime\prime}26\times2.^{\prime\prime}74$, $-5.25\degree$ & 128 & 28.1218, 32.1100 & J2218-0335 & 3C48 & 1.1649 & 28.1218 \\
{} & 19/10/2011 & 0.77\tablenotemark{d} & 24 & D & ($2.^{\prime\prime}86\times2.^{\prime\prime}48$, $-4.17\degree$) & 1024 & 28.184, 35.791 &  &  & 1.1714 & 35.3430 \\
{} & 21/10/2011 & 0.77\tablenotemark{d} & 25 & D & &  &  &  &  & 1.1505 &  \\
{} & 6/11/2015 & 0.51 & 22 & D & {} & {} & 28.1858, 35.9188 & {} & {} & 1.2462 & 27.7375 \\
{} & 8/11/2015 & 0.72 & 23 & D & {} & {} & {} & {} & {} & 1.2323 & {} \\
{} & 12/11/2015 & 0.52 & 24 & D & {} & {} & {} & {} & {} & 1.2137 & 27.7374 \\
{} & 13/11/2015 & 0.51 & 24 & D & {} & {} & {} & {} & {} & 1.2137 & {} \\
{} & 3/12/2015 & 0.51 & 25 & D & {} & {} & {} & {} & {} & 1.1867 & {} \\
\hline
B1359+154 & 5/7/2010 & 1.41 & 21 & D & $1.^{\prime\prime}73\times1.^{\prime\prime}46$, $25.38\degree$ & 128 & 27.2026, 32.1100 & J1415+1320 & 3C286 & 0.5424\tablenotemark{e} & 27.2026\\
{} & 25/8/2010 & 1.41 & 24 & D & ($1.^{\prime\prime}23\times1.^{\prime\prime}03$, $31.19\degree$) &  &  &  &  & 0.5354 & 27.2026 \\
{} & 15/10/2011 & 0.31\tablenotemark{d} & 25 & D & & 1024 & 27.251, 34.599 &  &  & 0.5390\tablenotemark{e} & 34.2790 \\
{} & 16/1/2012 & 0.51 & 23 & C & &  &  &  &  & 0.5378 & 34.2790
\enddata
\tablecomments{Columns with missing data denote repeat values that are unchanged from the previous observing track.}
\tablenotetext{a}{For D configuration observations the minimum and maximum baselines are generally 40.0\,m and 1.03\,km, respectively. For C configurations observations the minimum and maximum baselines are generally 78.0\,m and 3.39\,km, respectively. The exceptions are: the minimum baseline for the HS\,1002+4400 track is 44.8\,m, the minimum baselines for the first and third HE\,0230--2130 tracks are 45.1\,m, the minimum baselines for the C configuration tracks of J1543+5359 and J044307+0210 are 78.1\,m, the maximum baselines for the J044307+0210 and J22174+0015 tracks observed on 19/10/2011 are 1.49\,km, and the maximum baselines for third J044307+0210 track and the J22174+0015 tracks observed on 8/11/2015 and 3/12/2015 are 971\,m. These changes in baseline lengths reflect observing tracks where specific antennas were removed from the array or flagged out during data reduction.}
\tablenotetext{b}{The beam FWHM and position angle for the integrated line map (listed first) and the continuum map (listed second/in parentheses if different from the line map).}
\tablenotetext{c}{Phase calibrator model flux at listed frequency.}
\tablenotetext{d}{These tracks were affected by the correlator error where only the first second of the three second integration times were recorded; the data and listed integration times have been corrected to account for this problem.}
\tablenotetext{e}{Assumed source flux at specified frequency; also used for flux calibration.}
\end{deluxetable}
\normalsize
\end{turnpage}
\clearpage

\section{Results}
\label{sec:results}

\begin{figure*}
\plotone{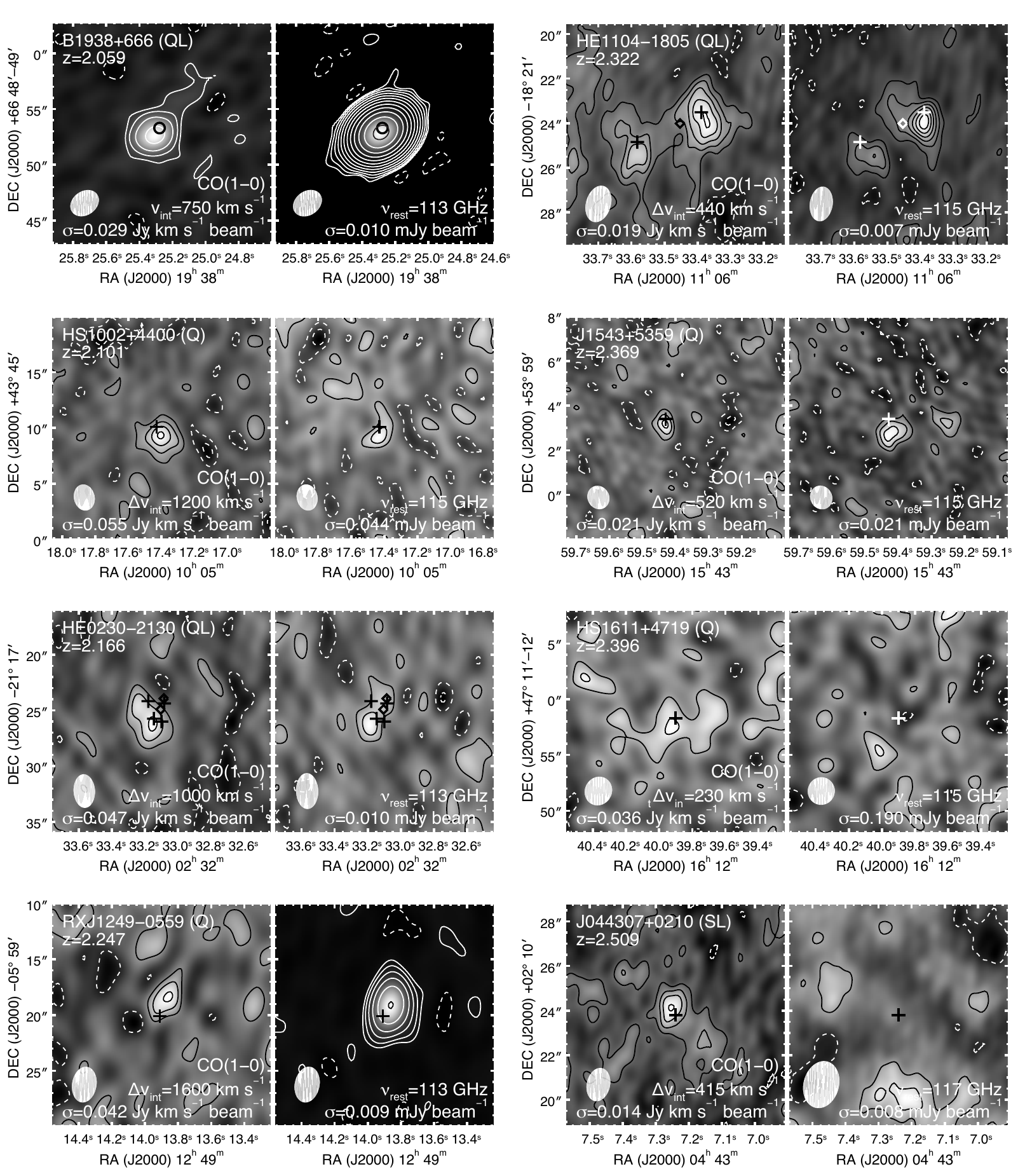}
\caption{Velocity-integrated \mbox{CO(1--0)} line maps (left) and continuum maps (right) for the sources observed in this survey in order of increasing redshift. Contours are multiples of $\pm1.5\sigma$, except for the integrated line map of B1938+666 and the continuum maps of B1938+666, RX\,J1249-559, MG\,0414+0534, and B1359+154 where the contours are powers of $\pm2\sigma$ (i.\/e.\/, $\pm2\sigma$, $\pm4\sigma$, $\pm8\sigma$, etc.); negative contours are dashed. Beam sizes are shown at lower left. Crosses mark the positions of the sources (from the \mbox{CO(3--2)} observations for the unlensed galaxies; see references in Table~\ref{tab:measure}) or images of the sources (from high-resolution optical/radio data for strongly lensed galaxies; \citealt{browne2003, kneib2000}; CASTLeS \href{https://www.cfa.harvard.edu/castles/}{https://www.cfa.harvard.edu/castles/}), and diamonds mark the position of the foreground lensing galaxies. For B1938+666, the circle marks the position and size of the Einstein ring. \label{fig:map1}}
\end{figure*}

\begin{figure*}
\plotone{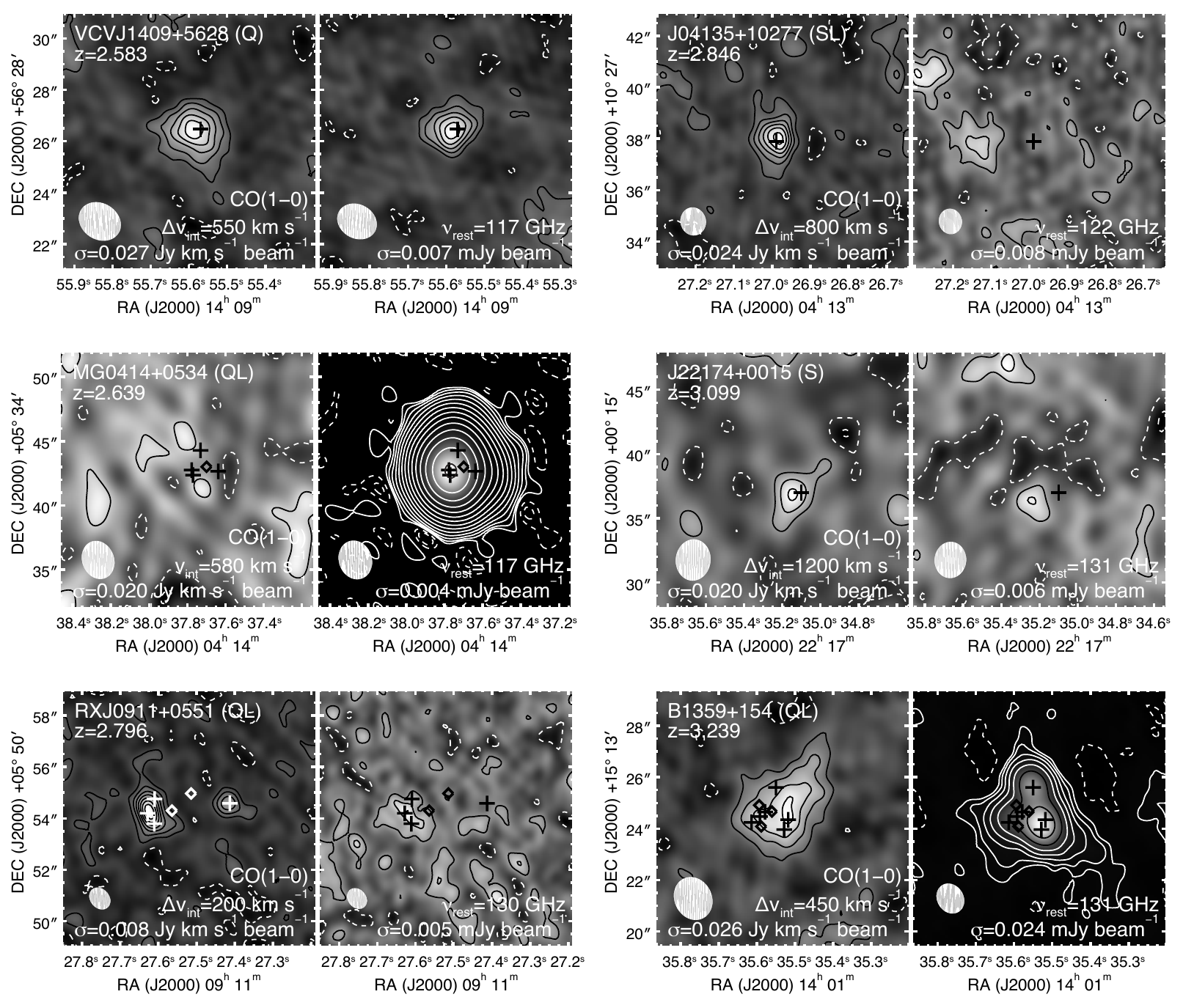}
\caption{Fig.~\ref{fig:map1} continued. \label{fig:map2}}
\end{figure*}

We successfully detected the \mbox{CO(1--0)} line in 13 objects (five of which are tentative) and did not detect the \mbox{CO(1--0)} line in one object; the integrated line maps are shown in Figures~\ref{fig:map1} and \ref{fig:map2}. We consider the sources successfully detected if (a) the peak of the \mbox{CO(1--0)} emission is spatially coincident with the \mbox{CO(3--2)} emission to within twice the position uncertainty and (b) the peak emission is at least $5\times$ the map noise; if the peak is between $3$--$5\sigma$ we consider the source to be tentatively detected. To be considered a (tentative) detection, we also require that the source emission be spatially distinct from any nearby noise peaks and comparable to or larger than the synthesized beam in size (listed in Table~\ref{tab:obs}). Offsets between the centroid positions of the \mbox{CO(1--0)} and \mbox{CO(3--2)} emission range from $0.^{\prime\prime}1$--$1.^{\prime\prime}9$ and the astrometric uncertainties on the \mbox{CO(3--2)} emission range from $0.^{\prime\prime}11$--$1.^{\prime\prime}7$. For strongly lensed objects with multiple images that lack previous high-resolution radio maps, we allow offsets of up to $1^{\prime\prime}$ since astrometric calibrations for optical data are less accurate. We discuss our single \mbox{CO(1--0)} non-detection, for MG\,0414+0534, in Section\,\ref{sec:mg0414}. In addition to the \mbox{CO(1--0)} measurements, we detected continuum emission in ten of the fourteen objects (Figures~\ref{fig:map1} and \ref{fig:map2}).

For most objects we lacked the S/N to precisely measure the full line profile. In these cases, we calculated the \mbox{CO(1--0)} line flux over the approximate full width half maximums (FWHMs) or full width zero intensities (FWZIs) measured for the \mbox{CO(3--2)} lines, choosing whichever velocity range retrieves the most flux from the source. In principle the FWZI maps should retrieve larger fluxes by definition, but in some cases the larger velocity range incorporates noise that reduces the measured flux in our marginally detected sources. While this method could bias our resulting \mbox{CO(3--2)}/\mbox{CO(1--0)} measurements towards lower values, we see no evidence for systematically low $r_{3,1}$ values in our new measurements. In addition, previous observations indicate that the \mbox{CO(1--0)} emission may be broader in velocity than the \mbox{CO(3--2)} emission \citep[e.\/g.\/,][]{hainline2006, ivison2011, riechers2011e, thomson2012}, which suggests that choosing larger velocity widths would more accurately capture the broad \mbox{CO(1--0)} emission. The measured integrated line fluxes and velocity integration widths are listed in Table~\ref{tab:measure}. With the exception of the sources with the narrowest FWHMs ($\sim200\,{\rm km\,s^{-1}}$), the measured line fluxes were robust (within the statistical uncertainties) to perturbations of $\sim100\,{\rm km\,s^{-1}}$ in line centroid and velocity integration width. For five sources, B1938+666, HE\,1104--1805, VCVJ1409+5628, RX\,J0911+0551, and J04135+10277, we were able to measure the \mbox{CO(1--0)} line profiles (Figures~\ref{fig:B1938spec}--\ref{fig:J04135spec}).

We discuss the individual sources in the following sections. We defer discussions comparing our new \mbox{CO(1--0)} measurements with existing (largely single-dish) literature values to Section~\ref{sec:analysis}, where we also evaluate the effects on the populations' CO line ratio measurements.

\begin{turnpage}
\begin{deluxetable*}{llcccccccccccc}
\tabletypesize{\tiny}
\tablewidth{0pt}
\setlength{\tabcolsep}{0pt}
\tablecaption{List of $z\sim2$--3 Galaxies with \mbox{CO(1--0)} and \mbox{CO(3--2)} Detections \label{tab:measure}}
\tablehead{{Source}	& {Type}\tablenotemark{a} & {$z$\tablenotemark{b}} & {$\mu$\tablenotemark{b}} & {$L_{\rm FIR}$\tablenotemark{b,c}} & {FWHM\tablenotemark{b}} & {$\Delta v_\text{int}$} & {$\nu_{\rm cont}$} & {$S_{\nu,{\rm cont}}$} & {$S_\text{1--0}\Delta v$} & {$L^\prime_\text{CO(1--0)}$\tablenotemark{c}} & {$S_\text{3--2}\Delta v$\tablenotemark{b}} & {$r_{3,1}$} & Ref.\tablenotemark{d}\\
{} & {} & {} & {} & {$(10^{12}\,L_\sun)$} & {${\rm(km\,s^{-1})}$} & {${\rm(km\,s^{-1})}$} & {(GHz)} & {(mJy)} & {${\rm (Jy\,km\,s^{-1})}$} & {${\rm (10^{10}\,K\,km\,s^{-1}\,pc^2)}$} & {${\rm (Jy\,km\,s^{-1})}$} & {} & {}}
\startdata
\multicolumn{12}{c}{New}\\
\hline
B1938+666 & Q/L & 2.0590 & 173\tablenotemark{e} & $0.17\pm0.02$ & $529\pm75$ & 750 & 114 & $56.27\pm5.63$ & $0.93\pm0.11$ & $0.12\pm0.01$ & $9.1\pm1.1$ & $1.09\pm0.19$ & 1, 2\\
HS\,1002+4400 & Q & 2.1015 & 1 & $9.62\pm1.83$ & $640\pm160$ & 1200 & 115 & $<0.13$\tablenotemark{f} & $0.53\pm0.15$ & $11.9\pm3.4$ & $1.7\pm0.3$ & $0.36\pm0.12$ & 3, 4\\
HE\,0230--2130 & Q/L & 2.1664 & 14.5 & $1.34\pm0.14$ & $666\pm121$ & 1000 & 114 & $0.046\pm0.012$ & $0.39\pm0.12$ & $0.64\pm0.19$ & $8.3\pm1.2$ & $2.35\pm0.79$ & 1, 2\\
RX\,J1249--0559 & Q & 2.2470 & 1 & $6.65\pm1.29$ & $1090\pm340$ & 1600 & 114 & $0.65\pm0.07$ & $0.21\pm0.05$ & $5.1\pm1.2$ & $1.3\pm0.4$ & $0.71\pm0.27$ & 3, 5\\
HE\,1104--1805 & Q/L & 2.3221 & 10.8 & $1.26\pm0.26$ & $441\pm81$ & 440 & 115 & $0.120\pm0.023$ & $0.56\pm0.09$ & $1.4\pm0.2$ & $7.5\pm1.2$ & $1.49\pm0.35$ & 1, 2\\
J1543+5359 & Q & 2.3698 & 1 & $7.31\pm2.12$ & $520\pm140$ & 520 & 115 & $0.11\pm0.02$ & $0.10\pm0.02$ & $2.9\pm0.7$ & $1.0\pm0.2$ & $1.07\pm0.32$ & 3, 4\\
HS\,1611+4719 & Q & 2.3961 & 1 & $8.85\pm1.35$ & $230\pm40$ & 230 & 115 & $<0.57$\tablenotemark{f} & $0.20\pm0.08$ & $5.8\pm2.2$ & $1.7\pm0.3$ & $0.92\pm0.39$ & 3\\
J044307+0210	 & S/L & 2.5090 & 4.4 & $1.51\pm0.36$ & $415\pm62$ & 415 & 117 & $<0.025$\tablenotemark{f} & $0.22\pm0.06$ & $1.5\pm0.4$ & $1.4\pm0.2$ & $0.70\pm0.20$ & 6, 7, 8\\
VCV\,J1409+5628 & Q & 2.5832 & 1 & $20.58\pm1.15$ & $311\pm28$ & 550 & 117 & $0.063\pm0.009$ & $0.27\pm0.07$ & $8.6\pm2.3$ & $2.3\pm0.2$ & $0.95\pm0.26$ & 4, 9\\
MG\,0414+0534 & Q/L & 2.6390 & 20 & $1.17\pm0.08$ & $580$ & 580 & 117 & $95.4\pm9.5$ & $<0.11$\tablenotemark{f} & $<0.19$\tablenotemark{f} & $2.6\pm0.4$ & $>2.56$\tablenotemark{f} & 2, 10 \\
RX\,J0911+0551 & Q/L & 2.7961 & 21.8 & $1.13\pm0.06$ & $129\pm18$ & 200 & 131 & $0.065\pm0.009$ & $0.22\pm0.04$ & $0.38\pm0.07$ & $2.22\pm0.44$ & $1.10\pm0.29$ & 2, 11\\
J04135+10277	 & S/L & 2.8460 & 1.6 & $17.75\pm1.99$ & $679\pm120$ & 800 & 123 & $<0.024$\tablenotemark{f} & $0.37\pm0.07$ & $8.6\pm1.7$ & $4.78\pm0.67$ & $1.45\pm0.35$ & 12, 13\\
J22174+0015 & S & 3.0990 & 1 & $4.52\pm1.20$ & $560\pm110$ & 1200 & 131 & $0.019\pm0.007$ & $0.099\pm0.023$ & $4.3\pm1.0$ & $0.7\pm0.2$ & $0.79\pm0.29$ & 14, 15\\
B1359+154 & Q/L & 3.2399 & 118\tablenotemark{e} & $0.12\pm0.02$ & $228\pm42$ & 450 & 131 & $10.7\pm1.1$ & $0.37\pm0.02$ & $0.15\pm0.03$ & $1.2\pm0.4$ & $0.36\pm0.14$ & 1, 2\\
\hline
\multicolumn{12}{c}{Literature}\\
\hline
J123549+6215 & S & 2.2020 & 1 & $7.66\pm2.31$ & $600\pm50$ & \nodata & \nodata & \nodata & $0.32\pm0.04$ & $7.78\pm0.97$ & $1.6\pm0.2$ & $0.56\pm0.10$ & 8, 16, 17\\
F10214+4724 & Q/L & 2.2856 & 17 & $2.71\pm0.27$ & $224\pm12$ & \nodata & \nodata & \nodata & $0.383\pm0.032$ & $0.58\pm0.05$ & $3.4\pm0.39$\tablenotemark{g} & $0.99\pm0.14$ & 2, 18, 19\\
HXMM01 & S/L & 2.3081 & 1.6 & $20\pm4$  & $980\pm200$ & \nodata & \nodata & \nodata & $1.7\pm0.3$ & $28.0\pm4.95$ & $9.8\pm1.7$ & $0.64\pm0.16$ & 20 \\
J2135-0102 & S/L & 2.3259 & 32.5 & $2.3\pm0.2$ & $600$ & \nodata & \nodata & \nodata & $2.16\pm0.24$\tablenotemark{g}  & $1.78\pm0.20$ & $13.2\pm1.3$\tablenotemark{g} & $0.68\pm0.10$ & 21, 22\\
J163650+4057	 & S & 2.3853 & 1 & $10.23\pm2.38$ & $710\pm50$ & \nodata & \nodata & \nodata & $0.34\pm0.04$ & $8.14\pm0.96$ & $2.3\pm0.3$ & $0.75\pm0.13$ & 8, 15, 23\\
J163658+4105	 & S & 2.4520 & 1 & $6.92\pm1.61$ & $800\pm50$ & \nodata & \nodata & \nodata & $0.37\pm0.07$ & $10.84\pm2.05$ & $1.8\pm0.2$ & $0.54\pm0.12$ & 8,15, 23\\
J123707+6214	 & S & 2.4876 & 1 & $4.34\pm1.38$ & $434\pm90$ & \nodata & \nodata & \nodata & $0.329\pm0.041$ & $9.91\pm1.23$ & $1.13\pm0.14$ & $0.38\pm0.07$ & 14, 24\\
J16359+6612 & S/L & 2.5156 & 45 & $0.43\pm0.09$ & $220\pm20$ & \nodata & \nodata & \nodata & $0.92\pm0.09$ & $0.63\pm0.06$ & $5.75\pm0.60$\tablenotemark{g} & $0.69\pm0.10$ & 25, 26, 27, 28 \\
Cloverleaf & Q/L & 2.5575 & 11 & $4.93\pm0.68$ & $416\pm6$ & \nodata & \nodata & \nodata & $1.387\pm0.144$\tablenotemark{g} & $3.97\pm0.41$ & $13.2\pm1.3$\tablenotemark{g} & $1.06\pm0.15$ & 17, 29\\
J14011+0252 & S/L & 2.5652 & 2.75 & $3.14\pm1.03$ & $190\pm11$ & \nodata & \nodata & \nodata & $0.32\pm0.04$ & $3.69\pm0.46$ & $2.8\pm0.3$ & $0.97\pm0.16$ & 28, 30, 31\\
J00266+1708 & S/L & 2.7420 & 2.41 & $3.21\pm0.75$ & $800$ & \nodata & \nodata & \nodata & $0.306\pm0.044$ & $4.50\pm0.65$ & $2.62\pm0.36$ & $0.95\pm0.19$ & 28, 32\\
J02399-0136 & S/L & 2.8076 & 25 & $0.46\pm0.05$ & $824\pm212$ & \nodata & \nodata & \nodata & $0.60\pm0.12$ & $0.89\pm0.18$ & $3.1\pm0.4$ & $0.57\pm0.14$ & 27, 28, 33\\
J14009+0252 & S/L & 2.9344 & 1.5 & $6.74\pm1.57$ & $412\pm24$ & \nodata & \nodata & \nodata & $0.31\pm0.04$\tablenotemark{g} & $8.22\pm1.06$ & $2.7\pm0.3$ & $0.97\pm0.12$ & 27, 28, 34\\
HLSW-01 & S/L & 2.9574 & 10.9 & $14.3\pm0.9$ & $348\pm18$ & \nodata & \nodata & \nodata & $1.01\pm0.10$ & $3.66\pm0.36$ & $9.7\pm0.5$ & $1.07\pm0.12$ & 35, 36\\
MG\,0751+2716 & Q/L & 3.1999 & 16 & $1.49\pm0.08$ & $400\pm50$ & \nodata & \nodata & \nodata & $0.525\pm0.070$ & $1.51\pm0.20$ & $4.6\pm0.5$  & $0.97\pm0.17$ & 2, 19, 37 \\
\hline
\multicolumn{12}{c}{Other (for comparison)}\\
\hline
cB58 & LBG/L & 2.7265 & 32 & $0.15\pm0.08$\tablenotemark{f} & $174\pm43$ & \nodata & \nodata & \nodata & $0.052\pm0.013$ & $0.057\pm0.014$ & $0.37\pm0.08$ & $0.79\pm0.26$ & 38, 39, 40\\
Cosmic Eye & LBG/L & 3.0743 & 28 & $0.83\pm0.44$\tablenotemark{f} & $190\pm24$ & \nodata & \nodata & \nodata & $0.077\pm0.013$ & $0.12\pm0.02$ & $0.50\pm0.07$ & $0.72\pm0.16$ & 40, 41, 42\\
HFLS3 & S/L & 6.3369 & 1.6 & $28.6\pm3.2$ & $977\pm160$ & \nodata & \nodata & \nodata & $0.074\pm0.024$ & $6.11\pm1.98$ & $0.717\pm0.094$ & $1.08\pm0.38$ & 43
\enddata
\tablenotetext{a}{{``S" denotes SMGs and ``Q" denotes AGN host galaxies, as classified in the literature. ``L" denotes if the galaxy is lensed. We also list two Lyman break galaxies (``LBG")  and a $z\sim6$ SMG for comparison.}}
\tablenotetext{b}{{Literature values; see list of references in the last column.}}
\tablenotetext{c}{Magnification-corrected; see previous column for assumed magnification factors.}
\tablenotetext{d}{{\bf References.} (1) \citealt{riechers2011b}; (2) \citealt{barvainis2002}; (3) \citealt{coppin2008}; (4) \citealt{omont2003}; (5) \citealt{page2001}; (6) \citealt{frayer2003}; (7) \citealt{smail2002}; (8) \citealt{tacconi2006}; (9) \citealt{beelen2004}; (10) \citealt{barvainis1998}; (11) \citealt{riechers_inprep}; (12) \citealt{hainline2004}; (13) \citealt{riechers2013a}; (14) \citealt{greve2005}; (15) \citealt{bothwell2013}; (16) \citealt{ivison2011}; (17) \citealt{chapman2003}; (18) \citealt{riechers2011f}; (19); \citealt{ao2008}; (20) \citealt{fu2013}; (21) \citealt{danielson2011}; (22) \citealt{ivison2010b}; (23) \citealt{kovacs2006}; (24) \citealt{riechers2011c}; (25) \citealt{sheth2004}; (26) \citealt{kneib2005}; (27) \citealt{thomson2012}; (28) \citealt{magnelli2012}; (29) \citealt{weiss2003}; (30) \citealt{downes2003}; (31) \citealt{sharon2013}; (32) \citealt{sharon2015}; (33) \citealt{genzel2003}; (34) \citealt{weiss2009a}; (35) \citealt{scott2011}; (36) \citealt{riechers2011d}; (37) \citealt{alloin2007}; (38) \citealt{seitz1998}; (39) \citealt{baker2004}; (40) \citealt{riechers2010}; (41) \citealt{dye2007}; (42) \citealt{coppin2007}; (43) \citealt{riechers2013b}}
\tablenotetext{e}{Although \citet{barvainis2002} suggest that the magnification factor should be a maximum of $\mu=20$.}
\tablenotetext{f}{$3\sigma$ limit; see discussions of the individual sources for the assumed flux distributions.}
\tablenotetext{g}{Published integrated line strengths include only the statistical uncertainty. Here we have added in quadrature an additional 10\% flux calibration uncertainty to the statistical uncertainties reported in the literature.}
\end{deluxetable*}
\normalsize
\end{turnpage}

\subsection{B1938+666}

\begin{figure}
\plotone{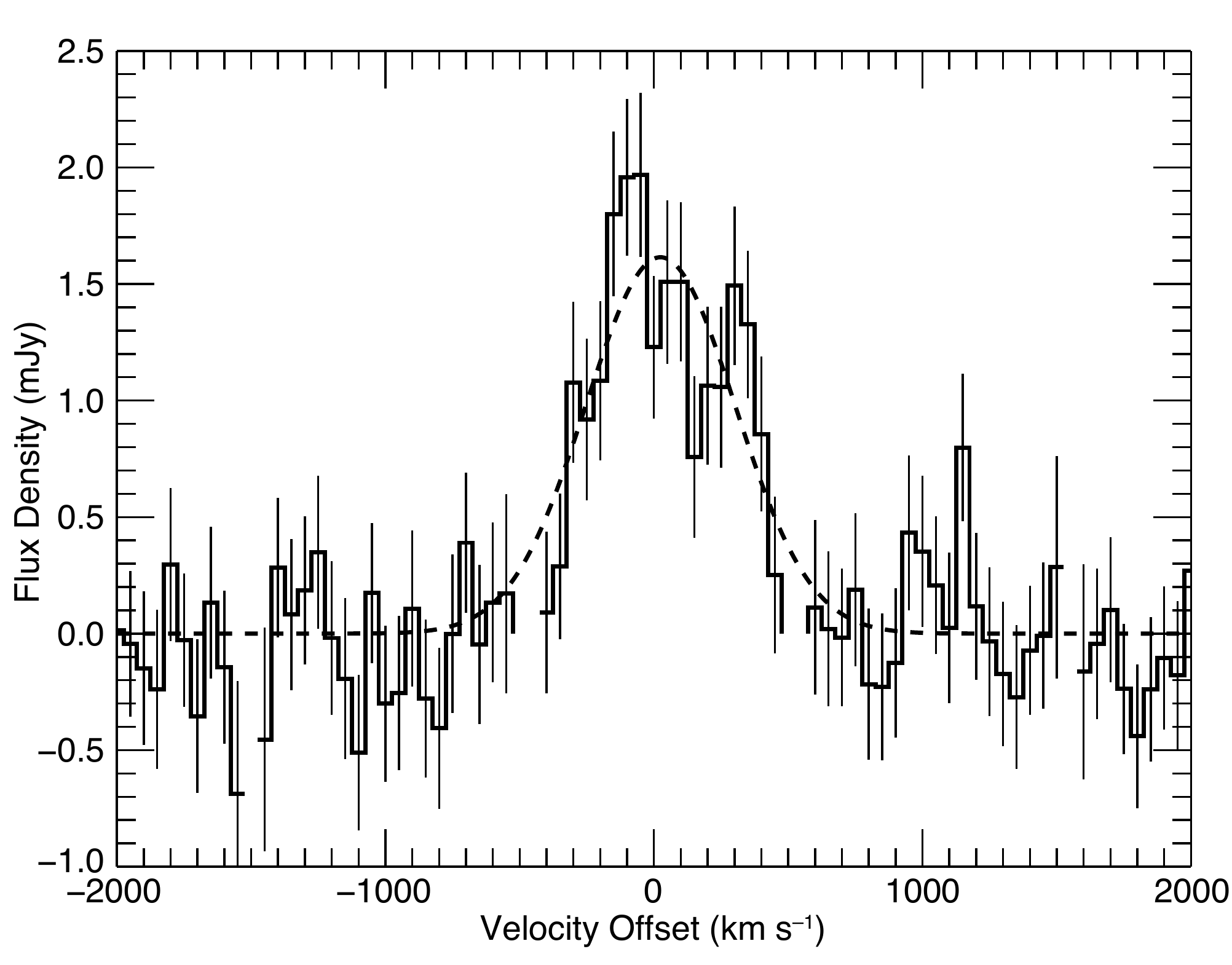}
\caption{\mbox{CO(1--0)} spectrum (thick solid histogram) and Gaussian fit to the line profile (dashed line) for B1938+666 plotted relative to the \mbox{CO(3--2)}-determined redshift ($z=2.0590\pm0.0003$) from \citet{riechers2011b}. Thin vertical lines denote the $\pm1\sigma$ uncertainty on the measured flux ($\pm0.33\,{\rm mJy}$, on average). Channel widths are $50\,{\rm km\,s^{-1}}$. Gaps in the spectral coverage are due to spectral window end channels that were flagged out during the data reduction. Our best-fit \mbox{CO(1--0)}-determined redshift is $z=2.0592\pm0.0003$. \label{fig:B1938spec}}
\end{figure}

We successfully detect \mbox{CO(1--0)} emission from the strongly-lensed radio-loud AGN host galaxy B1938+666 with a peak ${\rm SNR}=20.6$ (Figure~\ref{fig:map1}). We measure $S_{1-0}\Delta v=0.93\pm0.07(\pm0.09)\,{\rm Jy\,km\,s^{-1}}$ (where the latter $10\%$ uncertainty is associated with the flux calibration) at the position of the \mbox{CO(3--2)} emission from \citet{riechers2011b}. Gaussian fits to the line profile (Figure~\ref{fig:B1938spec}; reduced $\chi^2=0.77$) give a peak flux of $1.61\pm0.13\,{\rm mJy}$ and FWHM of $654\pm71\,{\rm km\,s^{-1}}$. The velocity centroid yields a CO(1--0)-determined redshift of $z_{1-0}=2.0592\pm0.0003$, which is consistent with the redshift of the CO(3--2) line ($z_{3-2}=2.0590\pm0.0003$) from \citet{riechers2011b}. The CO(1--0) emission is partially resolved, and elliptical Gaussian fits to the $uv$ data give a FWHM of $1.^{\prime\prime}17\pm0.^{\prime\prime}23$ and axis ratio consistent with unity ($uv$-continuum fits assuming a ring-shaped emission distribution do not converge). The source size is consistent with the diameter of the Einstein ring, $\sim0.^{\prime\prime}95$ \citep{king1997}.

We also detect $114\,{\rm GHz}$ (rest frame) continuum emission from B1938+666 (peak ${\rm SNR}=4970$; Figure~\ref{fig:map1}). We measure a flux of $S_{114}=56.271\pm0.004(\pm5.627)\,{\rm mJy}$ where the latter $10\%$ uncertainty is associated with the flux calibration. $uv$-continuum fits assuming a ring-shaped emission distribution do not converge. Assuming an elliptical Gaussian yields a FWHM of $\lesssim0.^{\prime\prime}5$, which is inconsistent with the diameter of Einstein ring. However, if the radio continuum emission is dominated by the northern arc \citep{browne2003}, then the a smaller source size would be expected.

\subsection{HS\,1002+4400}

We successfully detect \mbox{CO(1--0)} emission from the optically-bright quasar HS\,1002+4400 with a peak ${\rm SNR}=6.30$. We measure $S_{1-0}\Delta v=0.53\pm0.14(\pm0.05)\,{\rm Jy\,km\,s^{-1}}$ at the position of the \mbox{CO(3--2)} emission from \citet{coppin2008}. The \mbox{CO(1--0)} emission is unresolved. We do not detect significant $115\,{\rm GHz}$ (rest frame) continuum emission from the source (with the same synthesized beam as the \mbox{CO(1--0)} emission). Assuming a point source we derive a $3\sigma$ upper limit of $0.13\,{\rm mJy}$.

\subsection{HE\,0230--2130}

We tentatively detect \mbox{CO(1--0)} emission from strongly-lensed optically-bright quasar HE\,0230--2130 with a peak ${\rm SNR}=4.95$. We measure $S_{1-0}\Delta v=0.39\pm0.11(\pm0.04)\,{\rm Jy\,km\,s^{-1}}$ at the position of the CO(3--2) emission \citep{riechers2011b}. The \mbox{CO(1--0)} emission is partially resolved and detected in the southern pair of images (which are not resolved individually) and the north-eastern image. It is possible that we have out-resolved some of the flux (the \mbox{CO(3--2)} emission in \citet{riechers2011b} is partially resolved in the East-West direction with a larger $7.^{\prime\prime}3\times4.^{\prime\prime}2$ beam) and the emission from the north-western image is below our sensitivity limits. While the brightness ratio between the two detected CO(1--0) peaks (0.83) is not consistent with the optical ratios from the CASTLeS\footnote{\href{https://www.cfa.harvard.edu/castles/}{https://www.cfa.harvard.edu/castles/}} survey of lensed galaxies ($0.71$--$0.32$ depending on whether you assume the CO(1--0) originates in one or both of the southern images), if we assume the the optical image ratios, the expected emission from the north-western peak is well below the map noise.

We also tentatively detect $114\,{\rm GHz}$ (rest frame) continuum emission from HE\,0230--2130 with a peak ${\rm SNR}=4.38$. We measure $S_{114}=46\pm11(\pm5)\,{\rm \mu Jy}$ near the position of the southern two images. We do not assume the weak secondary peak is emission from the north-western image (or the lensing galaxies) despite its spatial coincidence since it is a $<3\sigma$ detection and should not be brighter than the other images (assuming the optical image ratios). However, it is possible that differential lensing may be affecting the measured image ratios. Assuming the optical image ratios both northern images would be within the noise of our continuum map.

\subsection{RX\,J1249--0559}

We tentatively detect unresolved \mbox{CO(1--0)} emission from the X-ray absorbed quasar RX\,J1249--0559 with a peak ${\rm SNR}=4.76$. We measure $S_{1-0}\Delta v=0.20\pm0.04(\pm0.02)\,{\rm Jy\,km\,s^{-1}}$ at the position of the \mbox{CO(3--2)} emission from \citet{coppin2008}. We also detect unresolved $114\,{\rm GHz}$ continuum emission from RX J1249-0559 (peak ${\rm SNR}=65.5$), obtaining $S_{114}=0.65\pm0.02(\pm0.07)\,{\rm mJy}$.

\subsection{HE\,1104--1805}

\begin{figure}
\plotone{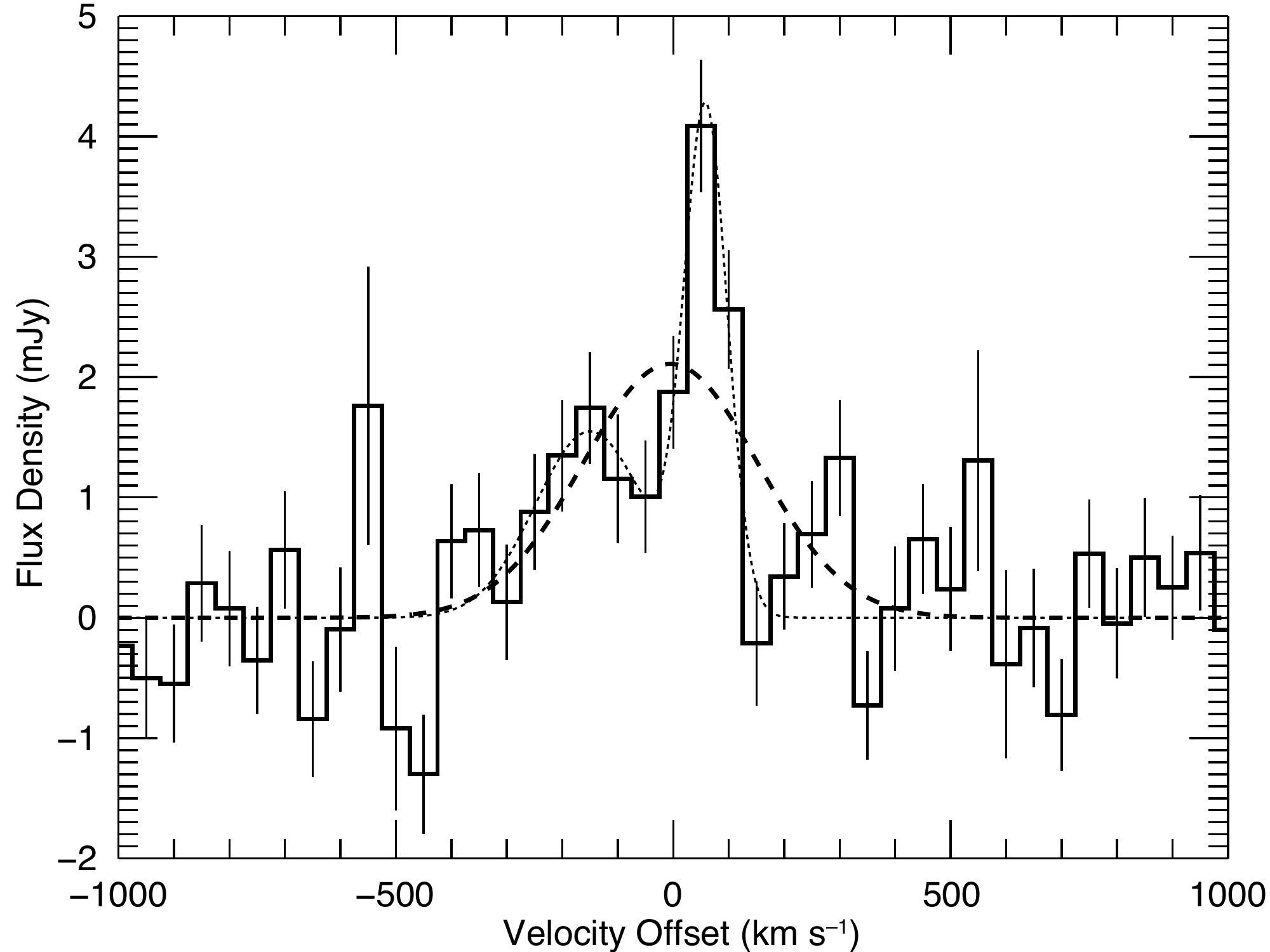}
\caption{\mbox{Continuum-subtracted CO(1--0)} spectrum (thick solid histogram) and Gaussian fits to the line profile (dotted and dashed lines) for HE1104--1805 plotted relative to the \mbox{CO(3--2)}-determined redshift ($z=2.3221\pm0.0004$) from \citet{riechers2011b}. Thin vertical lines denote the $\pm1\sigma$ uncertainty on the measured flux ($\pm0.51\,{\rm mJy}$, on average). Channel widths are $50\,{\rm km\,s^{-1}}$. We perform both single (dashed line) and double (dotted line) Gaussian fits to the spectrum. For the single Gaussian fit, our \mbox{CO(1--0)}-determined redshift is $z=2.3220\pm0.0002$. \label{fig:HE1104spec}}
\end{figure}

We successfully detect \mbox{CO(1--0)} emission from strongly-lensed optically-bright quasar HE\,1104--1805. The \mbox{CO(1--0)} emission is detected in both images (with peak SNRs of 7.68 and 5.80 for the western and eastern images, respectively) where the western image appears spatially extended. We measure $S_{1-0}\Delta v=0.56\pm0.08(\pm0.06)\,{\rm Jy\,km\,s^{-1}}$ at the position of the \mbox{CO(3--2)} emission from \citet{riechers2011b} using a $3^{\prime\prime}$ taper since some of the emission is potentially out-resolved (with taper, the synthesized beam FWHM is $4.^{\prime\prime}09\times3.^{\prime\prime}47$ at a position angle of $-12.11\degree$, and we retrieve $\sim25\%$ more flux). Using a single Gaussian to fit the line profile (Figure~\ref{fig:HE1104spec}; reduced $\chi^2=2.04$) gives a peak flux of $2.11\pm0.26\,{\rm mJy}$ and FWHM of $372\pm49\,{\rm km\,s^{-1}}$. The velocity centroid yields a \mbox{CO(1--0)}-determined redshift of $z_{1-0}=2.3220\pm0.0002$, which is consistent with the redshift of the \mbox{CO(3--2)} line ($z_{3-2}=2.3221\pm0.0004$) from \citet{riechers2011b}. There is, however, a conspicuous narrow peak in the line profile. If we perform a double Gaussian fit (reduced $\chi^2=1.36$, which is a significant improvement), we obtain a peak fluxes of $1.55\pm0.33\,{\rm mJy}$ and $4.11\pm0.58\,{\rm mJy}$ and FWHMs of $233\pm73\,{\rm km\,s^{-1}}$ and $91\pm16\,{\rm km\,s^{-1}}$. The two peaks are offset by $209\pm29\,{\rm km\,s^{-1}}$, where the redshift for broader/bluer peak is $z=2.3204\pm0.0003$. Both the single and double Gaussian fits produce integrated line fluxes slightly larger than our measured flux ($0.84\pm0.15\,{\rm Jy\,km\,s^{-1}}$ and $0.78\pm0.17\,{\rm Jy\,km\,s^{-1}}$, respectively), potentially due to resolved velocity structure, although the integrated line fluxes from the both Gaussian fits are consistent with our measured value at $\lesssim2\sigma$.

We also detect $115\,{\rm GHz}$ continuum emission from HE 1104-1805 at the positions of the western and eastern images (with peak SNRs of 10.2 and 4.6, respectively), obtaining $S_{115}=120\pm19(\pm12)\,{\rm \mu Jy}$. 

The average flux ratio between the two images in the optical and infrared from the CASTLeS\footnote{\href{https://www.cfa.harvard.edu/castles/}{https://www.cfa.harvard.edu/castles/}} survey of lensed galaxies is $\sim4.4$. However, the flux ratios based on the \mbox{CO(1--0)} and continuum maps are half the optical value. It is possible that the \mbox{CO(1--0)} and $115\,{\rm GHz}$ continuum emission is distributed differently from optical and infrared, causing differences in the effective lensing magnification and thus observed brightness ratios between the two images as a function of wavelength (i.e., differential lensing may be occurring). It is also possible that we have out-resolved some of the flux, particularly in the spatially extended western image, although that does not explain the difference in the image ratios for the more compact continuum emission.

\subsection{J1543+5359}

We tentatively detect \mbox{CO(1--0)} emission from the optically-bright quasar J1543+5359 with a peak ${\rm SNR}=4.95$. We measure $S_{1-0}\Delta v=0.10\pm0.02(\pm0.01)\,{\rm Jy\,km\,s^{-1}}$ at the position of the \mbox{CO(3--2)} emission from \citet{coppin2008}. The \mbox{CO(1--0)} emission is unresolved. We also detect unresolved $115\,{\rm GHz}$ continuum emission from J1543+5359 (peak ${\rm SNR}=5.21$), obtaining $S_{115}=0.11\pm0.02(\pm0.01)\,{\rm mJy}$.

\subsection{HS\,1611+4719}

The optically-bright quasar HS\,1611+4719 is tentatively detected with a peak ${\rm SNR}=3.41$ in the integrated \mbox{CO(1--0)} map. We measure $S_{1-0}\Delta v=0.20\pm0.07(\pm0.02)\,{\rm Jy\,km\,s^{-1}}$ at the position of the \mbox{CO(3--2)} emission from \citet{coppin2008}. We do not detect $115\,{\rm GHz}$ (rest frame) continuum emission from the source. Assuming a point source we derive a $3\sigma$ upper limit of $0.57\,{\rm mJy}$.

\subsection{J044307+0210}

We detect \mbox{CO(1--0)} emission from weakly-lensed SMG J044307+0210 with a peak ${\rm SNR}=6.12$ at the position of the \mbox{CO(3--2)} emission from \citet{tacconi2006}. There is also a second $3.34\sigma$ peak to the southwest, but it is unclear if that peak is associated with the \mbox{CO(1--0)} emission of J044307+0210. Using only the brighter of the two peaks, we measure $S_{1-0}\Delta v=0.12\pm0.03(\pm0.02)\,{\rm Jy\,km\,s^{-1}}$. However, if we assume that the emission is more extended, and extract the flux from the map with a $3^{\prime\prime}$ taper applied (beam FWHM of $4.^{\prime\prime}00\times3.^{\prime\prime}68$ at a position angle of $16.35\degree$) that includes the secondary peak, we then obtain $S_{1-0}\Delta v=0.22\pm0.05(\pm0.02)\,{\rm Jy\,km\,s^{-1}}$. This second flux is more in line with the value reported for the single dish measurement in \citet{harris2010}. We therefore assume the larger flux measurement from the tapered map in the subsequent analysis, which is the value recorded in Table~\ref{tab:measure}. We do not detect significant $117\,{\rm GHz}$ (rest frame) continuum emission from the source. Assuming a point source we derive a $3\sigma$ upper limit of $25\,{\rm \mu Jy}$.

\subsection{VCV\,J1409+5628}

\begin{figure}
\plotone{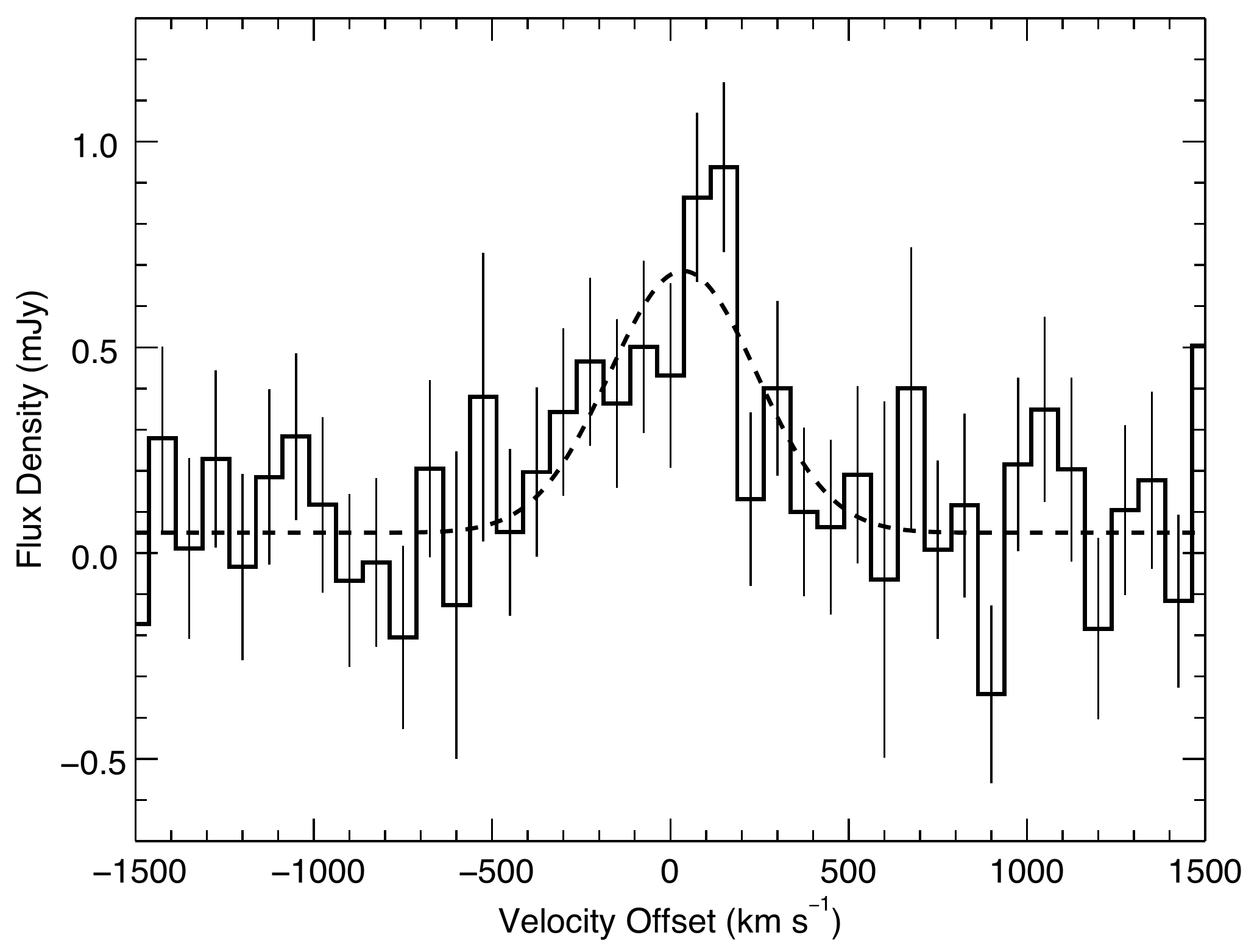}
\caption{\mbox{CO(1--0)} spectrum (thick solid histogram) and Gaussian fit to the line profile (dashed line) for VCVJ1409+5628 plotted relative to the \mbox{CO(3--2)}-determined redshift ($z=2.5832\pm0.0001$) from \citet{beelen2004}. Thin vertical lines denote the $\pm1\sigma$ uncertainty on the measured flux ($\pm0.23\,{\rm mJy}$, on average). Channels widths are $75\,{\rm km\,s^{-1}}$. The best-fit \mbox{CO(1--0)}-determined redshift is $z=2.5836\pm0.0005$. \label{fig:VCVJ1409spec}}
\end{figure}

We successfully detect \mbox{CO(1--0)} emission from the optically-bright radio-quiet quasar VCV\,J1409+5628 with peak ${\rm SNR}=8.47$. We measure $S_{1-0}\Delta v=0.27\pm0.07(\pm0.03)\,{\rm Jy\,km\,s^{-1}}$ at the position of the \mbox{CO(3--2)} emission from \citet{beelen2004}. The \mbox{CO(1--0)} emission is unresolved. Gaussian fits to the line profile (Figure\,\ref{fig:VCVJ1409spec}; reduced $\chi^2=0.83$) give a peak flux of $0.64\pm0.12\,{\rm mJy}$ and FWHM of $487\pm 101\,{\rm km\,s^{-1}}$. The velocity centroid yields a CO-determined redshift of $z_{\rm CO}=2.5836\pm0.0005$, which is consistent with the redshift of the \mbox{CO(3--2)} line ($z=2.5832\pm0.001$; \citealt{beelen2004}) and slightly offset from optically-determined redshifts \citep{korista1993}. We also detect unresolved $117\,{\rm GHz}$ continuum emission from VCV J1409+56289 (peak ${\rm SNR}=8.54$), obtaining $S_{117}=63\pm7(\pm 6)\,{\rm \mu Jy}$. We also note that VCV\,J1409+5628 is one of the two sources where we have included an observing track that lacks measurements of a primary flux calibrator.

\subsection{MG\,0414+0534}
\label{sec:mg0414}

The strongly-lensed radio-loud AGN MG\,0414+0534 is undetected in \mbox{CO(1--0)}. However, we do detect $117\,{\rm GHz}$ (rest frame) continuum emission from the source (peak ${\rm SNR}=17600$), obtaining $S_{117}=95.413\pm0.018(\pm9.541)\,{\rm Jy}$. The continuum emission is spatially resolved, and an elliptical Gaussian fit to the $uv$ continuum yields a major axis FWHM of $\sim1.^{\prime\prime}3$ and axis ratio of $\sim0.4$ that originates near the positions of the two brightest optical images of this quadruply-lensed source. This indicates that most of the $117\,{\rm GHz}$ continuum emission is from the eastern two images (consistent with the $5\,{\rm GHz}$ continuum emission; \citealt{browne2003}) which are not resolved individually.

Based on the \mbox{CO(3--2)} integrated line measurement from \citet{barvainis1998}, we should have detected \mbox{CO(1--0)} given the sensitivity of our map. If we assume thermalized excitation and that the line emission is distributed equally between the four images to establish a conservative limit on the \mbox{CO(1--0)} brightness, then each image should be at least $\sim3.5\sigma$; using the $5\,{\rm GHz}$ image flux ratios and accounting for the small separation of the brightest two images we would expect a $>10\sigma$ detection. We see no residual structure in the image to suggest that the bright continuum was poorly subtracted. We examine the data as function of frequency, both in the image plane and in the $uv$ data, and see no evidence of line emission at any of the observed frequencies. Observations of other CO lines may help clarify our non-detection. For an upper limit on the \mbox{CO(1--0)} line, we assume that the emission would be resolved, and originate from all four of the lensed images which would be individually unresolved. With no assumptions of the image brightness ratios, $95\%$ of the \mbox{CO(1--0)} flux should be contained within a box with an area $3.5$ synthesized beams (based on a Gaussian fit to the assumed emission pattern), which yields a $3\sigma$ upper limit of $<0.11\,{\rm Jy\,km\,s^{-1}}$.

\subsection{RX\,J0911+0551}

\begin{figure}
\plotone{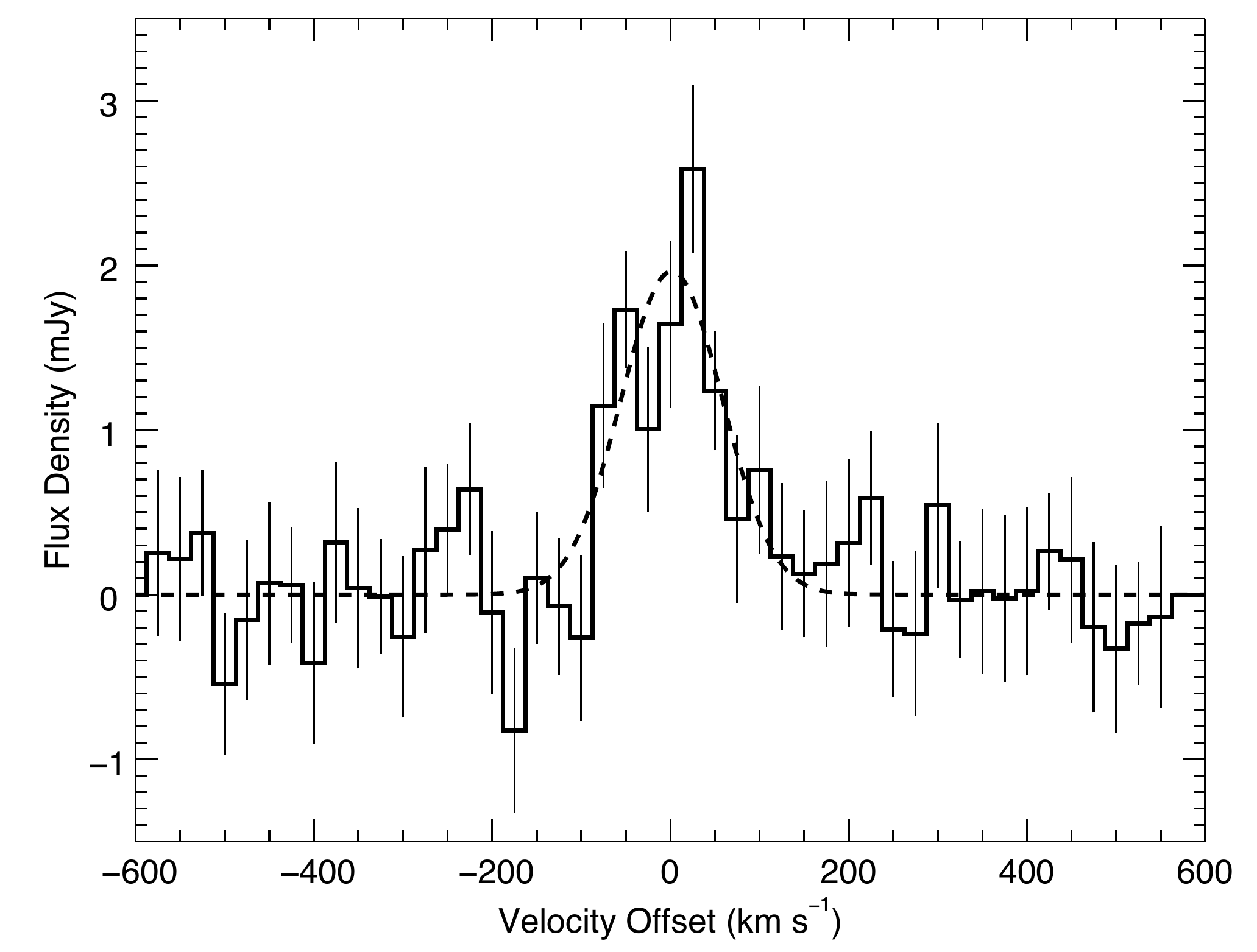}
\caption{\mbox{CO(1--0)} spectrum (thick solid histogram) and Gaussian fit to the line profile (dashed line) for RX\,J0911+0551 plotted relative to the \mbox{CO(3--2)}-determined redshift ($z=2.7961\pm0.0001$) from \citet{riechers_inprep}. Thin vertical lines denote the $\pm1\sigma$ uncertainty on the measured flux ($\pm0.41\,{\rm mJy}$, on average). Channels widths are $25\,{\rm km\,s^{-1}}$. The best-fit \mbox{CO(1--0)}-determined redshift is $z=2.7961\pm0.0001$. \label{fig:RXJ0911spec}}
\end{figure}

We detect \mbox{CO(1--0)} emission from the strongly-lensed broad absorption line quasar RX\,J0911+0551. RX J0911+0551 is partially resolved into its four images \citep[e.\/g.\/,][]{burud1998}---we have a peak ${\rm SNR}=11.8$ detection of the eastern three closely-spaced images and a $5.64\sigma$ peak detection of the fourth western image. We measure a total flux from the combined images of $S_{1-0}\Delta v=0.22\pm0.03(\pm0.02)\,{\rm Jy\,km\,s^{-1}}$. Gaussian fits to the line profile (Figure\,\ref{fig:RXJ0911spec}; reduced $\chi^2=0.63$) give a peak flux of $2.0\pm0.3\,{\rm mJy}$ and FWHM of $133\pm 23\,{\rm km\,s^{-1}}$. The velocity centroid yields a CO-determined redshift of $z_{\rm CO}=2.7961\pm0.0001$, which is consistent with the redshift of the \mbox{CO(3--2)} line \citep{riechers_inprep} and the \mbox{CO(7--6)} and C\,{\sc i}\,(${^3P}_2\rightarrow {^3P}_1$) lines \citep{weiss2012}.

We also tentatively detect $131\,{\rm GHz}$ continuum emission from the eastern components of RX J0911+0551 (${\rm SNR}=4.53$), obtaining $S_{131}=65\pm16(\pm7)\,{\rm \mu Jy}$. Assuming that the continuum flux ratio between the sum of the eastern images and the western image matches the average flux ratio determined in the optical and infrared from CASTLeS (7.2), the expected $130\,{\rm GHz}$ flux for the western image is $\sim9\,{\rm \mu Jy}$ (only $50\%$ larger than the map's RMS noise). However, the optical flux ratio is a factor of $\sim1.5$--$2$ greater than we observe in \mbox{CO(1--0)} (3.6) or has been observed in \mbox{CO(7--6)} (4.8; \citealt{weiss2012}). Using the \mbox{CO(1--0)} flux ratio, the expected $131\,{\rm GHz}$ flux for the western image is $\sim18\,{\rm \mu Jy}$ and should therefore be detected at the $3\sigma$ level. Using the \mbox{CO(7--6)} flux ratio, the expect $130\,{\rm GHz}$ flux for the western image is $\sim14\,{\rm \mu Jy}$ and should remain undetected ($<2\sigma$).The differences in the brightness ratios between the images in the optical and CO emission may be explained by differential lensing. Given the non-detection of western image, we favor the optical or higher-$J$ CO flux ratios (over the \mbox{CO(1--0)} flux ratios) since the the $130\,{\rm GHz}$ continuum emission may be associated with the AGN and not the extended molecular gas reservoir.

We also searched for the \mbox{CS(3--2)} $146.969\,{\rm GHz}$ line that the higher frequency IF pair was centered on. We do not detect any CS emission after exploring several possible line widths and derive a $3\sigma$ upper limit of $0.11\,{\rm Jy\,km\,s^{-1}}$ assuming the same spatially extended flux distribution and $200\,{\rm km\,s^{-1}}$ bin width as used for the \mbox{CO(1--0)} line measurement (beam FWHM of $0.^{\prime\prime}75\times0.^{\prime\prime}63$ at position angle $34.36\degree$).

\subsection{J04135+10277}

\begin{figure}
\plotone{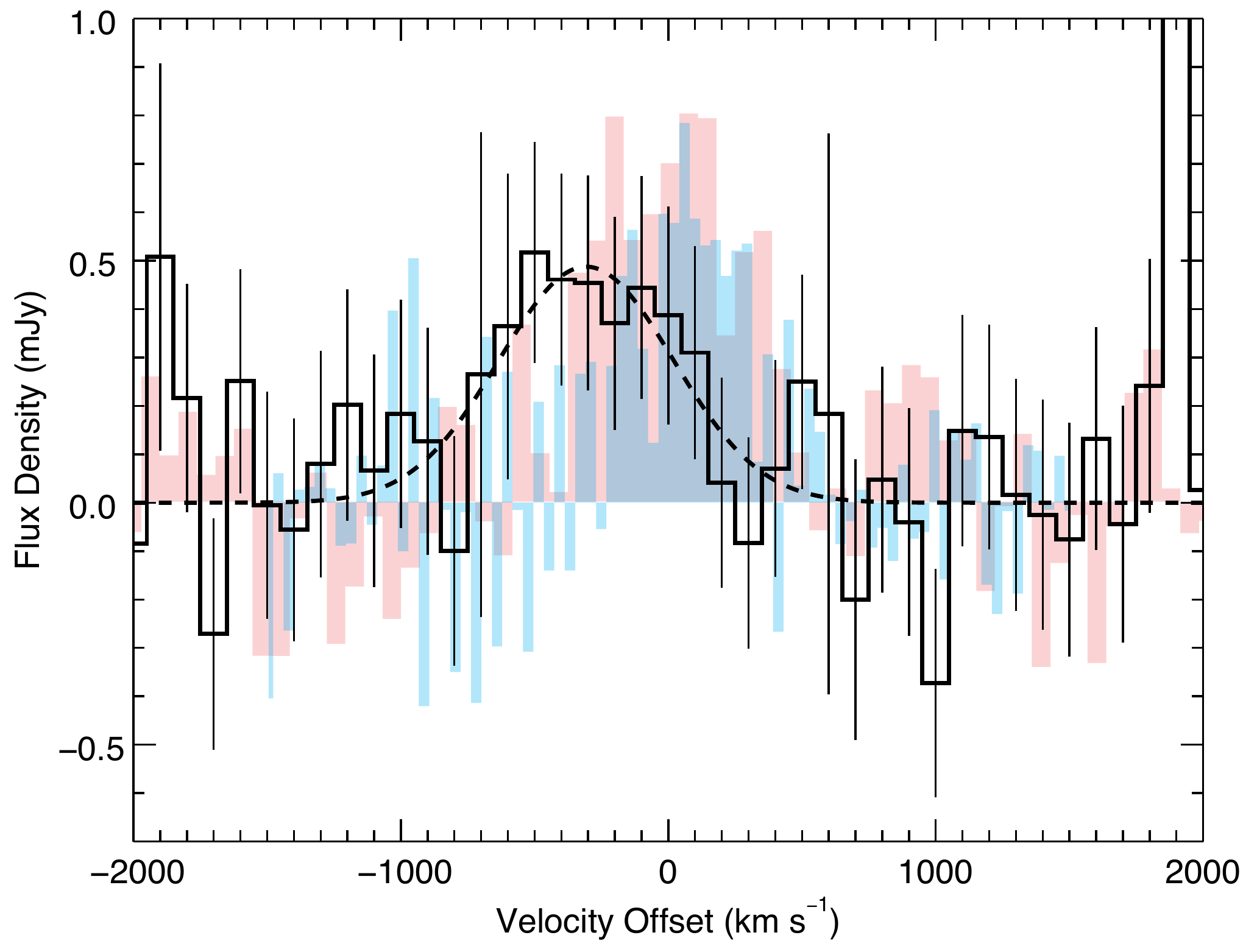}
\caption{\mbox{CO(1--0)} spectrum (thick solid histogram) and Gaussian fit to the line profile (dashed line) for J04135+10277 plotted relative to the \mbox{CO(3--2)}-determined redshift ($z=2.846\pm0.002$) from \citet{hainline2004}. Thin vertical lines denote the $\pm1\sigma$ uncertainty on the measured flux ($\pm0.28\,{\rm mJy}$, on average). Channel widths are $100\,{\rm km\,s^{-1}}$. In light blue we show the GBT spectrum of the \mbox{CO(1--0)} line from \citet{riechers2011f} scaled down by $50\%$, and in light red we show the \mbox{CO(3--2)} spectrum from \citet{riechers2013a} scaled down by a factor of nine. Our best-fit \mbox{CO(1--0)}-determined redshift is $z=2.842\pm0.001$. \label{fig:J04135spec}}
\end{figure}

We successfully detect \mbox{CO(1--0)} emission from the weakly-lensed SMG J04135+10277 with ${\rm SNR_{peak}}=9.6$. We measure $S_{1-0}\Delta v=0.37\pm0.06(\pm0.04)\,{\rm Jy\,km\,s^{-1}}$ at the position of the \mbox{CO(3--2)} emission from \citet{riechers2013a}, which is not being emitted by the nearby quasar-host galaxy as previous low-resolution observations assumed. Gaussian fits to the line profile (Figure\,\ref{fig:J04135spec}; reduced $\chi^2=0.57$) give a peak flux of $0.49\pm0.11\,{\rm mJy}$ and FWHM of $765\pm222\,{\rm km\,s^{-1}}$. The velocity centroid yields a CO-determined redshift of $z_{\rm CO}=2.8421\pm0.0013$, which is in slight tension with the \citet{hainline2004} $z_{\rm CO}=2.846\pm0.002$, and signficantly offset from the GBT-determined \mbox{CO(1--0)} redshift from \citet{riechers2011f} ($z=2.8470\pm0.0004$) as well as the \mbox{CO(3--2)}-determined redshift ($z=2.8458\pm0.0006$) from \citet{riechers2013a}. It is unclear why the interferometric \mbox{CO(1--0)} velocity centroid is $\sim400\,{\rm km\,s^{-1}}$ offset from previously determined redshifts; there are no obvious problems with the data or calibration. We note that the overall flux level is lower than the single-dish measurement as well as being offset in velocity, which suggests that we have perhaps resolved out some extended emission that may be biased towards the redder half of the line (although all measured FWHMs are consistent within their uncertainties). A similar unexplained offset has been observed in one other SMG, SMM\,J02399--0136 \citep{thomson2012}. The source appears slightly extended relative to the natural weighting beam, but fits to the $uv$-data do not converge on a source size.

We do not detect $123\,{\rm GHz}$ continuum emission from J04135+10277. Assuming a point source we obtain a $3\sigma$ upper limit of $24\,{\rm \mu Jy}$. We also searched for the \mbox{SiO(3--2)} $130.26861\,{\rm GHz}$ line that the higher frequency IF pair was centered on. We do no detect any SiO emission after exploring several possible line widths, and derive a $3\sigma$ upper limit of $0.077\,{\rm Jy\,km\,s^{-1}}$ for a point-like source (beam FWHM of $0.^{\prime\prime}94\times0.^{\prime\prime}85$ at a position angle of  $9.77\degree$) using the same $800\,{\rm km\,s^{-1}}$ bin width used for the \mbox{CO(1--0)} emission.

Finally, we look for emission from the nearby optical quasar ($z=2.837\pm0.003$; \citealt{knudsen2003}) originally assumed to be the source of the CO and FIR emission. We detect no continuum emission at the exact position of the optical quasar (J2000 $\alpha={\rm 4^h\,13^m\,27.28^s}$, $\delta=10\degree\,27^\prime\,41^{\prime\prime}.4$; \citealt{knudsen2003}), obtaining a $3\sigma$ upper limit of $24\,{\rm \mu Jy}$. However, $1.8^{\prime\prime}$ south of the optical quasar is a $4.8\sigma$ peak with $S_{122}=80\pm19(\pm8)\,{\rm \mu Jy}$, but this emission may be noise and/or not associated with the quasar. We also searched for \mbox{CO(1--0)} emission from the quasar and obtain a $3\sigma$ upper limit  of $0.066\,{\rm Jy\,km\,s^{-1}}$ assuming a point-like source and a $500\,{\rm km\,s^{-1}}$ line width.

\subsection{J22174+0015}

We tentatively detect the \mbox{CO(1--0)} line in the SMG J22174+0015 with ${\rm SNR_{peak}}=4.8$. We measure $S_{1-0}\Delta v=0.099\pm.021(\pm0.010)\,{\rm Jy\,km\,s^{-1}}$ at the position of the \mbox{CO(3--2)} emission from \citet{greve2005}. The \mbox{CO(1--0)} emission is unresolved. We also tentatively detect $131\,{\rm GHz}$ (rest frame) continuum emission from J22174+0015, but the continuum peak is weak (${\rm SNR}=3.1$) and slightly offset from the line emission. We obtain $S_{131}=19\pm6(\pm 2)\,{\rm \mu Jy}$.

We also searched for the \mbox{CS(3--2)} $146.969\,{\rm GHz}$ line that the higher frequency IF pair was centered on. We do not detect any CS emission after exploring several possible bin widths and derive a $3\sigma$ upper limit of $0.09\,{\rm Jy\,km\,s^{-1}}$ for a point-like source (beam FWHM of $2.^{\prime\prime}51\times2.^{\prime\prime}17$ at a position angle of $-0.70\degree$) assuming the same $1200\,{\rm km\,s^{-1}}$ bin width used for the \mbox{CO(1--0)} line.

\subsection{B1359+154}

We detect \mbox{CO(1--0)} emission from the strongly-lensed radio-loud AGN host galaxy B1359+154 with a peak ${\rm SNR}=6.4$. We measure $S_{1-0}\Delta v=0.37\pm0.07(\pm 0.04)\,{\rm Jy\,km\,s^{-1}}$ at the position of the \mbox{CO(3--2)} and \mbox{CO(4--3)} emission from \citet{riechers2011b}. We also detect $131\,{\rm GHz}$ continuum emission from B1359+154, obtaining $S_{131}=10.7\pm0.1(\pm 1.1)\,{\rm mJy}$ and $SNR_{\rm peak}=213$. The emission is partially resolved for both the CO and continuum maps; the continuum clearly shows three peaks of emission. Due to the complicated lensing configuration that creates six images of B1359+154 \citep[e.\/g.\/,][]{myers1999}, it is difficult to associate the $130\,{\rm GHz}$ peaks with individual images, but the three peaks roughly correspond to images A, B/C, and D/E/F and have the appropriate relative brightnesses. 

We also searched for the \mbox{CS(3--2)} $146.969\,{\rm GHz}$ line that the higher frequency IF pair was centered on. We do not detect any CS emission after exploring several possible line widths and derive a $3\sigma$ upper limit of $0.28\,{\rm Jy\,km\,s^{-1}}$ (assuming the same spatially extended flux distribution and $450\,{\rm km\,s^{-1}}$ bin width as for the \mbox{CO(1--0)} line measurement). The beam size for the CS line upper limit is $1.^{\prime\prime}09\times1.^{\prime\prime}01$ at a position angle of $52.33\degree$. Lastly, we also note that B1359+154 is one of the two sources where we have included observing tracks that lack measurements of a primary flux calibrator.

\section{Analysis}
\label{sec:analysis}

\subsection{Peculiar $r_{3,1}$ values and comparisons to previous measurements}

Before we compare the distribution of $r_{3,1}$ values to previous results and look for correlations between molecular gas excitation and other galaxy properties, we first compare our new \mbox{CO(1--0)} detections to any previously existing measurements and discuss the origins of some of our large $r_{3,1}$ values. Five galaxies (RX\,J0911+0551, J04135+10277, and J044307+0210 from our observations; J14011+0252 and J14009+0252 from the literature) have previous measurements of the \mbox{CO(1--0)} line, mostly from single-dish observations at the GBT (except RX\,J0911+0551). For RX\,J0911+0551, the previous VLA data was taken with the old narrow correlator yielding a line ratio $r_{3,1}\sim0.95$ \citep{riechers2011f} which is consistent with our new measurement of $r_{3,1}=1.01\pm0.29$. J04135+10277 has a previous \mbox{CO(1--0)} measurement using the GBT \citep{riechers2011f}, which gave $r_{3,1}=0.93\pm0.25$. However, higher angular resolution observations of the \mbox{CO(3--2)} line with the Combined Array for Research in Millimeter-wave Astronomy revealed that the CO emission was not associated with the nearby optically luminous quasar and is actually an SMG \citep{riechers2013a}. We use the SMG classification and our new VLA-determined \mbox{CO(1--0)} flux here, but note that our new measurement has significantly larger uncertainties ($r_{3,1}=1.45\pm0.35$; although this value of the line ratio is consistent with the previous measurement). SMG J044307+0210 has a previous measurement of $r_{3,1}=0.61\pm0.15$ from the GBT \citep{harris2010}, and our new measurement ($r_{3,1}=0.70\pm0.20$) is consistent with that value. Lastly, the SMGs J14009+0252 and J14011+0252 were both observed in \mbox{CO(1--0)} at the GBT, giving $r_{3,1}=0.67\pm0.08$ and $r_{3,1}=0.76\pm0.12$, respectively \citep{harris2010}. However, subsequent VLA observations in \citet{thomson2012} and \citet{sharon2013} gave $\sim25\%$ lower \mbox{CO(1--0)} fluxes, yielding $r_{3,1}=0.97\pm0.12$ for J14009+0252 and $r_{3,1}=0.97\pm0.16$ for J14011+0252. For J14011+0252 the two line ratios are consistent with one another, and for J14009+0252 the two line ratios are marginally inconsistent. Since interferometers generally provide better amplitude stability, and the GBT/Zpectromer used for the \mbox{CO(1--0)} detections is broadened by a $sinc(x)$ response function which leads to larger line widths and fluxes (see discussion of J14011+0252 in \citealt{sharon2013}), we therefore favor the VLA \mbox{CO(1--0)} measurements in the subsequent analyses. 

Three of our new detections (HE0230--2130, HE1104--1805, and J04135+10277) and our single line ratio limit (MG\,0414+0534) result in peculiarly large values of $r_{3,1}>1.4$, although they are consistent at the $1$--$2\sigma$ level with $r_{3,1}=1$ due to their large uncertainties. Values of $r_{3,1}>1$ are unlikely to occur under normal conditions for SMGs (where the ISM is dominated by molecular gas and emission is optically thick) but can occur when the emission is optically thin, when \mbox{CO(1--0)} is self-absorbed, when the CO emission is optically thick but emitted from an ensemble of small unresolved clouds, or when the source of the optically thick emission has a temperature gradient \citep[e.\/g.\/,][]{bolatto2000,bolatto2003}.  We also note that the new VLA measurements of the \mbox{CO(1--0)} line are systematically lower than (albeit consistent with) the previous GBT measurements, as might be expected if the interferometer is resolving out part of the emission.\footnote{We also note that the flux scales assumed for the original \citet{harris2010} GBT measurements differ from the Perley-Butler 2013 standards used in the VLA pipeline. However, the assumed calibration fluxes for 3C286 and 3C48 are larger for the VLA observations than for the GBT observations (by as much as $\sim10\%$), making the flux calibration issues an unlikely sources of our measured flux discrepancies.} While it is unlikely that many $z\sim2$ galaxies are more spatially extended than the largest angular scale accessible with the VLA at these frequencies/configurations ($\sim40^{\prime\prime}$--$50^{\prime\prime}$), many of our measurements have modest SNRs, which could yield weak extended emission undetected by our current observations. We suspect weak extended emission below our detection threshold is the explanation for two of the three sources with $r_{3,1}>1$ (and may be contributing to our non-detection). The third outlier, J04135+10277, is also among the sources with lower interferometric fluxes, despite being detected at $>9\sigma$. In this case, weak extended emission seems like an unlikely cause of the discrepancy between the interferometric and single-dish fluxes and some other origin is likely given its redshift discrepancy (discussed previously). In the subsequent analysis of the $r_{3,1}$ distributions, we consider the distributions both with and without the peculiar $r_{3,1}$ measurements, as well as the distributions using previous $r_{3,1}$ measurements for these sources.

Characterizations of strongly-lensed sources may also be affected by differential lensing: the variation of the magnification factor across a spatially extended source. Differential lensing tends to bias CO excitation measurements to more compact and therefore higher excitation (higher $r_{3,1}$) values (e.\/g.\/, \citealt{serjeant2012, hezaveh2012}). However, this is less likely to be important for the low-$J$ CO lines studied here, and there is little observational evidence supporting the existence of this effect in these transitions (although see F10214+4724; \citealt{deane2013b}). We do note that some of our most extreme line ratios are observed in our high-magnification ($\mu>10$) sources, but that does not explain values of $r_{3,1}>1$ and is degenerate with the high proportion of AGN host galaxies in the new observations (which might be expected to have $r_{3,1}\sim1$ based on previous line ratio measurements; \citealt{riechers2011f}).

\subsection{Is there a difference in $r_{3,1}$ values for SMGs and AGN host galaxies?}

\begin{figure*}
\plottwo{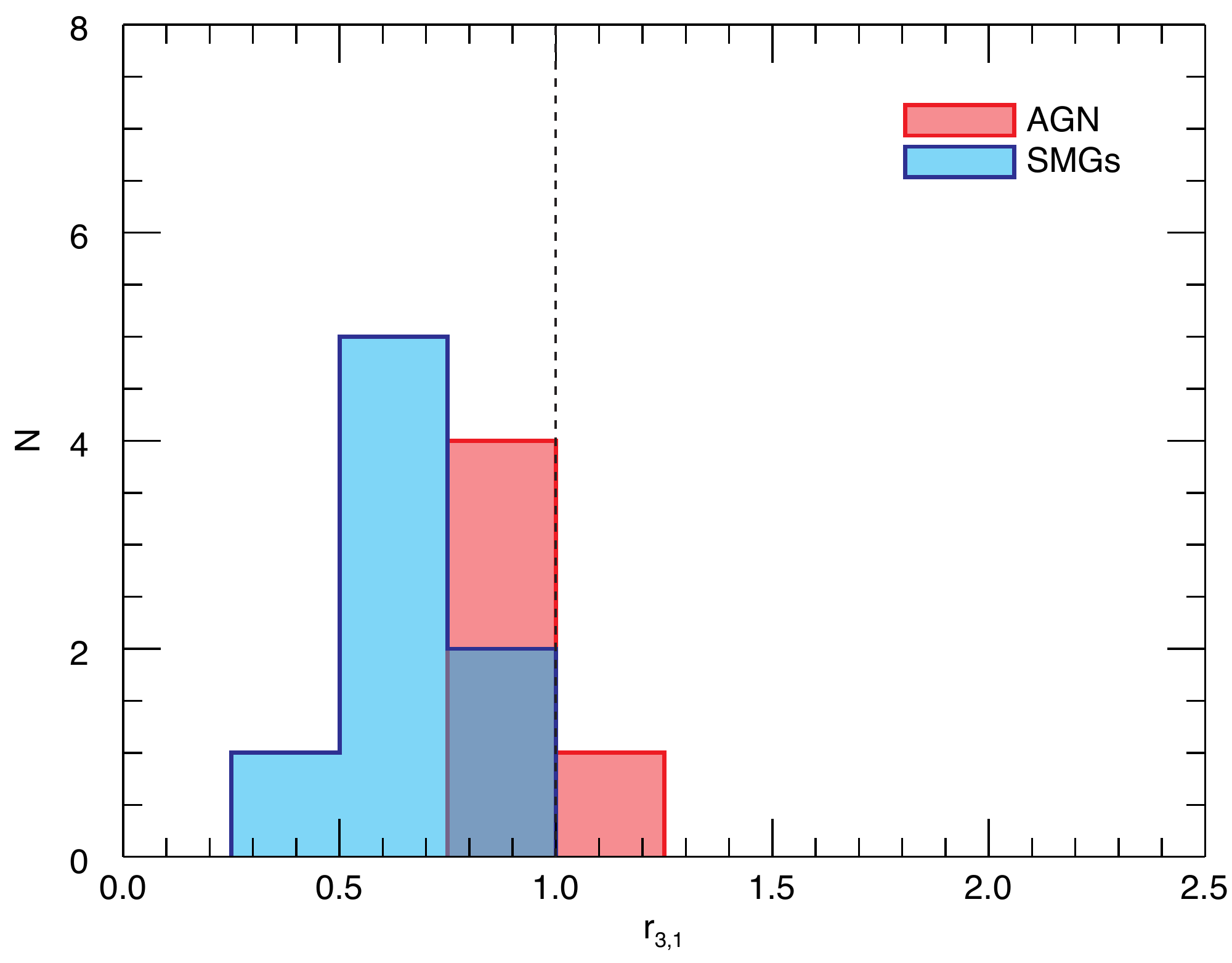}{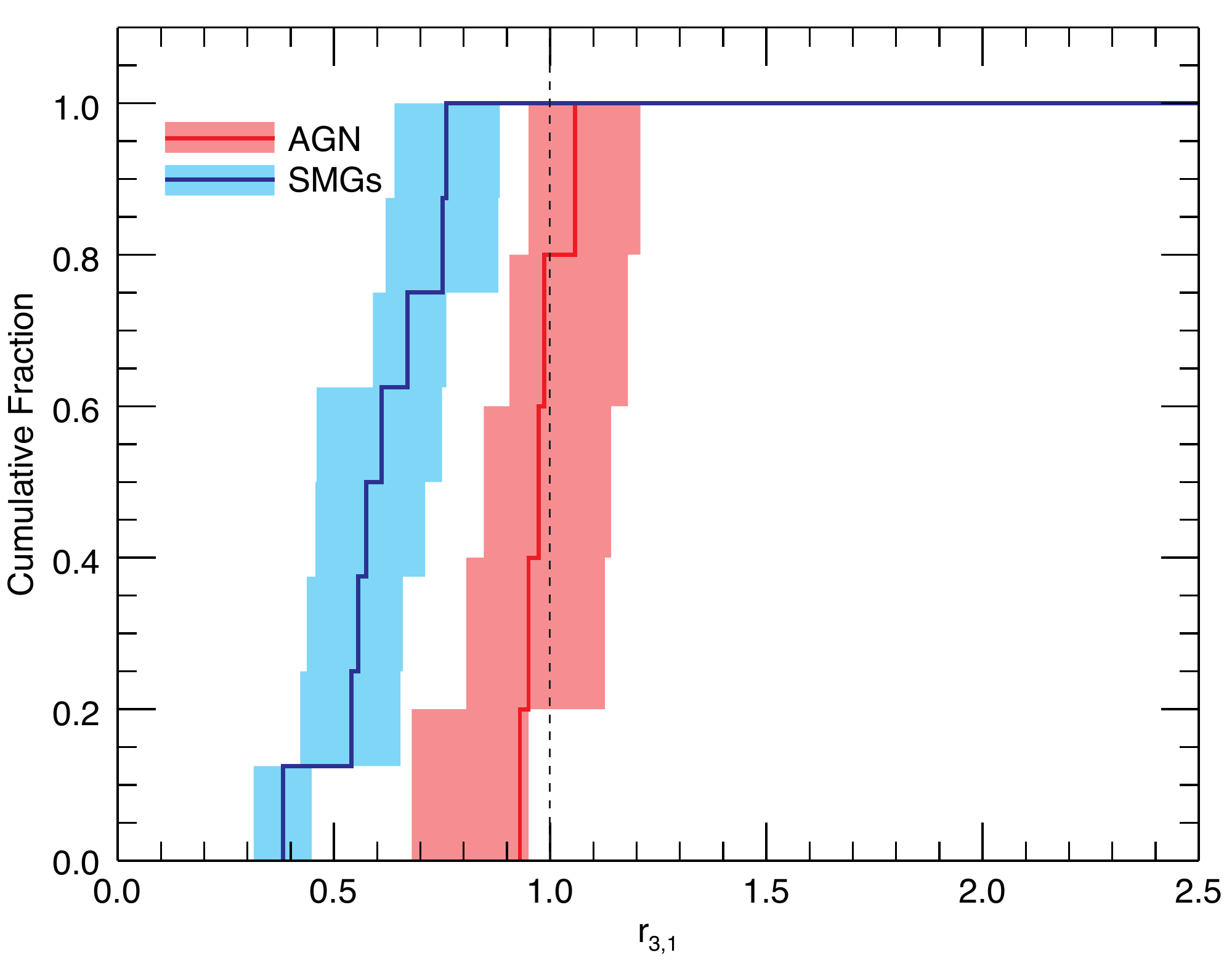}
\caption{Histogram (left) and cumulative distribution (right) of $r_{3,1}$ values for AGN host galaxies (red) and SMGs (blue) for the original detections that showed a clear difference between the two populations. Histogram bin sizes are $\Delta r_{3,1}=0.25$. Measured $r_{3,1}$ values are from \citet{harris2010, ivison2010a, ivison2011, riechers2011c} for the SMGs and \citet{riechers2011f} for the AGN host galaxies. For the cumulative distributions the shaded regions denote $\pm1\sigma$ regions for the distributions. The vertical dashed line shows $r_{3,1}=1$ for reference. \label{fig:r31histo_orig}}
\end{figure*}

\begin{figure*}
\plottwo{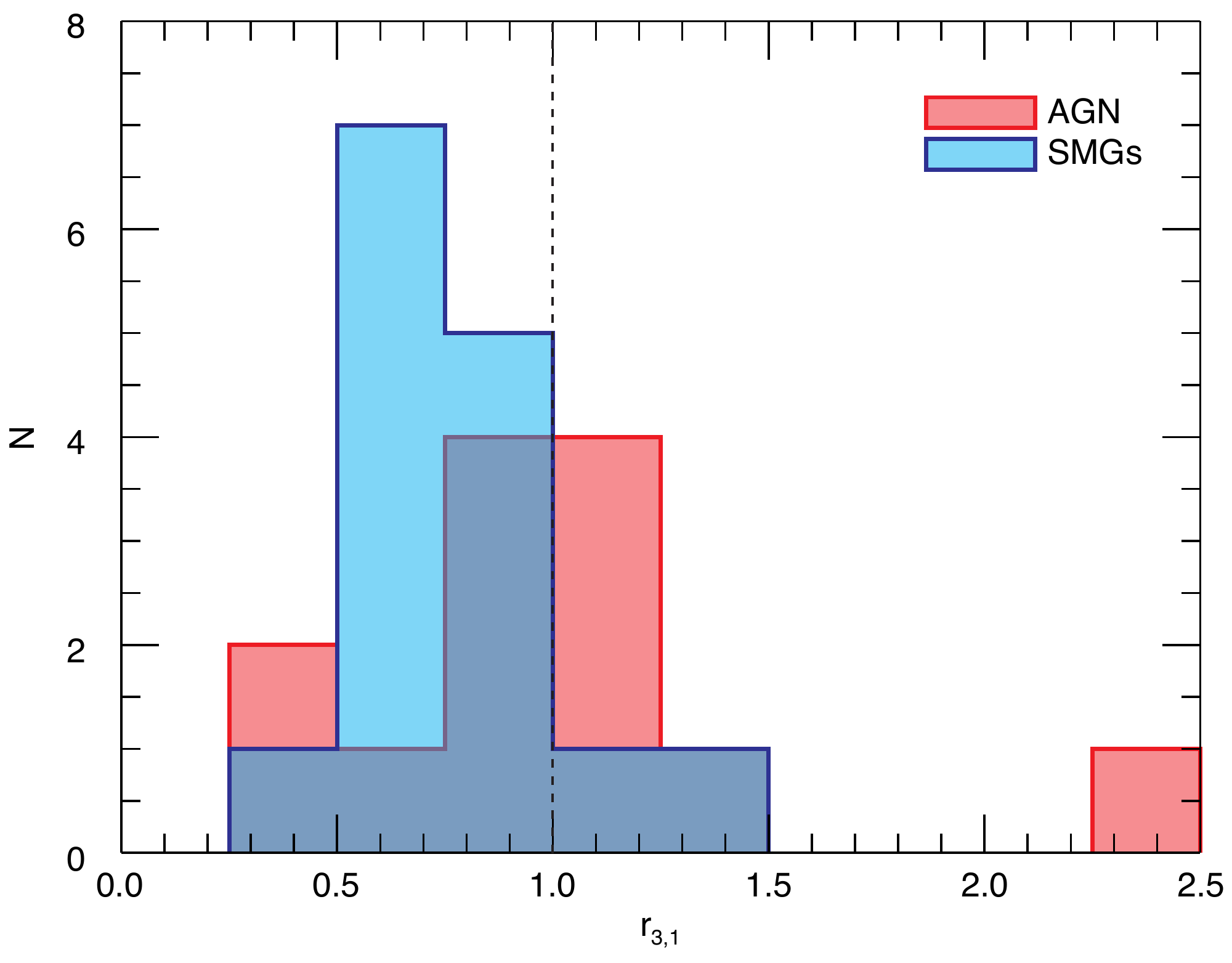}{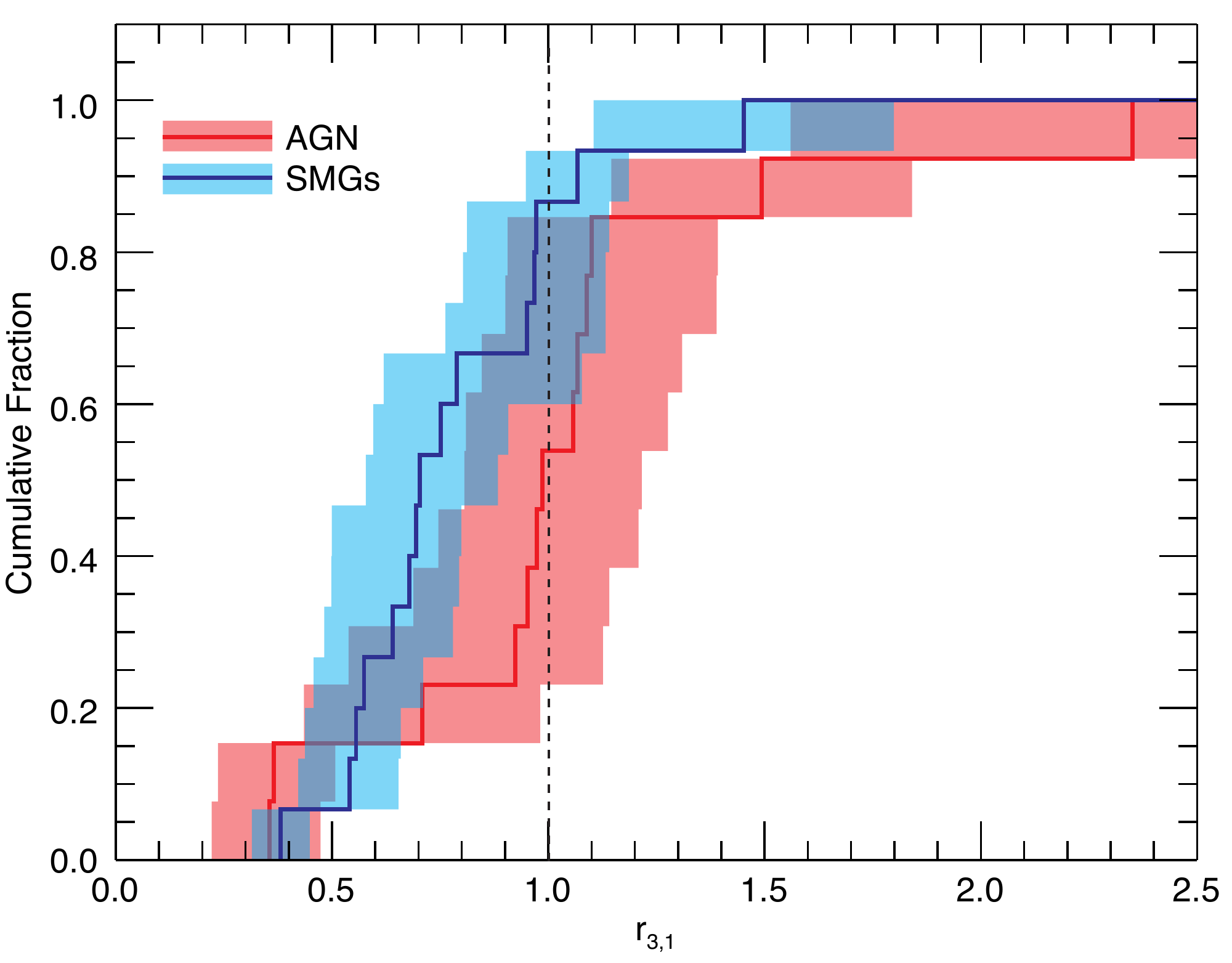}
\caption{Full updated histogram (left) and cumulative distribution (right) of $r_{3,1}$ values for AGN host galaxies (red) and SMGs (blue) including our new detections and literature values (which includes some updated classifications and \mbox{CO(1--0)} measurements). We perform a Student's t-test on the distribution of $r_{3,1}$ values and find no difference in the mean $r_{3,1}$ values  at the $\alpha=0.05$ significance level ($t=1.62$, $p=0.12$). We also perform a two-sample KS-test on the unbinned $r_{3,1}$ distributions, and find no difference in the shape of the distributions at the $\alpha=0.05$ significance level ($D=0.48$, $p=0.052$). Histogram bin sizes are $\Delta r_{3,1}=0.25$. Note that measured values of $r_{3,1}>1$ also have large uncertainties and are consistent with $r_{3,1}=1$. For the cumulative distributions the shaded regions denote $\pm1\sigma$ regions for the distributions. The vertical dashed line shows $r_{3,1}=1$ for reference. \label{fig:r31histo}}
\end{figure*}

Given the difference in $r_{3,1}$ values observed for SMGs and AGN host galaxies in \citet{harris2010, ivison2010a, ivison2011, riechers2011c, riechers2011f}, we look for this difference in our new expanded sample using \mbox{CO(3--2)} fluxes collected from the literature (Table~\ref{tab:measure}). In Figure~\ref{fig:r31histo_orig} we show the original distribution of $r_{3,1}$ values for AGN host galaxies and SMGs taken from the literature \citep{harris2010, ivison2010a, ivison2011, riechers2011c, riechers2011f} illustrating a clear difference in $r_{3,1}$ values for these populations. In Figure~\ref{fig:r31histo}, we show the updated distribution including our new $r_{3,1}$ measurements and any new/updated values from the literature. For the updated distributions of $r_{3,1}$ values, despite the tentative appearance of offsets for SMGs to lower $r_{3,1}$ values, we find no statistically significant difference between the population of SMGs and AGN host galaxies (assuming a significance threshold of $\alpha=0.05$ throughout). We perform a Student's t-test on the distribution of $r_{3,1}$ values and find $t=1.62$, which gives a probability of $p=0.12$ of measuring mean $r_{3,1}$ values at least as different as measured here, assuming the null hypothesis (that the SMGs and AGN have the same average $r_{3,1}$) is true. Using the cumulative distributions and a two-sample Kolmogorov-Smirnov test (KS-test) we obtain a test statistic of $D=0.48$, which gives a $p=0.052$ chance of measuring $r_{3,1}$ distributions at least as different as these given that AGN and SMGs have the same parent distribution, which is just above the significance threshold.

In order to examine the robustness of this result, we also evaluate the distribution of $r_{3,1}$ values while switching potential misclassifications or excluding particular measurements. Since there does appear to be some offset between the SMGs and AGN host galaxies in the cumulative distribution (albeit an offset that is not statistically significant), we particularly test physically motivated scenarios that might push the offset to statistical significance. Since three sources (one SMG and two AGN host galaxies) have unusually large $r_{3,1}$ values, likely due to emission below our detection threshold, we evaluate the binned and cumulative $r_{3,1}$ distributions with those sources removed. We find the probability of these two populations having different mean values or distributions decreases when we exclude the outliers, and obtain significance values of $p=0.18$ of measuring similar average $r_{3,1}$ values assuming the two populations have the same mean ($t=1.40$) and $p=0.086$ that the two populations have at least as similar distributions given that our measured values for AGN and SMGs are drawn from the same parent population ($D=0.47$). Five of our new measurements are only tentatively detected at the $3$--$5\sigma$ level (HE\,0230-2130, RX\,J1249-0559, J1543+5359, HS\,1611+4719, and J22174+0015). We perform both statistical tests with these five sources removed and do not find a statistically significant difference between the mean $r_{3,1}$ values of the two populations at the $\alpha=0.05$ level ($t=1.06$; $p=0.31$) nor between the two distributions ($D=0.52$; $p=0.063$). We also look for differences between the two distributions removing these five tentative detections and the remaining two outlier measurements, and we still find no difference between the mean and distributions of the two populations' $r_{3,1}$ values at the $\alpha=0.05$ level ($t=1.05$, $p=0.32$; $D=0.55$, $p=0.064$). We conclude that our marginal detections are not significantly influencing our comparisons between the two populations. 

Three objects (J00266+1708, J02399--0136, and J14009+0252, discussed further below) have somewhat ambiguous classifications as SMGs. We perform both statistical tests where we have swapped the classifications (SMGs to AGN host galaxies) for these three objects and find $p=0.12$ chance of measuring average $r_{3,1}$ values at least similar as these given the null hypothesis (that AGN and SMGs have the same mean $r_{3,1}$; $t=1.60$) and $p=0.04$ chance of measuring $r_{3,1}$ distributions at least as similar as obtained here ($D=0.50$). We also check whether removing the three $r_{3,1}>1.4$ measurements affects the resulting probabilities when we have switched these three classifications, and obtain similar results: $p=0.097$ significance for the similarity of the mean value ($t=1.73$) and $p=0.038$ significance for the similarity of the distributions ($D=0.53$). Switching the three objects' classifications from SMGs to AGN and without removing the three outliers is the only way to make the difference in distributions between the SMGs and AGN host galaxies statistically significant with our new measurements. Given this degree of manipulation and the limited sensitivities of some new detections, we conclude that we cannot reject the null hypothesis that $r_{3,1}$ values for SMGs and AGN host galaxies are the same with our present data, obtaining a global average $r_{3,1}=0.90\pm0.40$ (or $r_{3,1}=1.03\pm0.50$ for AGN host galaxies and $r_{3,1}=0.78\pm0.27$ for SMGs, which are consistent with the previous values to within the uncertainties). However, a good deal of the scatter on these average values is caused by the outliers where we have likely resolved out part of the flux; excluding the the three sources with $r_{3,1}>1.4$, we obtain an average $r_{3,1}=0.79\pm0.24$ (or $r_{3,1}=0.87\pm0.27$ for AGN host galaxies and $r_{3,1}=0.73\pm0.20$ for SMGs). The standard deviations of these populations' \mbox{CO(3--2)}/\mbox{CO(1--0)} ratios are likely somewhat inflated by the scatter introduced from the larger uncertainties on some our new detections; larger sample sizes and improved $r_{3,1}$ measurements for the outliers and weakly detected sources would help constrain the true means and distributions of these populations' $r_{3,1}$ values.

Given the previous \mbox{CO(1--0)} detections that exist for some of these sources, we also consider the effect of using older measurements on the distributions of $r_{3,1}$ values (even though individual galaxies' measurements are largely consistent with one another). Using \emph{all} of the older values reproduces the previously observed difference in $r_{3,1}$ values between SMGs and AGN host galaxies, both in the mean and distribution of $r_{3,1}$ values (again, assuming a significance threshold of $\alpha=0.05$); the Student's t-test yields $t=2.16$, which gives $p=0.048$ chance of measuring average $r_{3,1}$ values at least as different as we obtain given the null hypothesis that the two populations have the same mean $r_{3,1}$, and the KS-test yields $D=0.57$, which gives a $p=0.012$ chance of obtaining distributions at least as different as we measure given the null hypothesis that the AGN and SMG $r_{3,1}$ values are drawn from the same parent population. The likelihood of these differences reduce a bit if we remove the remaining two outliers with $r_{3,1}>1.4$ (HE1104--1805 and J04135+10277), giving $p=0.12$ and $p=0.037$ for the statistical tests of differences in the means and distributions, respectively. Since J04135+10277 is an outlier in $r_{3,1}$ \emph{and} has a previous single-dish measurement of the \mbox{CO(1--0)} flux which is noticeably discrepant from our new measurement, we also evaluate the $r_{3,1}$ distributions using the older single-dish value for J04135+10277 on its own (since there are no obvious problems or discrepancies between the other interferometric and single-dish measurements). The Student's t-test gives a $p=0.071$ ($t=1.94$) chance of obtaining average $r_{3,1}$ measurements at least as different as measured given the null hypothesis that the two populations share the same mean; the KS-test gives a $p=0.018$ ($D=0.55$) chance of measuring $r_{3,1}$ distributions as different as these given the null hypothesis that SMGs and AGN have the same $r_{3,1}$ distribution. Removing the remaining two galaxies with $r_{3,1}>1.4$ increases those likelihoods to a $p=0.22$ ($t=1.28$) chance of measuring at least as similar average $r_{3,1}$ values and a $p=0.074$ ($D=0.48$) probability of measuring $r_{3,1}$ distributions at least as similar. These results show that any differences in the mean between the two populations are largely driven by our outlier measurements where we have likely resolved out weak extended emission, and therefore there is no statistically significant differences in their average $r_{3,1}$ values. However, the KS tests' results are more ambiguous, and tentatively suggest there may be some difference in the $r_{3,1}$ distributions for SMGs and AGN host galaxies although that result is not uniformly statistically significant. The apparent disappearance of the difference in $r_{3,1}$ values between SMGs and AGN host galaxies has multiple causes, including new measurements and misclassifications. Improved S/N observations of weakly detected sources and sources with  $r_{3,1}>1.4$ would help resolve the remaining ambiguities in potential differences in these populations, as would an increased number of SMGs and AGN with $r_{3,1}$ measurements.

While intriguing, it is perhaps not surprising that the difference in $r_{3,1}$ values for SMGs and AGN largely disappears in our expanded sample. First, the excitation temperature for the \mbox{CO(3--2)} transition is not so high that it requires a particularly energetic source (like an AGN) to drive its excitation; a compact starburst and molecular gas reservoir could plausibly create the near-thermalized $r_{3,1}\sim1$ we observe in some SMGs' molecular gas \citep[e.\/g.\/,][]{narayanan2014}. Second, the division between AGN and SMGs may not be strict. It very possible that SMGs contain a dust-enshrowded AGN (which may also be a viewing angle effect) that is not identified in optical wavelengths. Several objects are suspected of potentially harboring an AGN based on their continuum emission at mid-IR wavelengths \citep[J00266+1708;][]{valiante2007, sharon2015}, large line widths \citep[J02399--0136 and J14009+0252; e.\/g.\/,][]{ivison1998, ivison2000, thomson2012}, or high-$J$ CO emission and radio excess \citep[HLSW-01; e.\/g.\/,][]{scott2011}, and these object contain a range of line ratios ($r_{3,1}\sim0.6$--$1$). As samples of $z\sim2$ galaxies with \mbox{CO(1--0)} detections expand beyond the initial small numbers that largely contained the best-studied and most well-characterized objects, one would expect to find more hybrid objects and galaxies with ambiguous classifications. In addition, several types of AGN host galaxies are considered in this study, including optically-selected quasars and highly-obscured AGN selected in the infrared, which may alter the distribution of $r_{3,1}$ values if there are systematically different molecular gas conditions in different types of AGN host galaxies. Lastly, galaxies' AGN may not be the dominant source of dust and gas heating, even if the AGN are providing some additional excitation for low-$J$ CO lines. We also note that observations of galaxies at low-$z$ have also measured a range of $r_{3,1}$ values for systems which do and do not host a central AGN \citep[e.\/g.\/,][]{mauersberger1999, yao2003, mao2010}.

\begin{figure*}
\plotone{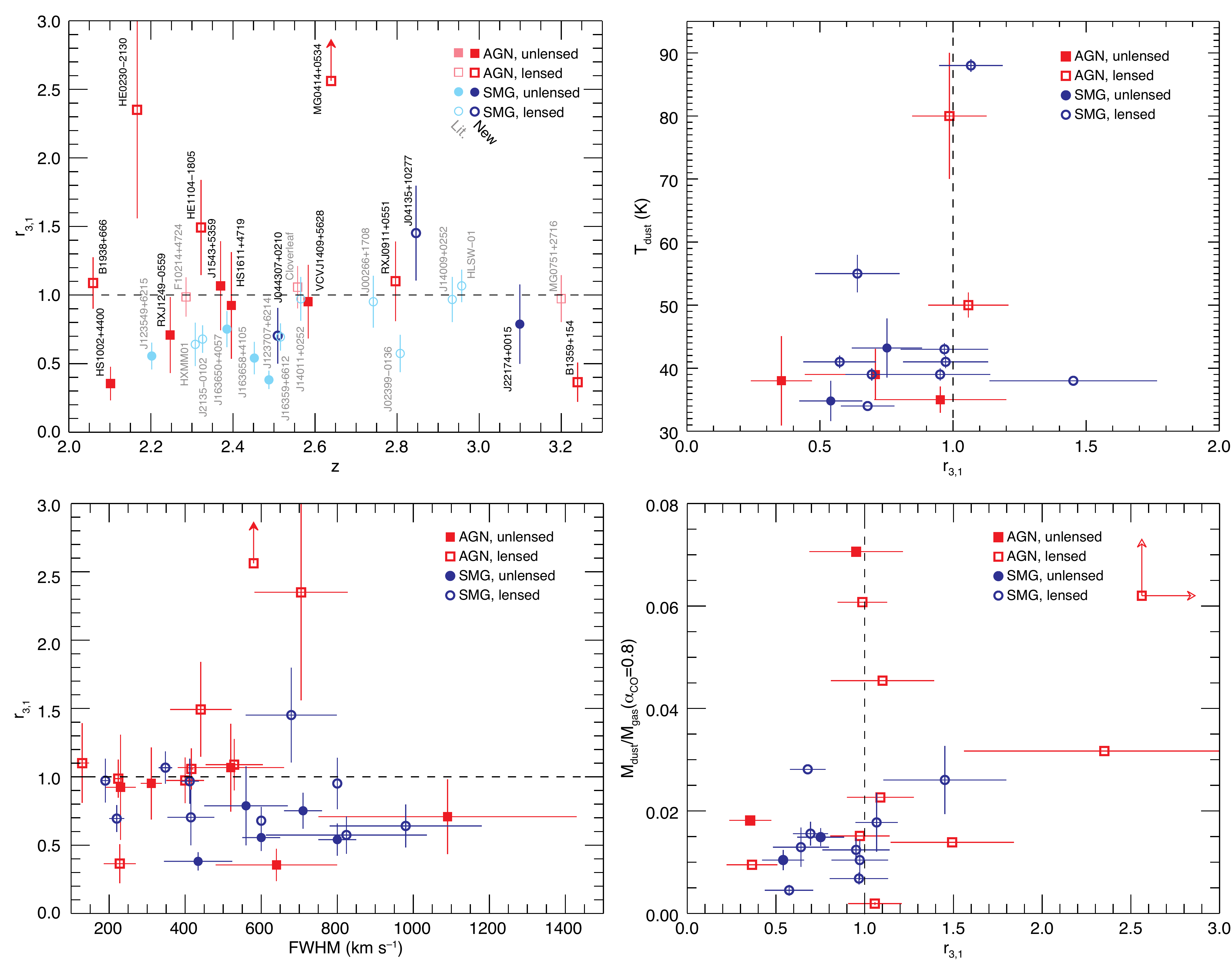}
\caption{Redshift (top left), dust temperature (top right), \mbox{CO(3--2)} line FWHM (bottom left), and dust-to-gas ratio (bottom right) as function of $r_{3,1}$ for the complete sample of SMGs (blue circles) and AGN (red squares). Gravitationally lensed objects are shown with hollow symbols and unlensed objects have filled symbols; all gravitationally lensed objects have been lensing-corrected. The $r_{3,1}=1$ line (dashed) and upper limits for new measurements are also shown. For the latter three panels, we adopt values from the literature (when available) which may not have published uncertainties (see Table~\ref{tab:measure} and \ref{tab:other}). In only the top left panel, literature values are in lighter shades of blue/red to highlight the new observations and sources are labeled by name. For the dust-to-gas mass we assume a CO luminosity to molecular gas mass conversion factor of $\alpha_{\rm CO}=0.8\,{\rm M_\sun\,(K\,km\,s^{-1}\,pc^2)^{-1}}$. We calculate Pearson's and Spearman's correlation coefficients to look for monotonic trends for these variables and find no significant correlationa at the $\alpha=0.05$ significance level (except for $r_{3,1}$ vs.~$z$ for SMGs and $r_{3,1}$ vs.~FWHM using both populations except sources with $r_{3,1}>1.4$).
\label{fig:r31nocorr}}
\end{figure*}

It may be that the excitation difference as probed by the \mbox{CO(3--2)}/\mbox{CO(1--0)} line ratio for these objects is actually a function of other physical parameters of the galaxies (Table~\ref{tab:other}). In subsequent sections we explore the excitation dependence of these galaxies' star formation rates and dynamical properties. We calculate Pearson's and Spearman's correlation coefficients to look for monotonic trends in $r_{3,1}$ with redshift, dust temperature, line FWHM, dust-to-gas mass ratio (Figure~\ref{fig:r31nocorr}), and dust mass (not pictured). We find no significant correlation for SMGs, AGN, or both population in aggregate except for $r_{3,1}$ vs.~$z$ for SMGs and $r_{3,1}$ vs.~FWHM for both populations in aggregate (but excluding the three sources with $r_{3,1}>1.4$). For the potential correlation with redshift, the likelihood of obtaining as tight of a trend assuming the null hypothesis of no correlation is true has a frequency of $\alpha<0.05$ for Spearman's $\rho=0.67$ ($p=0.0061$) and Kendall's $\tau=0.47$ ($p=0.015)$. For the potential correlation with the line FWHM (which uses the \mbox{CO(3--2)} FWHMs from the literature since we cannot measure the FWHM of the \mbox{CO(1--0)} line for all objects in our sample), the likelihood of obtaining as tight of a trend assuming the null hypothesis of no correlation is true has a frequency of $\alpha<0.05$ for Spearman's $\rho=-0.44$ ($p=0.029$), but only $p=0.055$ for Kendall's $\tau=-0.055$. Previous results for local galaxies \citep{mauersberger1999, yao2003, mao2010} show no correlation between $r_{3,1}$ and any value (except star formation efficiency; see subsequent section), but those studies obviously cannot address trends with redshift. Given the number of correlations we explore here, it is not surprising that we find one or two to be statistically significant; expanding the sample of $r_{3,1}$-measured SMGs would provide greater confidence on whether these trends persist or not.

\begin{deluxetable*}{lcccc}
\tablecaption{Other Galaxy Properties from the Literature Used Here \label{tab:other}}
\tablehead{{Source}	& {$T_{\rm dust}$} & {$M_{\rm dust}$} & {$\log(M_{\rm BH}/M_\sun)$} & {References}\\
{} & {(K)} & {$(10^8\,{\rm M_\sun})$} & {} & {}}
\startdata
\multicolumn{5}{c}{New}\\
\hline
B1938+666 & {\nodata} & {$0.21$\tablenotemark{a}} & {\nodata} & {\citet{barvainis2002}} \\
HS\,1002+4400 & {$38\pm7$} & {$17.30$} & {$10.14\pm0.20$} & {\citet{beelen2006,coppin2008}} \\
HE\,0230--2130 & {\nodata} & {$1.62$\tablenotemark{a}} & {$7.95\pm0.24$} & {\citet{barvainis2002,pooley2007}} \\
RX\,J1249--0559 & {$39\pm4$} & {\nodata} & {$9.76\pm0.20$} & {\citet{khan2015,coppin2008}} \\
HE\,1104--1805 & {\nodata} & {$1.53$\tablenotemark{a}} & {$9.38$} & {\citet{barvainis2002,peng2006}} \\
J1543+5359 & {\nodata} & {\nodata} & {$10.13\pm0.15$} & {\citet{coppin2008}} \\
HS\,1611+4719 & {\nodata} & {\nodata} & {$9.26\pm0.2$} & {\citet{coppin2008}} \\
J044307+0210	& {\nodata} & {\nodata} & {\nodata} & {\nodata} \\
VCV\,J1409+5628 & {$35\pm2$} & {$48.60$} & {$9.28\pm0.20$} & {\citet{beelen2006,coppin2008}} \\
MG\,0414+0534 & {\nodata} & {$0.93$\tablenotemark{a}} & {$9.04\pm0.17$} & {\citet{barvainis2002,pooley2007}} \\
RX\,J0911+0551 & {\nodata} & {$1.37$\tablenotemark{a}} & {$8.6\pm0.18$} & {\citet{barvainis2002,pooley2007}} \\
J04135+10277	& {$38$} & {$18\pm3$} & {\nodata} & {\citet{knudsen2003}} \\
J22174+0015 & {\nodata} & {\nodata} & {\nodata} & {\nodata} \\
B1359+154 & {\nodata} & {$0.11$\tablenotemark{a}} & {\nodata} & {\citet{barvainis2002}} \\
\hline
\multicolumn{5}{c}{Literature}\\
\hline
J123549+6215 & {\nodata} & {\nodata} & {\nodata} & {\nodata} \\
F10214+4724 & {$80\pm10$} & {$2.84$} & {$8.36\pm0.56$} & {\citet{ao2008,deane2013c}} \\
HXMM01 & {$55\pm3$} & {$29\pm7$} & {\nodata} & {\citet{fu2013}} \\
J2135-0102 & {$34$} & {$4.00$} & {\nodata} & {\citet{ivison2010b}} \\
J163650+4057	& {$43.2\pm4.7$} & {$9.69\pm0.17$} & {\nodata} & {\citet{kovacs2006}} \\
J163658+4105	& {$34.8\pm3.2$} & {$9.04\pm0.14$} & {\nodata} & {\citet{kovacs2006}} \\
J123707+6214	& {\nodata} & {\nodata} & {\nodata} & {\nodata} \\
J16359+6612 & {$39\pm1$\tablenotemark{b}} & {$0.78\pm0.09$\tablenotemark{b}} & {\nodata} & {\citet{magnelli2012}} \\
Cloverleaf & {$50\pm2$} & {$0.61\pm0.07$} & {$9.24\pm0.51$} & {\citet{weiss2003,pooley2007}} \\
J14011+0252 & {$41\pm1$} & {$3.07\pm1.08$} & {\nodata} & {\citet{magnelli2012}} \\
J00266+1708 & {$39\pm1$} & {$4.47\pm0.52$} & {\nodata} & {\citet{magnelli2012}} \\
J02399-0136 & {$41\pm1$} & {$0.32\pm0.04$} & {\nodata} & {\citet{magnelli2012}} \\
J14009+0252 & {$43\pm1$} & {$4.47\pm0.52$} & {\nodata} & {\citet{magnelli2012}} \\
HLSW-01 & {$88$} & {$5.20\pm1.60$} & {\nodata} & {\citet{scott2011}} \\
MG\,0751+2716 & {\nodata} & {$1.82$\tablenotemark{a}} & {\nodata} & {\citet{barvainis2002}}
\enddata
\tablenotetext{a}{Corrected to account for the cosmology, redshifts, and magnifications factors assumed here (see Table~\ref{tab:measure}).}
\tablenotetext{b}{Averaged over the three components.}
\end{deluxetable*}

\subsection{Excitation dependence of galaxies' star formation properties}

\begin{figure}
\epsscale{1.0}
\plotone{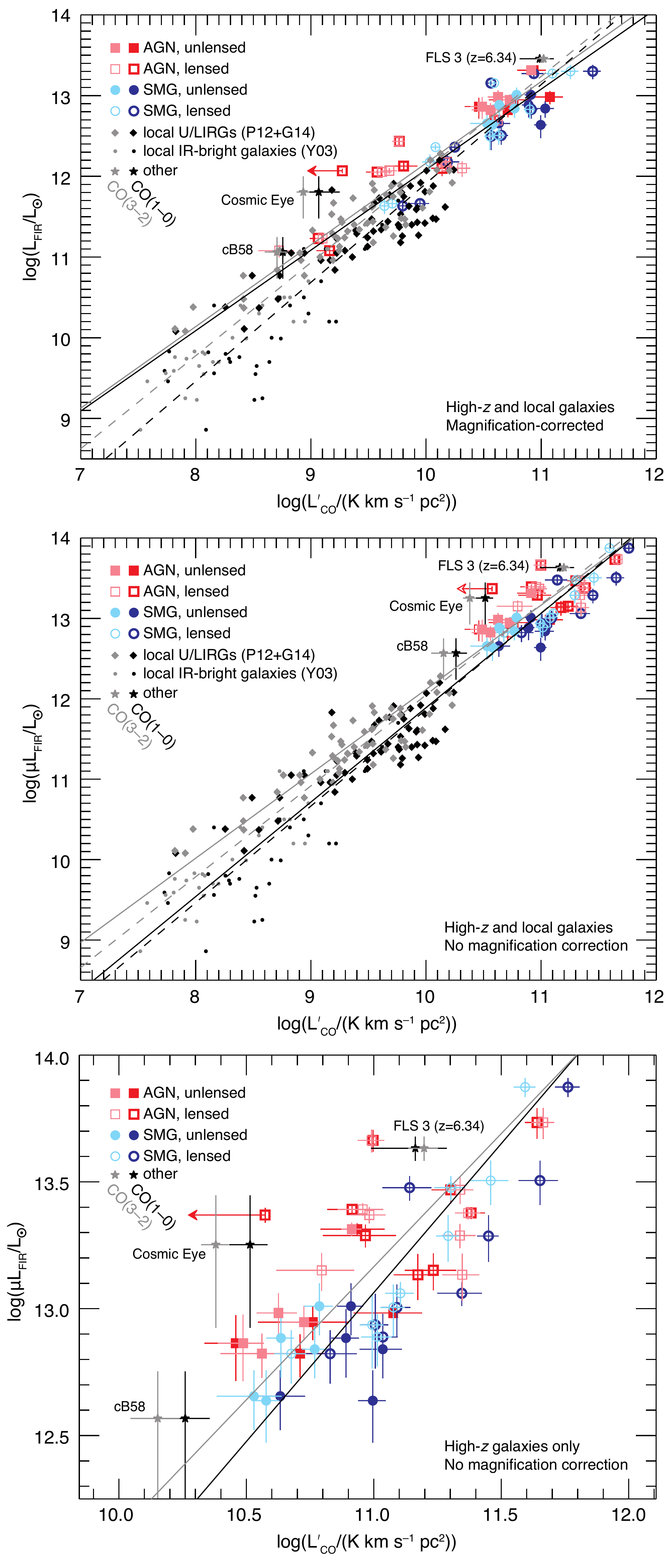}
\caption{Integrated Schmidt-Kennicutt relation where the high-redshift sources have either been magnification-corrected (top) or not magnification corrected (middle/bottom), where the bottom panel shows a zoom-in of only the high-$z$ sources. AGN host galaxies are shown as red squares, SMGs are blue circles, low-$z$ U/LIRGs from \citet{papadopoulos2012d}/\citet{greve2014} are diamonds, low-$z$ IR-bright galaxies from \citet{yao2003} are dots, and other reference high-$z$ sources are the labeled stars. For the SMGs and AGN, hollow symbols denote objects that are gravitationally lensed. Darker colored symbols/lines use the \mbox{CO(1--0)}-determined CO line luminosities, while the lighter colored symbols/lines use the \mbox{CO(3--2)}-determined CO line luminosities. Solid lines are fits to the Schmidt-Kennicutt relation using only the high-$z$ objects, while dashed lines include the low-$z$ sources; the fits do not use the three labeled comparison objects. We find no significant difference in either the slope or offset for the integrated Schmidt-Kennicutt relation between any of the high-$z$ sub-populations, with or without correction for lensing magnification, or between the correlation using the \mbox{CO(3--2)} and \mbox{CO(1--0)} lines. All of the best-fit indices are consistent with a slope of $N=1$ at the $\lesssim2\sigma$ level unless we include the \citet{yao2003} IR-bright sample (see Table~\ref{tab:SKrel}).
\label{fig:SKrel}}
\end{figure}

Since we do not have spatially resolved measurements of the SFR or molecular gas for most objects in our new observations, nor for most literature objects, we evaluate the effects of CO excitation on the integrated form of the Schmidt-Kennicutt relation. By looking at the correlation between $L_{\rm FIR}$ and $L^\prime_{\rm CO}$ we can also avoid uncertainty in the gas mass conversion factor, $\alpha_{\rm CO}$; the traditionally assumed bi-modal values of $\alpha_{\rm CO}$ are suspected to strongly affect comparisons between galaxy populations, including comparisons between galaxies at different redshifts. In addition, the luminosity-luminosity correlation may be better for probing variations in $\alpha_{\rm CO}$ with gas physical conditions (e.\/g.\/, gas temperature and density; \citealt{Bolatto2013} and references therein) which also set the relative CO line strengths. We analyze the integrated Schmidt-Kennicutt relation (Figure~\ref{fig:SKrel}) using both the SMGs in our sample and the galaxies known to host a bright central AGN even though the AGN may be contributing additional luminosity that is not associated with star formation. 

In Table~\ref{tab:SKrel} we list both the offset and the index to our fits for AGN and SMGs, analyzed both separately and together, and in combination with local U/LIRGs from \citet{papadopoulos2012d}/\citet{greve2014} and the low-$z$ IR-bright galaxies from \citet{yao2003}, which generally reach lower luminosities\footnote{Some sources are repeated between the two low-$z$ samples, in which case we use the \citet{papadopoulos2012d,greve2014} values.} All fits are from an ordinary least squares bisector linear regression. Since the molecular gas measurements and FIR luminosities are collected from across the literature and use a variety of methods, we neglect measurement uncertainties in the linear fit and assume an equal weight for every measurement, neglecting all upper limits. We also examine potential differences in the Schmidt-Kennicutt relation for these sources as a function of gas excitation and check that including certain subsets of sources do not bias our results (particularly the \mbox{CO(3--2)}-detected source that only has a \mbox{CO(1--0)} upper limit, the two AGN host galaxies B1938+666 and B1359+154, which potentially have very low intrinsic luminosities, and the Cloverleaf and J2135-0102, which have high luminosities prior to magnification correction). Since the lensing magnification factors are an additional source of uncertainty, particularly for the largest magnifications, we also evaluate the Schmidt-Kennicutt relation without corrections for gravitational lensing to illustrate the potential range of effects caused by applying incorrect magnification factors.

We note that \citet{greve2014} uses the $50$--$300\,{\rm \mu m}$ integrated flux to determine $L_{\rm FIR}$ while \citet{yao2003} uses the $40$--$1000\,{\rm \mu m}$ range. Since the two samples use different techniques for fitting the dust SED, one cannot use a simple rescaling to correct between the two wavelength regimes. For typical assumptions of dust temperatures and modified black body indices, the correction only ranges between $0$--$10\%$ for these two low-$z$ samples, but can be as high as $\sim30\%$ to correct to $L_{\rm FIR}$ wavelength coverage assumed for many of the high-$z$ galaxies studied here ($40$--$400\,{\rm \mu m}$). A more uniform determination of the FIR luminosity across populations, including corrections for AGN contamination (like in \citealt{greve2014}; only five galaxies on our high-$z$ sample are shared between our analyses) would improve the robustness these results.

\begin{deluxetable*}{lcccccc}
\tablewidth{0pt}
\tablecaption{Fits to the integrated Schmidt-Kennicutt relation \label{tab:SKrel}}
\tablehead{{Sources} & \multicolumn{2}{c}{CO(1--0)} & \multicolumn{2}{c}{CO(3--2) match\tablenotemark{a}} & \multicolumn{2}{c}{CO(3--2) all\tablenotemark{b}} \\
{} & {offset} & {slope} & {offset} & {slope} & {offset} & {slope}}
\startdata
\multicolumn{7}{c}{Magnification-corrected}\\
\hline
SMGs & {$0.42\pm1.10$} & {$1.14\pm0.11$} & {$1.10\pm0.67$} & {$1.10\pm0.07$} & {\nodata} & {\nodata} \\
AGN & {$2.02\pm0.89$} & {$1.02\pm0.09$} & {$2.09\pm0.60$} & {$1.01\pm0.06$} & {$2.21\pm0.54$} & {$1.00\pm0.05$} \\
AGN (w/o bias)\tablenotemark{c} & {$3.67\pm1.27$} & {$0.86\pm0.12$} & {$1.43\pm1.95$} & {$1.08\pm0.18$} & {$2.01\pm1.42$} & {$1.02\pm0.13$} \\
SMGs+AGN & {$2.11\pm0.77$} & {$1.00\pm0.07$} & {$2.05\pm0.61$} & {$1.01\pm0.06$} & {$2.18\pm0.58$} & {$1.00\pm0.06$} \\
SMGs+AGN (w/o bias)\tablenotemark{c} & {$2.35\pm1.20$} & {$0.98\pm0.11$} & {$1.40\pm1.00$} & {$1.07\pm0.09$} & {$1.75\pm0.93$} & {$1.04\pm0.09$} \\
SMGs+AGN+U/LIRGs & {$1.14\pm0.43$} & {$1.08\pm0.04$} & {$1.87\pm0.30$} & {$1.02\pm0.03$} & {\nodata} & {\nodata} \\
SMGs+AGN+all low-$z$ & {$-0.34\pm0.45$} & {$1.22\pm0.05$} & {$0.56\pm0.27$} & {$1.15\pm0.03$} & {\nodata} & {\nodata} \\
\hline
\multicolumn{7}{c}{No magnification correction}\\
\hline
SMGs & {$-1.93\pm2.43$} & {$1.35\pm0.22$} & {$-0.34\pm1.56$} & {$1.22\pm0.14$} & {\nodata} & {\nodata} \\
SMGs (w/o bias)\tablenotemark{d} & {$-2.36\pm3.25$} & {$1.34\pm0.29$} & {$-0.14\pm2.02$} & {$1.20\pm0.19$} & {\nodata} & {\nodata} \\
AGN & {$2.89\pm1.36$} & {$0.94\pm0.12$} & {$4.55\pm0.92$} & {$0.79\pm0.09$} & {$4.49\pm0.91$} & {$0.80\pm0.09$} \\
AGN (w/o bias)\tablenotemark{d} & {$2.36\pm1.93$} & {$0.99\pm0.18$} & {$4.34\pm1.50$} & {$0.81\pm0.14$} & {$4.21\pm1.44$} & {$0.82\pm0.14$} \\
SMGs+AGN & {$0.15\pm1.61$} & {$1.17\pm0.15$} & {$1.62\pm1.13$} & {$1.05\pm0.10$} & {$1.59\pm1.11$} & {$1.05\pm0.10$} \\
SMGs+AGN (w/o bias)\tablenotemark{d} & {$-0.28\pm2.12$} & {$1.21\pm0.19$} & {$1.53\pm1.45$} & {$1.06\pm0.13$} & {$1.44\pm1.41$} & {$1.07\pm0.13$} \\
SMGs+AGN+U/LIRGs & {$0.90\pm0.38$} & {$1.10\pm0.04$} & {$1.71\pm0.27$} & {$1.04\pm0.03$} & {\nodata} & {\nodata} \\
SMGs+AGN+all low-$z$ & {$-0.14\pm0.37$} & {$1.20\pm0.04$} & {$0.74\pm0.23$} & {$1.13\pm0.02$} & {\nodata} & {\nodata} \\
\hline
\multicolumn{7}{c}{Low-redshift sources only}\\
\hline
{IR-bright (Y03)} & {$-0.89\pm1.15$} & {$1.27\pm0.13$} & {$-0.24\pm0.63$} & {$1.23\pm0.07$} & {\nodata} & {\nodata} \\
{U/LIRGs (P12+G14)} & {$2.63\pm0.54$} & {$0.92\pm0.06$} & {$3.10\pm0.36$} & {$0.89\pm0.04$} & {\nodata} & {\nodata} \\
{all} & {$-0.20\pm0.64$} & {$1.21\pm0.07$} & {$0.53\pm0.39$} & {$1.15\pm0.04$} & {\nodata} & {\nodata}
\enddata
\tablenotetext{a}{Uses only \mbox{CO(3--2)} measurements for sources with \mbox{CO(1--0)} detections.}
\tablenotetext{b}{Uses all \mbox{CO(3--2)} measurements.}
\tablenotetext{c}{Excludes the AGN B1938+666 and B1359+154 which are outliers and would dominate the linear fit.}
\tablenotetext{d}{Excludes the AGN Cloverleaf and/or the SMG J2135-0102 which are outliers and would dominate the linear fit.}
\end{deluxetable*}
\normalsize

In short, we find no significant difference in either the slope or offset for the integrated Schmidt-Kennicutt relation between any of the high-$z$ sub-populations, with or without correction for lensing magnification, or between the correlation using the \mbox{CO(3--2)} and \mbox{CO(1--0)} lines. All of the best-fit indices are consistent with a slope of $N=1$ at the $\lesssim2\sigma$ level. In addition, the offset (or y-intercept) of our best-fit functions are consistent and have considerable uncertainty. One might expect the offset to be different between the correlations using the \mbox{CO(3--2)} and \mbox{CO(1--0)} line luminosities since we do not apply an excitation correction, but the expected offset is only $N\times\log(r_{3,1})$, which is at most a difference in the offset of $\sim0.1$--$0.3$ and well within our uncertainties. We do find some tension between the offsets and slopes measured for the high-$z$ SMGs and AGN (for the \mbox{CO(1--0)} line and the \mbox{CO(3--2)} line without magnification corrections), but the inconsistencies are $\lesssim2\sigma$. As expected, removing corrections for lensing magnification adds considerable scatter to the best-fit relationships of the high-$z$ samples.

The most conspicuous difference we find in our analysis of the integrated Schmidt-Kennicutt relation comes with the inclusion of the low-$z$ more ``normal" IR-bright galaxies from \citet{yao2003}. Including the low-$z$ IR-bright galaxies we get a super-linear slope of $N\sim1.15$--$1.2$. Whether there is an offset (i.e. a change in star formation efficiency) between starburst galaxies and quiescent galaxies or simply a non-unity slope in the Schmidt-Kennicutt relation is a long-standing debate \citep[e.\/g.\/,][]{daddi2010,genzel2010}. However, observed offsets between starburst and quiescent star-forming galaxies have largely been attributed to different assumptions about gas-mass conversion factors. Here we have made no assumptions on conversion factor yet still find offsets/index-changes depending on which galaxies we include in the linear fits. This result indicates that assumed gas mass conversion factors are not the sole and \emph{artificial} cause of high SFEs for U/LIRGs and SMGs, which has also been seen in some spatially resolved measurements \citep[e.\/g.\/,][]{sharon2013} (although this result does not rule out that different galaxy populations may have different CO-to-${\rm H_2}$ abundances). Although the starbursts and normal galaxies analyzed here prefer different Schmidt-Kennicutt indices, the \citet{yao2003} galaxies dominate linear fits to combined populations due to the large scatter and uncertainties on the high-$z$ galaxies and different luminosity regimes covered by these populations. The scatter and different luminosity ranges make it difficult to distinguishing between the scenario where starbursts (and AGN) are offset from the Schmidt-Kennicutt relation for local galaxies and the scenario where starbursts are not offset but instead extend the super-linear local relation to higher luminosities. Interestingly, linear fits to the SMGs on their own are mostly consistent with fits to the \citet{yao2003} sample. More consistent measurements of both the FIR and CO luminosities across populations would help clarify these results (as illustrated by the differing indices determined using the same U/LIRG data in \citealt{greve2014} vs.~\citealt{kamenetzky2015}), as would adding populations of high-$z$ normal galaxies \citep[e.\/g.\/,][]{tacconi2013} to the analysis. Given the relatively small differences in fluxes that likely arise due to the different wavelength ranges used to calculate $L_{\rm FIR}$ for the two low-$z$ samples, and that a larger Schmidt-Kennicutt index arises only when using the \cite{yao2003} sample, we find it unlikely that difference in $FIR$ definitions dominates the change in Schmidt-Kennicutt indices. More reliable stellar mass measurements may also help determine if there are truly different star formation mechanisms; \citet{sargent2014} find that true outliers in star formation efficiency (SFE) largely correspond to outliers in specific star formation rate relative to the (redshift-appropriate) galaxy main sequence.

\begin{figure*}
\epsscale{1}
\plottwo{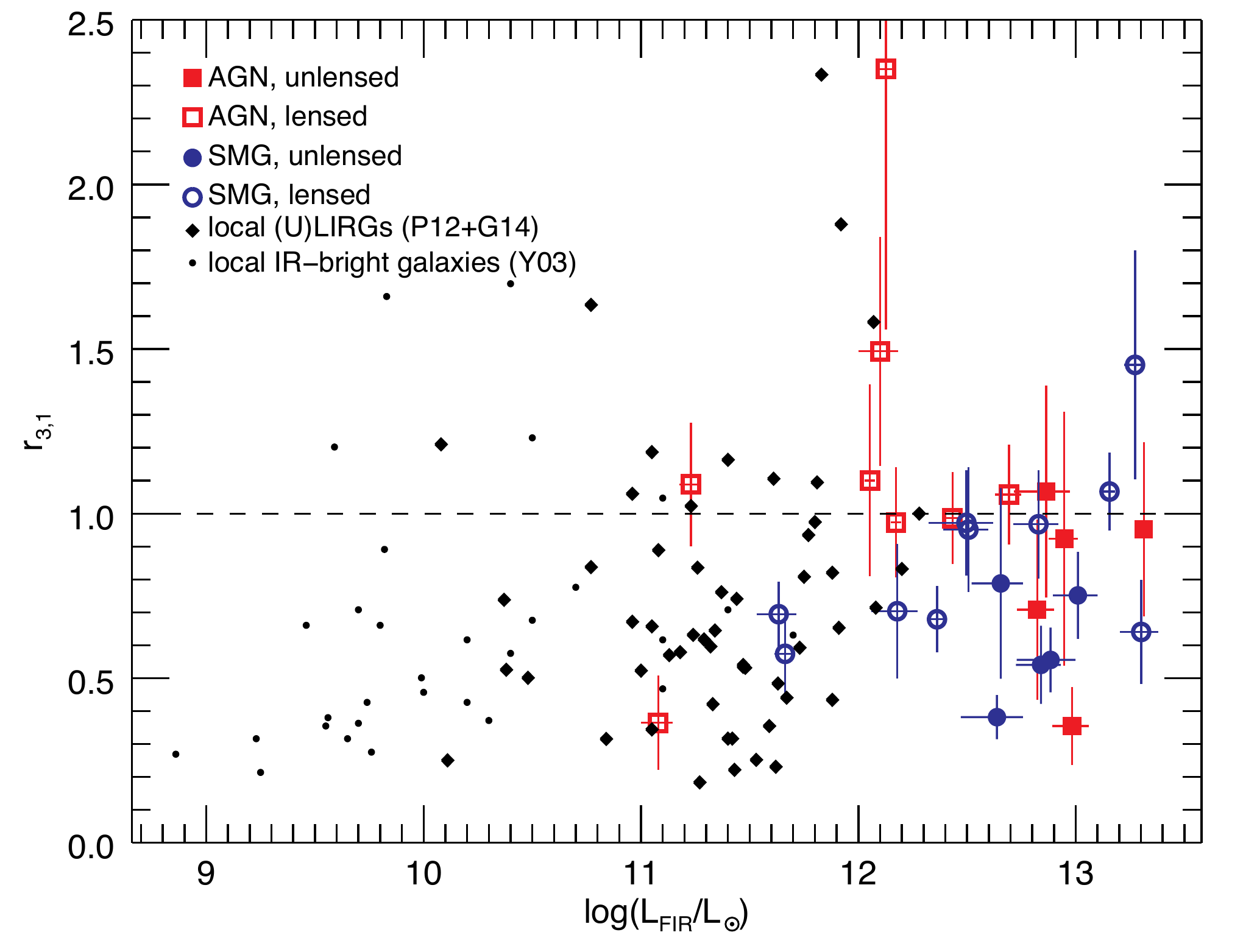}{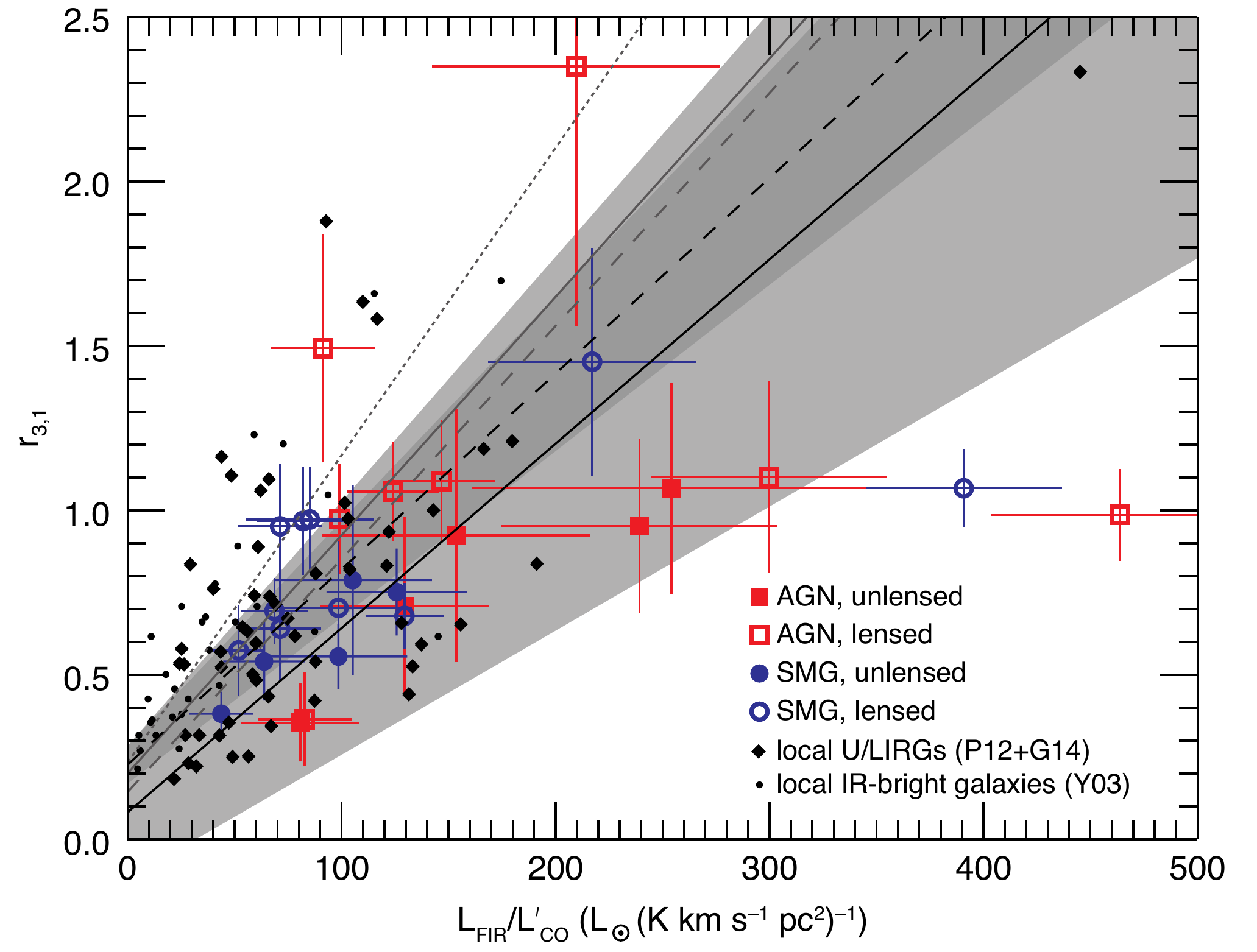}
\caption{The \mbox{CO(3--2)}/\mbox{CO(1--0)} line ratio as a function of the FIR luminosity (left) and a proxy for the star formation efficiency, $L_{\rm FIR}/L^\prime_{\rm CO}$ (right), for AGN host galaxies (red squares), SMGs (blue circles), low-$z$ U/LIRGs from \citet{papadopoulos2012d}/\citet{greve2014} (black diamonds), and low-$z$ IR-bright galaxies from \citet{yao2003} (black dots). Gravitationally lensed sources have hollow symbols while unlensed objects have solid symbols. On the left we show the $r_{3,1}=1$ line for clarity (dashed). We find no correlation between the FIR luminosity and $r_{3,1}$ at the $\alpha=0.05$ significance level. We evaluate the Spearman's and Kendall's rank correlation coefficients for the sub-populations for $r_{3,1}$ vs.~$L_{\rm FIR}/L^\prime_{\rm CO}$ and show the best linear fits to correlations that are significant at the $\alpha<0.05$ level: the high-$z$ populations together (black solid line with $\pm1\sigma$ fit uncertainties in the light shaded region; $\rho=0.67$, $p=9.3\times10^{-5}$; $\tau=0.48$, $p=3.2\times10^{-4}$), the high-$z$ and low-$z$ galaxies together (black dashed line with $\pm1\sigma$ fit uncertainties in the dark shaded region; $\rho=0.66$, $p=4.5\times10^{-15}$; $\tau=0.46$, $p\sim0$), the low-$z$ galaxies on their own (gray solid line; $\rho=0.63$, $p=1.5\times10^{-10}$; $\tau=0.44$, $p\sim0$), the local U/LIRG sample from \citet{papadopoulos2012d}/\citet{greve2014} on their own (gray dashed line; $\rho=0.54$, $p=02.4\times10^{-5}$; $\tau=0.37$, $p=7.5\times10^{-5}$), and the \citet{yao2003} sample of local IR-bright galaxies on their own (gray dotted line; $\rho=0.82$, $p=2.4\times10^{-8}$; $\tau=0.65$, $p=4.8\times10^{-7}$). MG\,0414+0534, which only has limits on the CO(1--0) emission, is excluded. \label{fig:r31sf}}
\end{figure*}

While there is theoretical motivation for the Schmidt-Kennicutt index to change with the excitation of the gas phase tracer \citep[e.\/g.\/,][]{narayanan2008d,narayanan2011}\footnote{However, these theoretical predictions are for measurements of the gas and star formation \emph{surface densities}, and we only analyze total luminosities here.}, the observed lack of difference in the index of the integrated Schmidt-Kennicutt relation when using \mbox{CO(1--0)} and \mbox{CO(3--2)} as the tracer is consistent with some previous analyses \citep[e.\/g.\/,][]{greve2014}, but not all (e.\/g.\/, \citealt{yao2003, bayet2009}; see also \citealt{kamenetzky2015}). The lack of excitation difference is best demonstrated by looking for a correlation between the FIR luminosity and the low-$J$ CO excitation (Figure~\ref{fig:r31sf}). As seen previously for the low-$z$ IR-bright galaxies \citep[e.\/g.\/,][]{mauersberger1999, yao2003, mao2010}, we find no correlation between the FIR luminosity and $r_{3,1}$ (although see \citealt{kamenetzky2015}). A likely explanation for the lack of trend is that in many cases both CO lines are largely tracing the same physical regions that emit in the infrared, and the sources with a strong spatial division between the multi-phase components of the ISM may add additional scatter. 

The only significant trend we do find with excitation and the galaxies' star-forming properties is with SFE as traced by $L_{\rm FIR}/L^\prime_{\rm CO(1-0)}$ (Figure~\ref{fig:r31sf}). We evaluate the Spearman's and Kendall's rank correlation coefficients for the SMGs, AGN, all high-$z$ sources, the two low-$z$ populations, and all sources together. With the exception of the high-$z$ AGN-host galaxies evaluated on their own, for all other populations and combinations of populations we find a likelihood of $p<0.05$ of getting correlations as strong as observed given the null hypothesis of no correlation between $r_{3,1}$ and SFE (the AGN evaluated separately have $p>0.05$). In fact, the correlations are highly significant and we find a $p\lesssim0.0003$ likelihood of finding a correlation at least as strong as these given the null hypothesis of no correlation, except for the SMGs evaluated on their own (which give $\rho=0.64$/$p=0.010$ and $\tau=0.49$/$p=0.012$ for the Spearman and Kendall rank correlation coefficients, respectively). We perform an orthogonal least squares bisector fit to the high-$z$ galaxies in Figure~\ref{fig:r31sf} and find a slope of $(5.6\pm1.8)\times10^{-3}$ and intercept of $0.08\pm0.20$ (neglecting lower limits). Combining both the high-$z$ and low-$z$ sources, we find a slope of $(5.9\pm0.9)\times10^{-3}$ and intercept of $0.23\pm0.06$ (neglecting lower limits), which is consistent with the best-fit line for high-$z$ sources. We note that the fit to our data is unweighted since we want to make a fair comparison between the samples and do not have uncertainties on the SFEs for all low-$z$ galaxies. However, given that the uncertainties are correlated and heteroscedastic, the fit is somewhat biased towards the more uncertain values of large $r_{3,1}$ and large SFEs. Excluding the outlier sources with $r_{3,1}>1.4$ decreases the slope and increases the offset, but the fit uncertainties still have considerable overlap when considering either the high-$z$ galaxies or galaxies at all redshifts combined.

The correlation between $r_{3,1}$ and SFE in Figure~\ref{fig:r31sf} is in line with the theoretical results of \citet{narayanan2014}. They find that the CO spectral line energy distributions of galaxies can be described as a function of a single parameter, the star-formation rate surface density, where higher excitation molecular gas states were achieved with higher star formation surface densities. The physical explanation for their correlation is that higher SFR density regions provide more sources of energy to inject into the molecular gas in that volume. Similarly, if there is a higher SFR per unit of molecular gas mass (i.\/e.\/, a larger $L_{FIR}/L^\prime_{\rm CO}$ ratio), then we would expect to get larger fractions of the molecular gas in higher excitation states (i.\/e.\/, larger $r_{3,1}$ values). The correlation between $r_{3,1}$ and SFE might also imply that dense molecular gas (which naturally produces larger $r_{3,1}$ values since a larger fraction of the gas is above the critical density necessary to excite CO into the $J=3$ state) is the component of the molecular ISM actively forming stars, and therefore has a larger SFR per unit of molecular gas than low-density gas (where there is additional mass not involved in star formation). In principle, this correlation indicates that there is in fact an excitation-dependent offset in our measured Schmidt-Kennicutt relation, but it is buried in the scatter of the relation (or perhaps is hidden by errors in the assumed magnification factors) so we cannot identify the offset directly. 

We compare the best-fit linear relations between the SFE and $r_{3,1}$ for the high-$z$ and low-$z$ galaxy correlations in Figure~\ref{fig:r31sf} and find that the slope of the \citet{yao2003} low-$z$ IR-bright correlation is marginally inconsistent with those involving the high-$z$ sources. While the offsets are consistent, we find slopes of $0.0093\pm0.0012$ for the \citet{yao2003} low-$z$ IR-bright galaxies, $0.0034\pm0.0012$ for the SMGs, and $0.0056\pm0.0018$ for the combined high-$z$ galaxies. As seen in the Schmidt-Kennicutt relation, the high-$z$ galaxies appear to have higher SFEs than in local galaxies, however that would result in a difference offset in the $r_{3,1}$-SFE correlation, not a difference in slope. The difference in slope implies that there is less excitation per unit increase of SFE for high-$z$ galaxies than for the low-$z$ galaxies. This could arise due to the difference in slope in the Schmidt-Kennicutt relation between the two populations (when evaluated separately), or it could be a resolution effect if the low-$z$ IR-bright galaxies have a significant extended cold-gas component traced by \mbox{CO(1--0)} that was not accounted for with the primary beam correction applied in \citet{yao2003}.

We find that the average gas consumption time scale (the inverse of the SFE) for SMGs and AGN host galaxies are broadly consistent \footnote{We assume a CO-to-${\rm H_2}$ conversion factor of $\alpha_{\rm CO}=0.8\,{\rm M_\sun\,(K\,km\,s^{-1}\,pc^2)^{-1}}$ and determine the SFRs from $L_{\rm FIR}$ using the equations from \citet{kennicutt2012} (note that we do not correct to the $3$-$1100\,{\rm \mu m}$ $L_{\rm TIR}$ used there).}; we obtain $\tau=38\pm18\,{\rm Myr}$ for AGN host galaxies, $\tau=63\pm28\,{\rm Myr}$ for SMGs, and $\tau=52\pm27\,{\rm Myr}$ for both populations combined. While the uncertainties for the SFRs are inhomogeneously determined and perhaps underestimated, if we trust those uncertainties and calculate weighted average properties of the two populations, we find $\tau=17\pm1\,{\rm Myr}$ for AGN host galaxies, $\tau=20\pm1\,{\rm Myr}$ for SMGs, and $\tau=18.4\pm0.9\,{\rm Myr}$ for both populations combined. 

Mid-IR excesses are known to arise in the dusty tori of AGN \citep[e.\/g.\/,][]{weiss2003,beelen2006}, and it is plausible that highly obscured AGN would have some FIR emission that does not arise from dust heated by star formation. \citet{leipski2013} found that among $1.2\,{\rm mm}$-detected $z>5$ quasars, $25$--$60\%$ of the FIR emission is associated with dust heating by star formation (as opposed to emission from a dusty torus surrounding the AGN). Since this cold dust heated by star formation dominates at the longest wavelengths, FIR luminosities determined from small numbers of photometric data points on the Rayleigh-Jeans tail of the cold dust peak more accurately trace SFRs than single-component fits to the full dust SED (which does not account for the warm/hot dust emission from the AGN; see also quasars without FIR detections in \citealt{leipski2014}). \citet{kirkpatrick2015} break down the IR SED of low- and high-$z$ U/LIRGs into warm and cold components, and find that warm dust component gets hotter and contributes a larger fraction to the total infrared luminosity with increasing AGN activity (as diagnosed by mid-IR PAH spectroscopy). Therefore, SED fits that do not account for two dust components will likely attribute too much IR flux to star formation. It is likely that the FIR measurements and implied SFRs for some galaxies in our sample are similarly affected, causing the slightly elevated SFEs observed in the AGN host galaxies. Uniformly obtained measurements of the FIR luminosity are necessary to accurately capture these galaxies' SFRs. 

\begin{figure}
\epsscale{1}
\plotone{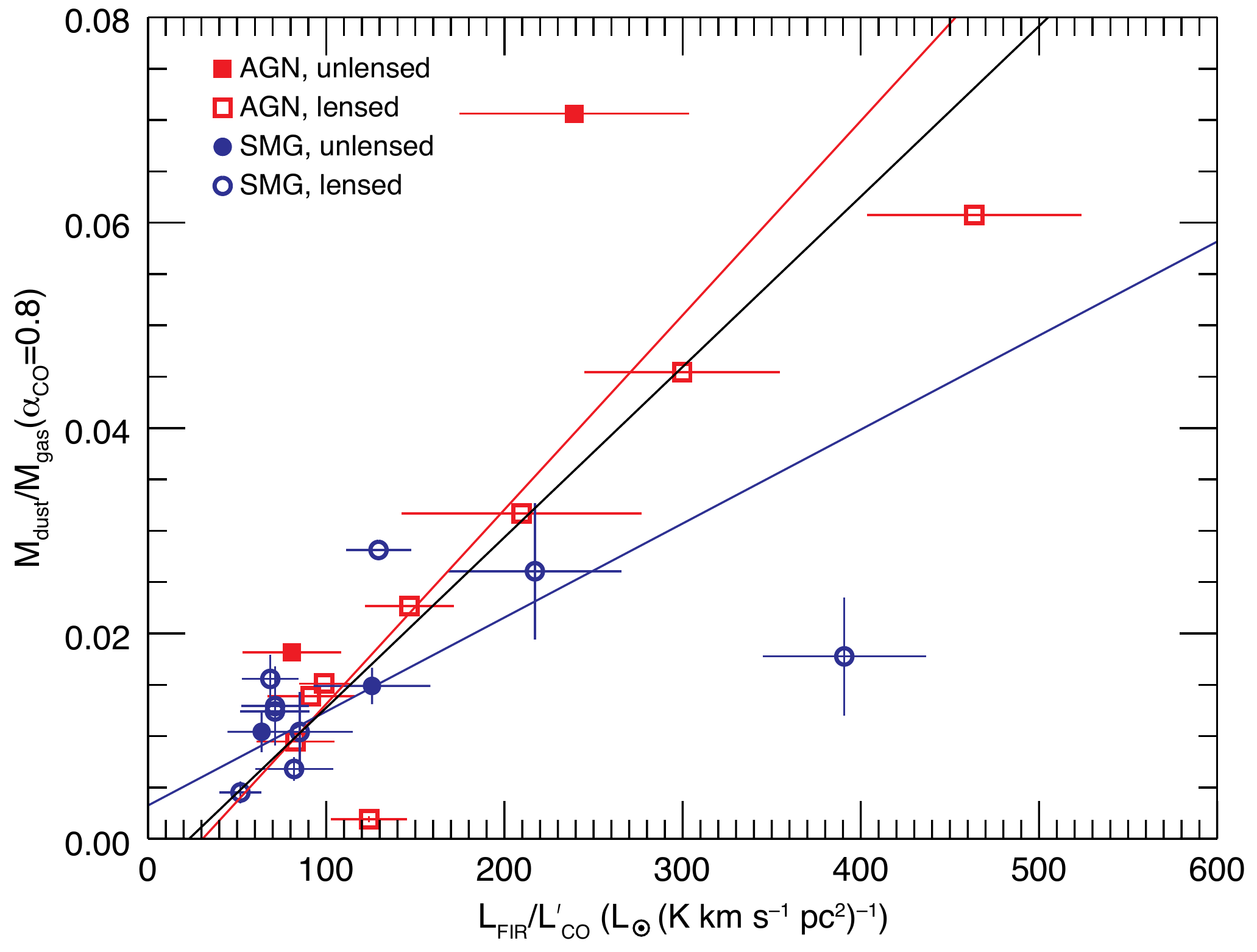}
\caption{The dust-to-gas mass ratio as function of $L_{\rm FIR}/L^\prime_{\rm CO}$, a proxy for the SFE, for the AGN host galaxies (red squares) and SMGs (blue circles) in our sample. Gravitationally lensed sources have hollow symbols while unlensed objects have solid symbols. We calculate the Spearman's and Kendall's rank correlation coefficients for  each population and find $p<0.05$ likelihood of obtaining a correlation for all. We show the best fit linear correlations for the AGN host galaxies (red line; $\rho=0.77$, $p=0.0092$; $\tau=0.60$, $p=0.015$), SMGs (blue line; $\rho=0.67$, $p=0.023$; $\tau=0.49$, $p=0.036$), and both populations combined (black line; $\rho=0.72$, $p=2.4\times10^{-4}$; $\tau=0.55$, $p=4.6\times10^{-4}$). For the dust-to-gas mass we assume a molecular gas mass conversion factor of $\alpha_{\rm CO }= 0.8\,{\rm M_\sun\,(K\,km\,s^{-1}\,pc^2)^{-1}}$. MG\,0414+0534, which only has limits on the CO(1--0) emission, is excluded.
\label{fig:d2gsfe}}
\end{figure}

We also evaluate SFE as a function of the dust-to-gas mass ratio in Figure~\ref{fig:d2gsfe}. We calculate the Spearman's and Kendall's rank correlation coefficients for the SMGs, the AGN host galaxies, and both populations combined, and for all cases we find $p<0.05$ likelihood of obtaining a correlation at least as strong as observed assuming the null hypothesis of no correlation between SFE and dust-to-gas ratio (for the SMGs, Kendall's rank correlation coefficient gives $p>0.05$). However, this correlation is not unexpected given that the gas mass is proportional to $L^\prime_{\rm CO}$ and the dust masses are either proportional to the same continuum flux measurements extrapolated to determine the FIR luminosity \citep[e.\/g.\/,][]{barvainis2002} or luminosity-weighted values since they are determined from fits to the spectral energy distribution \citep[e.\/g.\/,][]{magnelli2012}. Therefore, since the dust-to-gas mass ratios and SFEs are largely proportional to one another, we do not ascribe much significance to the correlation.

\subsection{$L^{\prime}_{\rm CO}$-FWHM correlation}

\citet{harris2012}, \citet{bothwell2013}, and \citet{goto2015} find a correlation between the CO line luminosity and the CO line FWHM for SMGs. However, \citet{aravena2016} does not find a significant correlation for their sample of dusty star forming galaxies, nor does \citet{carilli2013b} when evaluating a variety of different star-forming galaxy populations. A correlation between the line brightness and FWHM could occur if the CO line is tracing the gravitational potential occupied by the molecular gas, and especially if the molecular gas is dominating the dynamical mass. Some scatter would be expected in this correlation due to the unknown inclinations of any large-scale dynamical structures (such as disks), galaxies' merger states, or outflows (although emission from outflows is likely much weaker than the emission from the bulk of the gas). This correlation has some potential predictive power for determining the magnification for gravitationally lensed objects \citep{harris2012} or measuring cosmological distances for unlensed sources (if the scatter in the relation is reduced; \citealt{goto2015}). For strongly-lensed sources, differential lensing can also affect the shape of emission line profiles \citep{riechers2008a,sharon2016b} and thus the measured line FWHMs. 

\begin{figure*}
\epsscale{1}
\plottwo{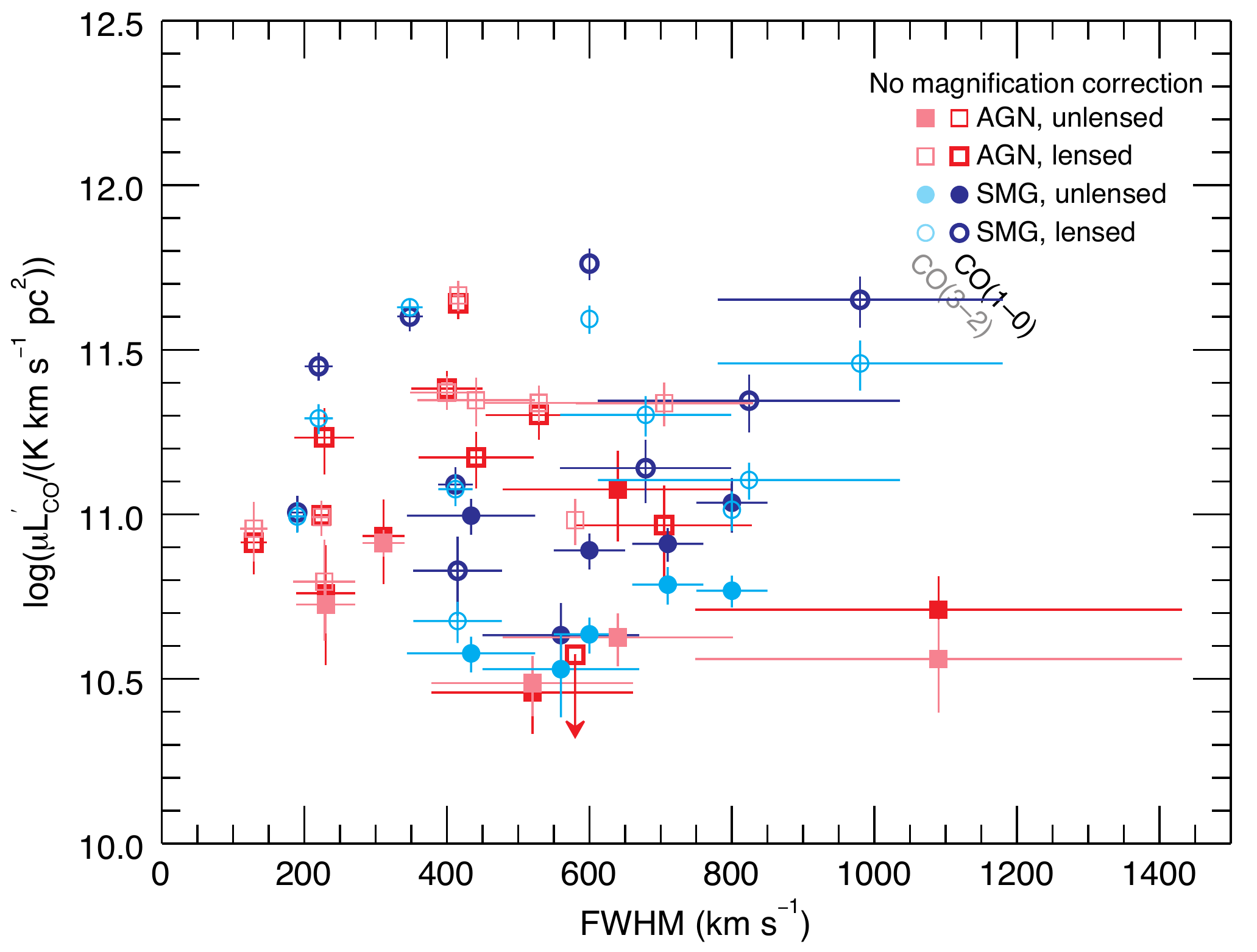}{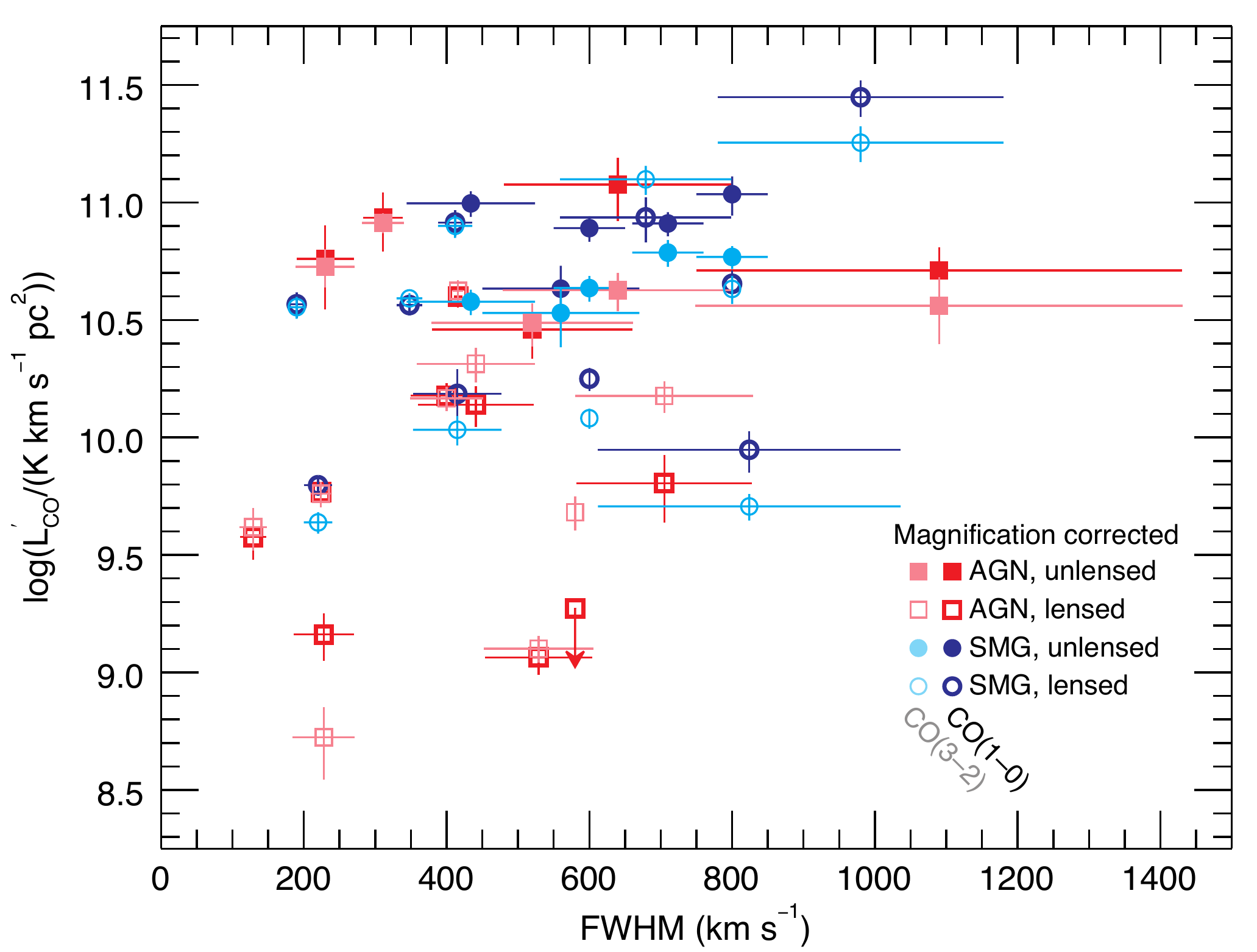}
\caption{The CO line luminosity as a function of line FWHM where the line luminosity has not been magnification corrected (left) and has been magnification corrected (right). AGN host galaxies are in red and SMGs are in blue; gravitationally lensed galaxies are hollow symbols and unlensed objects are solid. We show the \mbox{CO(1--0)} line luminosity in the darker shades of red/blue and the \mbox{CO(3--2)} line luminosity in lighter shades of red/blue. We calculate the Spearman's and Kendall's rank correlation coefficients and find no significant trend between the line luminosity and FWHM, regardless of CO line used, population evaluated, and magnification correction at a significance level of $\alpha=0.05$.\label{fig:COvsFWHM}}
\end{figure*}

We evaluate the possibility of correlation between the CO line luminosity and FWHM of our sample in Figure~\ref{fig:COvsFWHM}, and show both the gravitational lensing correct and uncorrected line luminosities. We calculate the Spearman's and Kendall's rank correlation coefficients (neglecting all upper limits) and find no significant trend between the line luminosity and FWHM, regardless of CO line used, population evaluated (i.\/e.\/ SMGs, AGN, or both together), and magnification correction (at a significance level of $\alpha=0.05$). The only exception is for the \mbox{CO(1--0)}-determined line luminosities for all galaxies (SMGs and AGN) with magnification corrections applied, where we find that the probability of finding a monotonic correlation between the variables at least as extreme as observed to be $p<0.05$ (assuming a null hypothesis of no correlation; Kendall's $\tau=0.30$, $p=0.024$; Spearman's $\rho=0.43$, $p=0.022$). However, this correlation is largely the product of the low-luminosity AGN host galaxy outliers, which have large uncertainties in their magnification factors \citep[e.\/g\/,][]{barvainis2002}, so we do not ascribe much significance to this result. When we compare the trend in FWHM with CO line luminosity, instead of finding a shallow correlation before lensing-correction and a steeper correlation after lensing correction \citep[e.\/g.\/,][]{harris2012}, we find that the lensed objects fill in a wedge-shaped region. The distribution of sources suggests that we are probing a wider range of intrinsic luminosities per FWHM than the {\it Herschel}-selected sample of \citet{harris2012}, likely due to the heterogeneity of our sample which is not selected via a brightness cutoff and may probe a variety of dynamical states (which is in line with the results of \citealt{carilli2013b}; see also \citealt{aravena2016}). One potential limitation of this analysis is that we use the same FWHMs for the analysis of both CO lines (due to the fact that our new \mbox{CO(1--0)} detections largely lack the required sensitivity to measure line shapes), and it may be that sources with a spatially extended cold molecular gas phase would have a larger FWHM for the \mbox{CO(1--0)} line than the \mbox{CO(3--2)} line (e.\/g.\/, \citealt{ivison2011,thomson2012}; for comparisons to higher-$J$ CO line widths, see also \citealt{hainline2006,riechers2011e}).

\subsection{Excitation and evidence for a SMG-quasar transition}

Much of the circumstantial evidence for the SMG-quasar transition is based on the similarity of their properties (their FIR luminosities and inferred star formation rates, molecular gas masses, and temporal coincidence) and the supposed analogous transition for low-$z$ U/LIRGs (which are more clearly major mergers) and quasars. These results also show broadly similar properties between SMGs and AGN host galaxies, including their molecular gas excitation and their trends in $r_{3,1}$ with different physical properties. Previous work demonstrating a difference in the $r_{3,1}$ ratio for SMGs and AGN host galaxies weaken with our expanded sample since we do not find a statistically significant difference, although tighter uncertainties on some objects would help solidify this result. If AGN are providing an additional higher excitation phase of molecular gas (rather than weaker forms of SMG-quasar transition where changes in gas excitation correlated with AGN activity due to some other evolutionary process), that phase may not be the bolometrically dominant part of the system (i.e., it may be a very small fraction of the total molecular gas mass). Such multi-phase models would result in measurements of global $r_{3,1}$ values near to those of the dominant star-forming gas phase, which can have a range in excitation conditions depending on the density of star formation relative to the molecular gas distribution. In addition, it is possible that dusty star-forming galaxies like SMGs contain a heavily enshrouded AGN, disguising the AGN's effects on the CO line ratios. Therefore the similarities in gas excitation may reinforce the similarities between the populations of SMGs and AGN host galaxies, but it does not reveal the evolutionary mechanisms that may connect the two populations as a temporal sequence.

The average $r_{3,1}\sim0.8$--$0.9$ we find for these populations reduces the uncertainties in the total molecular gas mass (on average) caused by observing in \mbox{CO(3--2)} to the $\sim10$--$20\%$ level, which is near typical flux calibration uncertainties and well within the typically observed scatter of analyses like the Schmidt-Kennicutt relation. However, there is still a wide distribution in $r_{3,1}$ values, and extrapolating total molecular gas masses from mid-$J$ CO lines may produce much larger errors for individual systems.

\begin{figure}
\epsscale{1}
\plotone{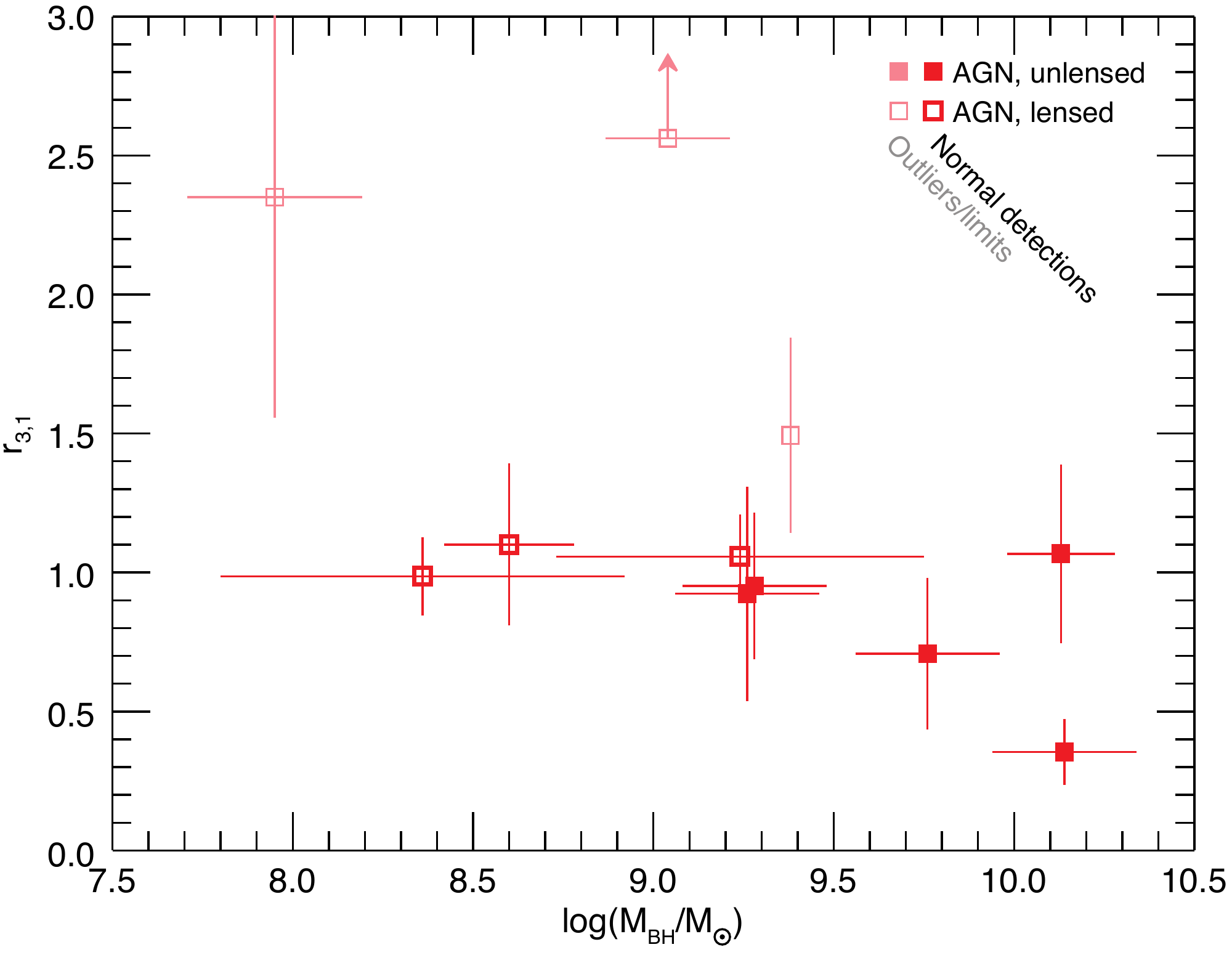}
\caption{The \mbox{CO(3--2)}/\mbox{CO(1--0)} line ratio as a function of the Black hole mass and the best-fit line (solid) excluding the outlier in $r_{3,1}$ and lower limits (shown in light red). Gravitationally lensed AGN host-galaxies are hollow symbols and unlensed objects are solid. We calculate the Spearman's and Kendall's rank correlation coefficients and find no correlation between the black hole mass and $r_{3,1}$ at the $\alpha=0.05$ significance level.
\label{fig:r31vsMBH}}
\end{figure}

Eleven of the AGN host galaxies in our sample also have black hole mass measurements (none of the SMGs have black hole measurements); we therefore look for correlations between the molecular gas properties and black hole masses for these systems. We find no correlation between the black hole mass and $r_{3,1}$ (Figure~\ref{fig:r31vsMBH}). Since SMGs have a similar range in $r_{3,1}$ values, we suspect  that including the lower mass black holes of the SMGs (relative to their total masses, which appear to be consistent between these populations to the extent we can determine from our unresolved sample; e.\/g.\/, \citealt{coppin2008,alexander2008}) would not produce any correlations. We find no clear difference in AGN types (i.e. optical vs. IR-bright, radio-loud vs. radio-quiet) with measured $r_{3,1}$ value or black hole mass.

Independent of the molecular gas excitation, there is a hint of higher SFEs (shorter gas depletion timescales) for the AGN compared to the SMGs. The higher SFEs appears to be due to larger average FIR luminosities for the AGN host galaxies over the same range of molecular gas masses in the SMGs, which may be due to a dusty torus contribution to the FIR luminosity that is not actually tracing star formation \citep[e.\/g.\/,][]{weiss2003,beelen2006,leipski2013,kirkpatrick2015}. If we force the linear fit between the FIR and CO luminosities in the Schmidt-Kennicutt relation to have the same index ($N=1$) for both SMGs and AGN host galaxies (in order to control for additional uncertainties in the index of the power law), the difference between the offsets measures the fraction of the FIR luminosity for the AGN host galaxies that is in excess of what would be predicted for pure starburst systems (the SMGs). Under these assumptions, on average $\sim10\%$ of the AGN host galaxies' FIR luminosity is associated with the central AGN rather than the star-forming gas. This fraction is lower than previous analyses of the Schmidt-Kennicutt relation \citep{riechers2011b} and analyses of the mid-IR emission from SMGs and AGN host galaxies \citep[e.\/g.\/,][]{pope2008,coppin2010}, but more uniform and reliable FIR luminosity measurements are necessary to make a fair comparison.

\section{Conclusions}
\label{sec:concl}

We present \mbox{CO(1--0)} observations obtained at the VLA for 14 $z\sim2$ galaxies, including 11 AGN host galaxies (six lensed and five unlensed) and three SMGs (two lensed and one unlensed), with existing \mbox{CO(3--2)} measurements. We successfully detect 13 objects (five of which are tentative) and obtain one new upper limit on the \mbox{CO(1--0)} line. We also detect continuum emission from ten galaxies. We combine this sample with an additional 15 $z\sim2$ galaxies from the literature that have both \mbox{CO(1--0)} and \mbox{CO(3--2)} measurements: three lensed AGN host galaxies and 12 SMGs (eight lensed and four unlensed). In contrast to previous work that showed a systematic difference between the \mbox{CO(3--2)}/\mbox{CO(1--0)} line ratios for SMGs and AGN host galaxies (which have $r_{3,1}\sim0.6$ and $r_{3,1}\sim1.0$, respectively; e.\/g.\/, \citealt{swinbank2010a, harris2010, ivison2011, riechers2011f}), we find no statistically significant difference between either the average $r_{3,1}$ values or distribution of $r_{3,1}$ values for SMGs and AGN host galaxies, obtaining a global average $r_{3,1}=0.90\pm0.40$ (or $r_{3,1}=1.03\pm0.50$ for AGN host galaxies and $r_{3,1}=0.78\pm0.27$ for SMGs, which are consistent with the previous values to within the uncertainty). The lack of a statistically significant difference between the distribution of $r_{3,1}$ values of the SMGs and AGN host galaxies is not as robust; switching the class of three SMGs which may have central AGN or using prior single-dish \mbox{CO(1--0)} measurements can return a significant offset between the distributions for the two populations. The likely disappearance of the $r_{3,1}$ differences is due to multiple factors, including revised \mbox{CO(1--0)} measurements and re-classifications of sources. The expanded sample likely also includes ``hybrid" or misclassified sources such as SMGs with buried AGN that have not been detected, or galaxies where the AGN may not be the dominant contributor to the emission at long wavelengths. High star formation rate densities relative to the molecular gas distribution could also produce high $r_{3,1}$ values \citep[e.\/g.\/,][]{narayanan2014}, broadening the observed excitation distribution. This result is in line with observations of different galaxy populations at low-$z$ (including normal galaxies, IR-bright galaxies, and/or U/LIRGs) or which show a range of $r_{3,1}$ values in star-forming systems both with and without central AGN \citep{mauersberger1999, yao2003, mao2010}.

Using this sample of matched \mbox{CO(1--0)}- and \mbox{CO(3--2)}-detected sources, we look for correlations between the low-$J$ CO excitation and other galaxy properties. We evaluate the integrated Schmidt-Kennicutt relation (the correlation between the $L_{\rm FIR}$ and $L^\prime_{\rm CO}$) for our sample and find no significant difference between the slope and offset of the relation when using the different CO lines, in line with some previous results from aggregated samples which did not have both lines for every object  \citep[e.\/g.\/,][]{greve2014}, although results in the literature are mixed \citep[e.\/g.\/,][]{yao2003, bayet2009,kamenetzky2015}. We find that the index of the integrated Schmidt-Kennicutt relation is consistent with $N\sim1$, even when we include low-$z$ U/LIRGs from \citet{papadopoulos2012d}/\citet{greve2014} in our analysis. However, when we include the more normal low-$z$ IR-bright sample from \citet{yao2003} in our analysis (which also has both \mbox{CO(1--0)} and \mbox{CO(3--2)} detections), the index increases to $N\sim1.2$; this index is similar to the low-$z$ objects on their own, indicating that they dominate the fit due to the large scatter in high-$z$ measurements. The similarity of the Schmidt-Kennicutt index for the two CO lines in our sample is more cleanly illustrated by the lack of correlation between $r_{3,1}$ and the FIR luminosity. Instead, we find that $r_{3,1}$ correlates with SFE, where sources with large SFE have large $r_{3,1}$ values, as found at low-$z$ in \citet{yao2003}. This is consistent with the theoretical results of \citet{narayanan2014}, where they find that the only effective predictor of CO excitation was the SFR surface density, indicating that it is density of energy sources per unit of molecular gas that dictates the gas excitation.

We also use this sample to further test observed correlations between the CO line FWHM and the line luminosity \citep[e.\/g.\/,][]{harris2012, bothwell2013, goto2015} which has potential predictive power for estimating the magnification by gravitational lensing \citep{harris2012}. We do not find a statistically significant correlation and find that our sample spans a larger range in line luminosities after correcting by the known lensing magnification, likely due to the inhomogeneous way in which our sample is selected when compared to \citet{harris2012}, and is in line with the results of \citet{carilli2013b} and \citet{aravena2016}.

Generally we find no significant difference between the molecular gas properties of the AGN and SMGs in our sample, reinforcing the similarities of these two high-$z$ populations. Confidence in our result would be improved with higher sensitivity \mbox{CO(1--0)} measurements for our marginal detections and uniformly measured values of the FIR luminosities (which are gathered from the literature). Since the lack of a dependence of $r_{3,1}$ on the galaxy type may be due to variations in the relative contributions of star-forming regions and AGN tori to the long wavelength properties of these galaxies, it may be that correctly accounting for both heating mechanisms will retrieve the original trend in $r_{3,1}$ values (since high CO excitation is known to originate near AGN), but with larger scatter for galaxies lacking AGN since high star formation densities can also drive high $r_{3,1}$ values.

\acknowledgments{We thank the anonymous referee for their helpful comments. The National Radio Astronomy Observatory is a facility of the National Science Foundation operated under cooperative agreement by Associated Universities, Inc.}

{\it Facilities:} \facility{VLA}

\bibliographystyle{apj}

\end{document}